\documentclass[a4paper,10pt]{article}
\pdfoutput=1 

\usepackage{jheppub} 

\usepackage[T1]{fontenc} 
\usepackage{lmodern} 

\usepackage{xcolor}
\usepackage{mathrsfs,wasysym}
\usepackage{booktabs}
\usepackage{slashed}
\usepackage{cancel}

\DeclareMathOperator{\csch}{csch}

\newcommand{\kt}{k_{\rm t}}

\newcommand{\MSbar}{\overline{\text{MS}}}
\newcommand{\sss}{\scriptscriptstyle}
\newcommand{\C}   {c}
\newcommand{\Cm}  {\widetilde{c}}
\newcommand{\DC}  {\Delta \C}
\newcommand{\DCm} {\Delta \Cm}
\newcommand{\DkC} [1]{\Delta_#1 \C}
\newcommand{\DkCm}[1]{\Delta_#1 \Cm}
\newcommand{\nf}{{[n_f]}}
\newcommand{\nfo}{{[n_f+1]}}
\newcommand{\as}{\alpha_s}
\newcommand{\ashat}{\hat\alpha_s}
\newcommand{\asb}{\bar\alpha_s}

\newcommand{\Ord}{\mathcal{O}}

\newcommand{\Mell}{\mathcal{M}}

\newcommand{\mz}{m_Z}
\newcommand{\mheavy}{m_{q_{n_f+1}}}
\newcommand{\muf}{\mu_{\sss\rm F}}
\newcommand{\mur}{\mu_{\sss\rm R}}

\newcommand{\Li}{\mathrm{Li}}

\newcommand{\abs}[1]{\left| #1 \right|}

\let\originalleft\left
\let\originalright\right
\renewcommand{\left}{\mathopen{}\mathclose\bgroup\originalleft}
\renewcommand{\right}{\aftergroup\egroup\originalright}

\def\beq{\begin{equation}}  
\def\eeq{\end{equation}}
\def\({\left(}
\def\){\right)}
\def\[{\left[}
\def\]{\right]}

\allowdisplaybreaks

\definecolor{tpurple}{RGB}{128,0,128}
\definecolor{darkgreen}{RGB}{0,200,0}
\definecolor{dark}{RGB}{200,0,0}

\title{\boldmath Towards  parton distribution functions with small-$x$ resummation: HELL 2.0}

\author[a]{Marco Bonvini,}
\affiliation[a]{Dipartimento di Fisica, Sapienza Universit\`a di Roma and INFN, Sezione di Roma 1,\\ Piazzale Aldo Moro~5, 00185 Roma, Italy}
\author[b]{Simone Marzani}
\affiliation[b]{Dipartimento di Fisica, Universit\`a di Genova and INFN, Sezione di Genova,\\ Via Dodecaneso 33, 16146, Italy}
\author[c]{and Claudio Muselli}
\affiliation[c]{Tif Lab, Dipartimento di Fisica, Universit\`a di Milano and INFN, Sezione di Milano,\\ Via Celoria 16, 20133 Milano, Italy}
\preprint{\href{https://arxiv.org/abs/1708.07510}{1708.07510}}

\emailAdd{marco.bonvini@roma1.infn.it}
\emailAdd{simone.marzani@ge.infn.it}
\emailAdd{claudio.muselli@mi.infn.it}

\abstract{
In global fits of parton distribution functions (PDFs) a large fraction of data points, mostly from the HERA collider,
lies in a region of $x$ and $Q^2$ that is sensitive to small-$x$ logarithmic enhancements.
Thus, the proper theoretical description of these data requires the inclusion of small-$x$ resummation.
In this work we provide all the necessary ingredients to perform a PDF fit to deep-inelastic scattering (DIS) data
which includes small-$x$ resummation in the evolution of PDFs and in the computation of DIS structure functions.
To this purpose, not only we include the resummation of DIS massless structure functions, but we also consider the production
of a massive final state (e.g.\ a charm quark), and the consistent resummation of mass collinear logarithms through the
implementation of a variable flavour number scheme at small $x$.
As a result, we perform the small-$x$ resummation of the matching conditions in PDF evolution at heavy flavour thresholds.
The resummed results are accurate at next-to-leading logarithmic (NLL) accuracy and matched, for the first time, to next-to-next-to-leading order (NNLO).
Furthermore, we improve on our previous work by considering a novel all-order treatment of running coupling contributions. 
These results, which are implemented in a new release of \texttt{HELL}, version \texttt{2.0},
will allow to fit PDFs from DIS data at the highest possible theoretical accuracy, NNLO+NLL,
thus providing an important step forward towards precision determination of PDFs and consequently precision phenomenology at the LHC and beyond.

}

\begin{document}

\maketitle
\flushbottom

\section{Introduction}

The outstanding quality of the CERN Large Hadron Collider (LHC) data continuously challenges the particle physics theoretical community to perform refined calculations with uncertainties comparable to the ones of the experimental results. Consequently, perturbative predictions for LHC processes nowadays often include radiative corrections in QCD at next-to-next-to-leading order (NNLO) and, in some cases, N$^3$LO~\cite{Anastasiou:2015ema,Dreyer:2016oyx}.
These remarkable calculations have a tremendous impact on many aspects of LHC phenomenology.
This can be seen, for instance, in the context of the determination of parton distribution functions (PDFs),
where the inclusion in the fit of data points describing the transverse momentum of Drell-Yan lepton pairs~\cite{Boughezal:2017nla}
or various differential distributions in top pair production~\cite{Czakon:2016olj}
has recently become possible thanks to the existence of fully differential calculations at NNLO.
Moreover, the recent completion of the NNLO QCD corrections to jet production~\cite{Currie:2016bfm,Currie:2017eqf},
paves the way for global determinations of parton densities which are truly NNLO. 
On the other hand, it is well known that fixed-order calculations fail to provide reliable results in regions of phase space which are characterized by the presence of two or more disparate energy scales. In such cases, large logarithmic contributions appear to any order in perturbation theory, and they must be accounted for to all orders. 
Resummed calculations have also seen remarkable progress in recent years, and for many observables the resummation of next-to-next-to-leading (NNLL) logarithms of soft-collinear origin has become standard, even reaching N$^3$LL for a few observables~\cite{Moch:2005ba,Bonvini:2016frm,Bonvini:2014joa,Catani:2014uta,Bizon:2017rah,Abbate:2010xh,Hoang:2015hka,Bonvini:2014tea,Ahmed:2015qda,Ahmed:2016otz}. 

Traditionally, resummation is not included in the calculations that are employed in PDF fits with the argument that the observables which are considered are rather inclusive and therefore not much sensitive to logarithmic enhancements. However, the LHC is exploring a vast kinematic region in both the momentum transferred $Q^2$ and the Bjorken variable $x=Q^2/S$, where $\sqrt{S}$ is the centre-of-mass energy. It is therefore important to assess the role of logarithmic corrections both in the small-$x$, i.e.\ high-energy, regime and at large $x$, i.e.\ in the threshold region. For instance, the production of a lepton pair via the Drell-Yan mechanism, which is measured by the LHCb collaboration in the forward region and at low values of the leptons' invariant mass, probes values of $x$ down to $10^{-5} \div10^{-6}$. At the other end of the spectrum, searches for new resonances at high mass are sensitive to PDFs in the $x\sim10^{-1}$ region.

In the past, some of us included threshold resummation in PDF fits~\cite{Bonvini:2015ira} (see also Ref.~\cite{Corcella:2005us}) and performed dedicated studies that included threshold resummation in both coefficient functions and PDFs  in the context of the production of heavy supersymmetric particles~\cite{Beenakker:2015rna}.
It has to be noted that the inclusion of threshold resummation in PDF fits did not pose particular challenges because in the widely used $\MSbar$ scheme the splitting functions that govern DGLAP evolution are not enhanced at large $x$~\cite{Korchemsky:1988si,Albino:2000cp}, and the effect of threshold resummation can be included through a $K$-factor. The situation is radically different if we consider small-$x$ resummation, where both coefficient functions and splitting functions receive single-logarithmic corrections to all orders in perturbation theory.

High-energy resummation of PDF evolution is based on the BFKL equation~\cite{Lipatov:1976zz,Fadin:1975cb,Kuraev:1976ge,Kuraev:1977fs,Balitsky:1978ic,Fadin:1998py}.
However, the correct inclusion of LL and NLL corrections to DGLAP splitting functions is far from trivial. 
This problem received great attention in 1990s and early 2000s by various groups, see Refs.~\cite{Salam:1998tj,Ciafaloni:1999yw,Ciafaloni:2003kd,Ciafaloni:2003rd,Ciafaloni:2007gf}, Refs.~\cite{Ball:1995vc,Ball:1997vf,Altarelli:2001ji,Altarelli:2003hk,Altarelli:2005ni,Altarelli:2008aj}, and Refs.~\cite{Thorne:1999sg,Thorne:1999rb,Thorne:2001nr,White:2006yh} (for recent work in the context of effective theories, see~\cite{Rothstein:2016bsq}).
High-energy resummation of partonic cross sections is based on the so-called $\kt$-factorization theorem~\cite{Catani:1990xk,Catani:1990eg,Collins:1991ty,Catani:1993ww,Catani:1993rn,Catani:1994sq,Ball:2007ra,Caola:2010kv}, which has been used to compute high-energy cross sections for various processes:
deep-inelastic scattering (DIS)~\cite{Catani:1994sq},
heavy quark production~\cite{Ball:2001pq},
direct photon production~\cite{Diana:2009xv,Diana:2010ef},
Drell-Yan~\cite{Marzani:2008uh},
and Higgs production~\cite{Hautmann:2002tu,Marzani:2008az,Caola:2011wq}.
The formalism has been subsequently extended to rapidity~\cite{Caola:2010kv} and transverse momentum distributions~\cite{Forte:2015gve,Marzani:2015oyb}.
Despite the wealth of calculations, it has proven very difficult to perform resummed phenomenology. Recently, some of us overcame these difficulties and developed a framework and a public code named \texttt{HELL} (High-Energy Large Logarithms)~\cite{Bonvini:2016wki}, which is based on the formalism developed by Altarelli, Ball and
Forte (ABF)~\cite{Ball:1995vc,Ball:1997vf,Altarelli:2001ji,Altarelli:2003hk,Altarelli:2005ni,Altarelli:2008aj}, but does contain significant improvements.

In this paper we further improve on our recent work, with the goal of providing  all the necessary theoretical ingredients and numerical tools to perform a PDF fit
which includes small-$x$ resummation in both PDF evolution and in DIS partonic coefficient functions,
including the correct treatment of the transitions at the heavy flavour thresholds, at NLL accuracy matched, for the first time, to NNLO. 
This is important because DIS data still represent the backbone of any PDF fits, and HERA data~\cite{Abramowicz:2015mha} do explore the small-$x$ region.
Achieving this target requires two ingredients which are the main results of this work.
The first is the construction of a so-called ``variable flavour number scheme'' at the resummed level~\cite{White:2006xv},
which provides coefficient functions with power-behaving mass effects and resummation of mass collinear logarithms,
as well as describing the transition of PDF evolution at heavy quark thresholds.
The second is the matching of the NLL resummed splitting functions to their fixed-order counterparts, which is realized for the first time up to NNLO.
This requires the expansion in power of $\as$ of the resummed result, which proved to be non-trivial because of the way the resummed kernels are constructed.
We also correct an error present in the our previous paper~\cite{Bonvini:2016wki} that we inherited from the original ABF work~\cite{Altarelli:2005ni}, which has however a small phenomenological impact.
More importantly, we derive a new way to perform the resummation of running coupling effects, which streamlines the construction of the resummed evolution kernels and solves an issue in the construction of its $\as$ expansion.
All these improvements are included in a new release of \texttt{HELL}, version \texttt{2.0}, which is publicly available at the webpage
\href{https://www.ge.infn.it/~bonvini/hell/}{\texttt{www.ge.infn.it/$\sim$bonvini/hell}}.

We point out that the investigation of the impact of the resummation on the NNLO fixed-order splitting functions is potentially of great phenomenological interest. It is well known that, due to accidental zeros, the effect of small-$x$ logarithms in the evolution is mild at NLO
but stronger at NNLO, and will be even stronger (with two extra logarithms) at N$^3$LO. On top of this, comparison of theoretical predictions with experimental data suggests that fixed-order NLO theory
describes the small-$x$ region better than NNLO, see e.g.~\cite{Caola:2009iy}.
Indeed, the final resummed anomalous dimensions tend to have a shape which is somewhat closer to NLO than to NNLO, as the small-$x$ growth is greatly reduced once resummation is performed, see e.g.~\cite{Dittmar:2005ed}.
Therefore, having the possibility of using NNLO theory stabilized at small $x$ with high-energy resummation can provide a significant improvement in the description of the data.
This is indeed observed in preliminary applications of our results~\cite{Rottoli:2017ifw}.
Furthermore, a reliable theory of DIS at low $x$ finds interesting applications beyond collider physics. For instance, it is a key ingredient in the description of ultra high-energy neutrinos from cosmic rays, see e.g.~\cite{CooperSarkar:2011pa,Gauld:2015kvh,Gauld:2015yia}.

This paper is organized as follows. In Sect.~\ref{sec:dis} we discuss the implementation of a variable flavour number scheme
in the context of small-$x$ resummation, while the details of resummation of DGLAP evolution and its expansion are collected
in the two following sections.
Specifically, in Sect.~\ref{sec:RCnew} we present our new realization of the resummation of running coupling contributions,
and later in Sect.~\ref{sec:expansion} we perform the perturbative expansion of the various ingredients and construct our final
resummed and matched splitting functions.
We present our numerical results in Sect.~\ref{sec:results}, before concluding in Sect.~\ref{sec:conclusions}. In Appendix~\ref{sec:DISoffshell} we provide analytical results for the off-shell DIS partonic cross section with mass dependence,
while in Appendix~\ref{sec:appRes} we collect some technical details on the actual implementation of the resummation in \texttt{HELL}.

\section{Resummation of deep-inelastic scattering structure functions}\label{sec:dis}

The standard way of describing  the deep-inelastic scattering of an electron off a proton is to express the cross section
in terms of structure functions that depend on the Bjorken variable $x$ and the momentum transfer $Q^2$,
\beq \label{eq:struc_func_def}
F_a(x,Q^2)= x \sum_{i} \int_x^1  \frac{dz}{z}\, C_{a,i}\(\frac{x}{z}, \frac{m_c^2}{Q^2}, \frac{m_b^2}{Q^2},\frac{m_t^2}{Q^2},\as \) f_{i} (z,Q^2),
\eeq
where sum runs over all active flavours, i.e.\ all the partons for which we consider a parton density.
The number of active partons depends on the choice of factorization scheme and will be discussed in Sect.~\ref{sec:HQsmallx}.
Note that the coefficient functions also depend on the charges of the quarks that strike the off-shell boson (left understood),
as well as on the heavy-quark masses, as indicated.
For simplicity, in the above equation, we have set both renormalization and factorization scales equal to the hard scale, $\mur^2=\muf^2=Q^2$, so $\as=\as(Q^2)$.
Finally, the index $a$ denotes the type of structure function under consideration.
In our case, we will mostly consider either $F_2$ or the longitudinal structure function $F_L$.
Note that when $Q^2 \sim \mz^2$, we also have a non-negligible contribution from the parity-odd structure function $F_3$.
This contribution is not logarithmically enhanced at small-$x$ in the massless case~\cite{Catani:1994sq};
this remains true for neutral current DIS with massive quarks,
however, in charged current DIS with massive quarks a small-$x$ logarithmic enhancement appears.
To our knowledge, the resummation of these logarithms in $F_3$ was never considered so far.

In order to diagonalize the convolution in Eq.~\eqref{eq:struc_func_def} we consider Mellin moments of the structure functions
\beq
F_a(N,Q^2) \equiv \int_0^1 dx\, x^{N-1} F_a(x,Q^2)= \sum_i C_{a,i}\(N, \frac{m_c^2}{Q^2}, \frac{m_b^2}{Q^2},\frac{m_t^2}{Q^2},\as \) f_i(N,Q^2),
\eeq
where, as often done in the context of small-$x$ resummation, we have introduced a non-standard definition for the moments of the coefficient functions and of the parton densities:
\begin{subequations}
\begin{align}
C_{a,i}\(N, \frac{m_c^2}{Q^2}, \frac{m_b^2}{Q^2},\frac{m_t^2}{Q^2},\as \)
&= \int_0^1 d z\,  z^{N}\, C_{a,i}\(z, \frac{m_c^2}{Q^2}, \frac{m_b^2}{Q^2},\frac{m_t^2}{Q^2},\as \), \\
f_i\(N, Q^2 \)&= \int_0^1 d z\,  z^{N}\, f_i\(z, Q^2 \).
\end{align}
\end{subequations}
The last equation implies that the DGLAP anomalous dimensions are defined as
\beq \label{eq:an_dim}
\gamma_{ij}(N,\as)=  \int_0^1 d z\,  z^{N} \, P_{ij}\(z ,\as\),
\eeq
where $P_{ij}$ are the usual Altarelli-Parisi splitting functions. In particular, in momentum space, the leading logarithmic (LL) behaviour at small-$z$ to any order $n>0$ in perturbation theory is $\as^n P^{(n-1)}_{ij}\sim \as^n \frac{1}{z} \ln^{n-1}\frac{1}{z}$. In Mellin space these logarithms are mapped into poles in $N=0$, which results in the following LL behaviour for the anomalous dimension: $\as^n \gamma^{(n-1)}_{ij}\sim \( \as/N\)^n$. In practice, not every entry of the anomalous dimension matrix is LL at small-$x$.
On the other hand, the behaviour of the DIS coefficient functions in Mellin space is $\as^n C^{(n)}_{a,i}  \sim \as \( {\as}/{N}\)^{n-1}$, i.e.\ the enhancement is at most next-to-leading logarithmic (NLL). Note that some care has to be taken when considering the LO contribution $C^{(0)}_{a,i} $, which is an $\Ord(\as^0)$ constant in Mellin space, and hence formally LL.

In this paper we construct the NLL resummation of DIS structure functions at small-$x$. In order to achieve this goal, we resum the first two towers of logarithmic contributions to the splitting functions, while we consider only the first non-vanishing tower of logarithmic contributions to the partonic coefficient functions.
Note that in a previous work~\cite{Bonvini:2016wki} we called this NLL resummation in DIS coefficient functions just LL, underlining the fact that it is the leading non-vanishing logarithmic enhancement. We refer to the counting of this previous work as \emph{relative} logarithmic counting (i.e.\ relative to the leading non-vanishing logarithmic enhancement), while the one adopted here as \emph{absolute} counting (i.e.\ using the overall powers of $\as$ and $1/N$).

Henceforth, unless explicitly stated, we are going to work in Mellin space and leave the dependence on the $N$ variable, as well as the other variables, understood.
As common in studies of DIS, we perform a flavour decomposition.
For our purposes it is enough to separate the structure functions into a singlet and non-singlet component
\beq\label{eq:FDIS}
F_a= F_a^{\rm S}+ F_a^{\rm NS}.
\eeq
The singlet structure functions contain both gluon and quark (singlet) contributions
\beq
F_a^{\rm S} = C_{a,g} \, f_g + C^\text{S}_{a,q} \, f_S ,
\eeq
where 
\beq
f_S = \sum_{i=1}^{n_f}\[f_{q_i}+f_{\bar q_i}\],
\eeq
$n_f$ denoting the number of active quark flavours.
The non-singlet structure function instead reads, see e.g.~\cite{Catani:1994sq,Buza:1995ie},
\beq\label{eq:Fnonsinglet}
F_a^{\rm NS} = C^{\rm NS}_{a,q} \,\sum_{i=1}^{n_f}e_i^2\(f_{q_i} +f_{\bar q_i}-\frac1{n_f}f_S \),
\eeq
where $e_i$ is the electric charge of the quark $q_i$, i.e.\ its coupling to the photon.\footnote
{Here we assume that the DIS interaction is mediated by a photon, thus ignoring a possible
contribution from the $Z$ boson, or the charged-current case in which a $W$ is exchanged.
This choice makes the discussion here and in the following somewhat simpler, and allows to focus on the details of the resummation.
Most of what follows does not depend on this assumption, as small-$x$ resummation affects just the singlet:
in particular, the small-$x$ logarithmic content of the singlet and pure-singlet coefficients, Eq.~\eqref{eq:CPS}, is identical (for a discussion about small-$x$ enhancements in the non-singlet sector see, for instance, Ref.~\cite{Blumlein:1995jp}).
The generalization to a generic vector boson is rather straightforward, and will be presented in Sect.~\ref{sec:Vgen}.
}
Furthermore, we can collect the terms proportional to the singlet PDF, obtaining
\begin{align}\label{eq:FPS}
F_a &= C_{a,g}\, f_g + C^{\rm PS}_{a,q}\, f_S
+ C^{\rm NS}_{a,q}\, \sum_{i=1}^{n_f}e_i^2\(f_{q_i}+f_{\bar q_i}\),
\end{align}
where we have defined the so-called pure-singlet coefficient function,
\beq\label{eq:CPS}
C^{\rm PS}_{a,q} = C_{a,q}^\text{S} - \langle e^2 \rangle C^{\rm NS}_{a,q},
\eeq
being $\langle e^2 \rangle$ the average squared charge.

In this study we consider resummed structure functions matched to their fixed-order counterparts. To this purpose, we find useful to introduce resummed contributions, defined as the all-order results minus its expansion to the fixed-order we are matching to,
\beq \label{eq:delta_def}
\Delta_n C = C^\text{res}- \sum_{k=1}^n \as^k \, C^{\text{res},(k)},
\eeq
where we are going to typically consider matching to NLO and NNLO structure functions, i.e.\ $n=1,2$, although DIS structure functions are also known, in the massless case, to three loops~\cite{Moch:2004xu,Vermaseren:2005qc,Moch:2008fj}.
Moreover, many of the formulae involving the resummed contribution that we derive in what follows hold regardless of the order in perturbation theory we are matching to. For these cases we adopt a simplified notation that omits the index $n$: $\Delta_n C \to \Delta C$.
Note also that we will sometimes write the full resummed expression $C^{\rm res}$ as $\Delta_0 C$.

\subsection{Factorization schemes in presence of massive quarks}\label{sec:HQsmallx}

In the context of collinear factorization, mass collinear singularities due to massless quarks must be factorized, such that perturbative coefficient functions are finite.
This is the case for the up, down and strange quarks that we always consider massless. 
For massive quarks, i.e.\ charm, bottom and top, collinear singularities are regulated by the quark mass and they manifest themselves as logarithms of the ratio $Q^2/m^2$. 
Despite the fact that the factorization of these contributions is not necessary in order to obtain finite cross-sections, if $Q^2\gg m^2$, these logarithms become large and their all-order resummation becomes desirable in order to obtain reliable perturbative predictions.
This resummation is obtained by factorizing the mass logarithms, in the very same way as done for the massless quarks.

Whether mass collinear logarithms are factorized or not for a given massive quark is a choice of factorization scheme.
A scheme where the collinear logarithms for the first $n_f$ lightest quarks are factorized is called a
scheme with $n_f$ active flavours. 
In such a scheme, collinear logarithms are resummed for \emph{light} quarks, while they are treated at fixed order for \emph{heavy} quarks,
thus defining the (relative) concept of light and heavy.
Note that while obviously a heavy quark is massive, a light quark could be either massless, e.g.\ up, down, strange, or massive, e.g.\ charm and bottom.
Thus, $n_f$ can be 3, 4, 5. Note that $n_f=6$ is not phenomenologically relevant, especially in the context of DIS,
because the hard scale of the process is at most comparable, but never much bigger, than the top mass and so the top will be always treated as a heavy flavour. 

In several cases, mostly in the contest of PDF fits where data span a large range in $Q^2$, it is convenient to define
so-called variable flavour number schemes (VFNS), where the number $n_f$ of active flavours varies as a function of $Q$,
such that the collinear resummation for a given massive quark is turned on only for scales where it is needed.
More specifically, a VFNS is a patch of factorization schemes with subsequent values of $n_f$, which switches
from a value $n_f$ to the next one ($n_f+1$) at a given ``heavy quark threshold'' $\mu_h$, typically chosen of the order of the heavy quark mass.
The relation between a scheme with $n_f$ active quarks and a scheme where the mass logarithms of the $(n_f+1)$-th flavour are resummed,
i.e.\ a scheme with $n_f+1$ active flavours, is at the core of the construction of a VFNS and provides
the ingredients to resum the collinear logarithms of the $(n_f+1)$-th flavour. For a recent review see Ref.~\cite{Ball:2015dpa}.

Let us consider DIS structure functions in a scheme with $n_f$ (massless or massive) active flavours.
As we increase the hard scale $Q$, we reach energy scales which are significantly bigger than the mass of the $(n_f+1)$-th quark flavour.
In this situation  the $n_f$-flavour scheme is no longer appropriate, as potentially large collinear logarithms $\log(Q^2/m^2)$ are left unresummed.
A more reliable framework is then provided by a factorization scheme in which $n_f+1$ flavours are considered active, i.e.\ they all participate
to parton evolution, having factorized, and hence resummed, their collinear behaviour~\cite{Aivazis:1993kh,Aivazis:1993pi,Collins:1997sr,Kramer:2000hn,Thorne:1997ga,Thorne:2006qt,Buza:1996wv,Cacciari:1998it,Forte:2010ta, Collins:1998rz, Guzzi:2011ew, Thorne:2008xf, Bonvini:2015pxa}.
The PDFs in the two schemes are related by matching conditions
\beq \label{eq:matching}
f_i^\nfo = \sum_{j=g,q_1,\bar q_1,\ldots, q_{n_f},\bar q_{n_f}} K_{ij}^\nf\; f_j^\nf,\qquad i=g,q_1,\bar q_1,\ldots, q_{n_f},\bar q_{n_f}, q_{n_f+1},\bar q_{n_f+1},
\eeq
where the sum runs over active flavours in the $n_f$ scheme, and $K_{ij}^\nf$ are matching functions. 
Note that we only consider here factorization schemes in which the matching coefficients depend on the heavy quark mass only through logarithms of $Q^2/\mheavy^2$.
In particular, we will consider only $\MSbar$-like schemes, where all the PDFs $f_i^\nfo$ in the $n_f+1$ scheme evolve through standard DGLAP equations, i.e.\ as they all were PDFs of massless quarks.
Note however that the heavy quark PDFs, being generated by the matching conditions Eq.~\eqref{eq:matching},
have a purely perturbative origin, and depend effectively on the heavy quark mass.

The structure functions Eq.~\eqref{eq:FPS} can be written in either scheme
\begin{align}\label{eq:Fscheme}
F_a
  &= F_a^\nf\equiv C^\nf_{a,g} f^\nf_g + C^{{\rm PS}\nf}_{a,q} f^\nf_S + C^{{\rm NS}\nf}_{a,q} \sum_{i=1}^{n_f}e_i^2\(f^\nf_{q_i}+f^\nf_{\bar q_i}\) \nonumber\\
  &= F_a^\nfo\equiv C^\nfo_{a,g} f^\nfo_g + C^{{\rm PS}\nfo}_{a,q} f^\nfo_S + C^{{\rm NS}\nfo}_{a,q} \sum_{i=1}^{n_f+1}e_i^2\(f^\nfo_{q_i}+f^\nfo_{\bar q_i}\).
\end{align}
To all orders in $\as$, the choice of factorization scheme is immaterial and the two expressions are identical.
Truncating the perturbative expansion of the coefficients to any finite order makes the two expressions different by higher order terms.
Requiring equivalence of the two expressions order by order allows to relate the various coefficients and to find the matching functions $K_{ij}^\nf$.
The coefficient functions in the $n_f$ scheme are computed in standard collinear factorization with $n_f$
active quarks, with the heavy quark(s) only appearing in the final state or through loops.
In the $n_f+1$ scheme the coefficient functions generally differ as the collinear logarithms due to the
heavy quark are also factorized. 
Their expressions, which include mass dependence, can be determined by the equality of the first and second line of Eq.~\eqref{eq:Fscheme}:
\begin{align}\label{eq:Cmatching}
  C^\nf_{a,g} &= C^\nfo_{a,g} K_{gg}^\nf + C^{{\rm PS}\nfo}_{a,q} (n_fK_{qg}^\nf+K_{hg}^\nf) + C^{{\rm NS}\nfo}_{a,q} \(\sum_{i=1}^{n_f}e_i^2 K_{qg}^\nf + e_{n_f+1}^2 K_{hg}^\nf\),
\end{align}
and similarly for the quark coefficient functions.
In the equation above, we have defined
\beq
K_{hg}^\nf \equiv K_{q_{n_f+1}g}^\nf + K_{\bar q_{n_f+1}g}^\nf
\eeq
and similarly for the gluon to light-quark matching function $K_{qg}^\nf$, since these functions do not depend on the specific flavour
but only on whether the final quark is light or heavy.

Eq.~\eqref{eq:Cmatching} and its quark counterparts can be solved to express the coefficient functions in the $n_f+1$ scheme in terms of those in
the $n_f$ scheme and the matching functions $K_{ij}^\nf$, as we shall see in the next section.
Furthermore, requiring that resummation of collinear logarithms due to the $n_f+1$ flavour is achieved in the $n_f+1$ scheme allows us to derive expressions for the matching functions as well. We will come back to this point in Sect.~\ref{sec:matchingfunctions}.
For a detailed and general discussion, not limited to small $x$, see e.g.~\cite{Ball:2015dpa,Bonvini:2015pxa}.

\subsubsection{Heavy-flavour schemes at small-$x$}
\label{sec:211}

\begin{figure}[t]
  \centering
  \includegraphics[trim=5.5cm 6.8cm 5.5cm 17cm,clip,width=0.7\textwidth]{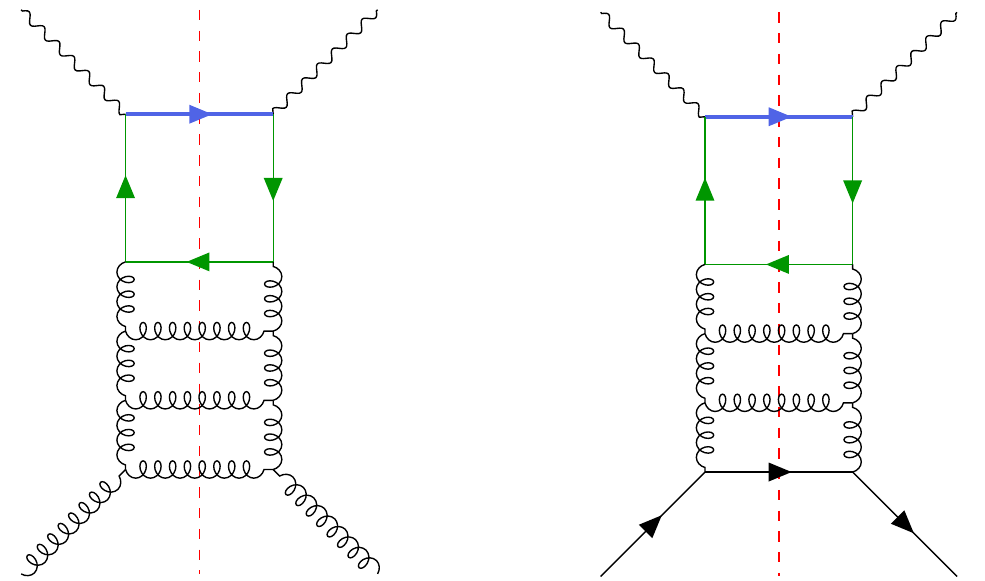}
  \caption{Representative diagrams that contribute to the DIS structure functions at NLL in the gluon channel (left) and quark channel (right).
    In the quark loop the flavour (and thus the mass) can change in the charged-current case (hence the different colours).}
  \label{fig:smallx}
\end{figure}

We now focus our discussion of heavy-quark factorization scheme to the contributions that are enhanced in the small-$x$ regime. 
This topic has been discussed at length in Refs.~\cite{White:2006xv,White:2006yh}, where explicit resummed results were presented in the DIS factorization scheme. 
Here, we extend the results to $\MSbar$-like schemes.

As previously mentioned, the resummed contribution $\Delta C$ can be either seen as a contribution to the singlet or to the pure singlet.
At the accuracy we are considering here, it always comes from a gluon ladder which ends with a quark pair production,
one of which is struck by the photon, as depicted in Fig.~\ref{fig:smallx}.
In the case of a quark initiated contribution, the quark immediately converts to a gluon, which then starts emitting.
Therefore, the resummed contribution to the singlet coefficient functions, denoted $\Delta C_{a,i}$, always has the form
(in the $n_f$ scheme)
\beq\label{eq:deltaCnf}
\Delta C_{a,i}^\nf = \sum_{k=1}^{n_f} e_k^2\, \DC_{a,i}(m_{q_k}) + \sum_{k=n_f+1}^{6} e_k^2 \,\DCm_{a,i}(m_{q_k}), \qquad i=g,q,
\eeq
where $\DC_{a,i}(m_{q_k})$ is the resummed contribution in the case of a light active flavour being struck by the photon,
and $\DCm_{a,i}(m_{q_k})$ is the resummed contribution in the case of a heavy flavour being struck by the photon.\footnote
{In charged-current DIS, where the photon is replaced by a $W$ boson, the quark flavour changes after hitting it,
and so does its mass. Therefore, in this case, the coefficient functions $\DC_{a,i}$ and $\DCm_{a,i}$
would also depend on the mass of the outgoing quark.}
Recall that being active does not necessarily imply being massless and indeed both massless and massive flavours contribute to $\DC_{a,i}$.
Crucially though, in this contribution collinear logarithms are factorized and resummed and, consequently, the zero mass limit of $\DC_{a,i}$ is finite. 
On the other hand, $\DCm_{a,i}(m_{q_k})$ only contains massive quarks and no resummation of mass logarithms has been performed. Thus, the massless limit of this type of contributions is logarithmically divergent. 
In some simplified approaches, the mass of the heavy-quark is immediately neglected once it becomes active:
this leads to what is sometimes called a zero-mass variable flavour number scheme (ZM-VFNS).
Note that the massless contribution is identical for each massless quarks, so in a ZM-VFNS we would have
\beq
 \sum_{k=1}^{n_f} e_k^2\, \DC_{a,i}(0) = \langle e^2 \rangle n_f \DC_{a,i}(0) , \qquad i=g,q.
\eeq
For this reason, a factor $n_f$ is usually included in the definition of the massless singlet coefficient function,
see e.g.\ Refs.~\cite{Catani:1994sq,Bonvini:2016wki}.
Here instead, we wish to retain the mass dependence of the active flavours, if present. 
To this purpose, we adopt a factorization scheme akin to S-ACOT~\cite{Collins:1997sr,Kramer:2000hn} or FONLL~\cite{Forte:2010ta}
(which are formally identical~\cite{Ball:2015dpa,Bonvini:2015pxa})
in which the mass dependence is retained in the coefficient functions.

We now perform a logarithmic counting on Eq.~\eqref{eq:Cmatching}.
None of the matching functions are LL, with the exception of the LO diagonal components, which are all equal to $1$ at $\Ord(\as^0)$, $K_{ii}^{(0)}=1$.
Furthermore, all coefficient functions are NLL, except the non-singlet LO coefficient of $F_2$,
which is $C^{\rm NS,(0)}_{2,q}=\Ord(1)$ and thus LL.
The leading non-trivial logarithmic contributions in the coefficient functions are then NLL,
and their resummed contributions, Eq.~\eqref{eq:deltaCnf}, are related in the two schemes by
\begin{align}\label{eq:CmatchingNLL}
  \Delta C^\nf_{a,g} &= \Delta C^\nfo_{a,g} + e_{n_f+1}^2 C^{\rm NS,(0)}_{a,q} \Delta K_{hg}(\mheavy), \nonumber\\
  \Delta C^\nf_{a,q} &= \Delta C^\nfo_{a,q} + e_{n_f+1}^2 C^{\rm NS,(0)}_{a,q} \Delta K_{hq}(\mheavy),
\end{align}
where $\Delta K_{ij}$ are the NLL resummed contributions to $K_{ij}^\nf$.
In the results above, we have neglected all contributions which are products of two NLL functions.
We have also dropped $K_{qg}^\nf$ (and $K_{qq}^\nf$), which start beyond NLL, as they are given by conversions
of gluons or quarks into quarks with the participation of the heavy quark, i.e.\ suppressed by at least two genuine powers of $\as$.
Note that, for simplicity, we are not indicating in $\Delta K_{ij}$ the label $^\nf$, but we emphasize its (logarithmic) dependence on the heavy quark mass.

Note that $C^{\rm NS,(0)}_{a,q}$ in Eq.~\eqref{eq:CmatchingNLL} is the one in the $n_f+1$ scheme,
so it is, in principle, an unknown of the problem.
However, the requirement that in the $n_f+1$ scheme the resummation of collinear logarithms is achieved, forces it to be equal to the value computed in the limit where the heavy flavour is massless, up to possibly power-behaving mass corrections.
This mass dependence can be arbitrarily fixed to be zero, as the original set of simultaneous equations, Eq.~\eqref{eq:Cmatching}, is undetermined: there are two more unknowns than equations~\cite{Ball:2015dpa,Bonvini:2015pxa}.
This choice is the one leading to S-ACOT/FONLL~\cite{Collins:1997sr,Kramer:2000hn,Forte:2010ta}
and TR~\cite{Thorne:1997ga,Thorne:2006qt} schemes, where we have
\begin{subequations}\label{eq:C0NSmassless}
\begin{align}
C^{\rm NS,(0)}_{2,q}(0)&=1, \\
C^{\rm NS,(0)}_{L,q}(0)&=0.
\end{align}
\end{subequations}
In alternative approaches, like ACOT~\cite{Aivazis:1993kh,Aivazis:1993pi,Collins:1998rz} or, equivalently, a new incarnation of FONLL~\cite{Ball:2015tna,Ball:2015dpa}, denoted here as FONLL$_\text{IC}$,
which has been introduced to account for a possible intrinsic component of the charm PDF,
the incoming heavy quark is treated as massive and therefore the mass-dependence is fully maintained in the coefficient function:
\begin{subequations}\label{eq:C0NSmassive}
\begin{align}
C^{\rm NS,(0)}_{2,q}(\mheavy)
&=\sqrt{1+4\mheavy^2/Q^2}\(\frac{2}{1+\sqrt{1+4\mheavy^2/Q^2}}\)^{N+1}, \\
C^{\rm NS,(0)}_{L,q}(\mheavy)
&=\frac{4\mheavy^2/Q^2}{\sqrt{1+4\mheavy^2/Q^2}}\(\frac{2}{1+\sqrt{1+4\mheavy^2/Q^2}}\)^{N+1}.
\end{align}
\end{subequations}
Note that the massless limit of these expressions reduces to the massless result, Eq.~\eqref{eq:C0NSmassless}.
We stress that for most applications the simpler S-ACOT/FONLL option is formally, and practically,
as good as the more complicated ACOT/FONLL$_\text{IC}$ approach and hence we focus on it in the following.
However, care must be taken when describing the charm structure functions in the case in which the charm PDF is fitted. In this case, the two approaches may lead to sizeable differences and the use of ACOT/FONLL$_\text{IC}$  might be advisable. 
We will come back to this point in Sect.~\ref{sec:FONLL}.

We now consider again the decomposition Eq.~\eqref{eq:deltaCnf}.
In the $n_f+1$ scheme it simply becomes
\beq\label{eq:Cnf+1massive}
\Delta C^\nfo_{a,i} = \sum_{k=1}^{n_f+1} e_k^2\, \DC_{a,i}(m_{q_k}) + \sum_{k=n_f+2}^{6} e_k^2 \,\DCm_{a,i}(m_{q_k}),
\qquad i=g,q,
\eeq
where $\DC_{a,i}(\mheavy)$ are the small-$x$ resummed contributions to the production of a heavy quark (pair)
in the scheme in which such heavy quark participates to the parton dynamics and its collinear logarithms have been factorized.
We now plug Eqs.~\eqref{eq:deltaCnf} and \eqref{eq:Cnf+1massive} into Eq.~\eqref{eq:CmatchingNLL}.
All terms involving the lightest $n_f$ flavours and the heaviest $n_f+2,\ldots,6$ flavours cancel out,
leaving only a relation between the coefficient functions for the $n_f+1$ flavour:
\beq
e_{n_f+1}^2 \,\DCm_{a,i}(\mheavy) = e_{n_f+1}^2 \,\DC_{a,i}(\mheavy) + e_{n_f+1}^2 C^{\rm NS,(0)}_{a,q} \Delta K_{hi}(\mheavy),
\qquad i=g,q.
\eeq
The squared charge is the same in all terms and cancels out, thereby suggesting that this result will hold in general for more generic couplings,
such as when the $Z$-boson exchange plays a role, as will be discussed in Sect.~\ref{sec:Vgen}.
We then find the (collinearly factorized) massive coefficients in the $n_f+1$ scheme for $F_2$,
assuming Eq.~\eqref{eq:C0NSmassless},
\begin{align}\label{eq:CmatchingF2}
  \DC_{2,g}(\mheavy) &= \DCm_{2,g}(\mheavy) - \Delta K_{hg}(\mheavy), \nonumber\\
  \DC_{2,q}(\mheavy) &= \DCm_{2,q}(\mheavy) - \Delta K_{hq}(\mheavy),
\end{align}
and for $F_L$
\begin{align}\label{eq:CmatchingFL}
  \DC_{L,g}(\mheavy) &= \DCm_{L,g}(\mheavy), \nonumber\\
  \DC_{L,q}(\mheavy) &= \DCm_{L,q}(\mheavy).
\end{align}
Given the matching function $\Delta K_{ij}$ at NLL accuracy, the above results completely fix the relation of the NLL resummed contributions to the coefficient functions in $n_f$ and $n_f+1$ schemes. In particular, Eq.~\eqref{eq:CmatchingFL} shows that the resummed contribution to $F_L$ is the same in either schemes.

\subsubsection{Computation of matching functions}
\label{sec:matchingfunctions}

In the derivation above the matching functions $K^\nf_{ij}$ (or $\Delta K_{ij}$) are assumed to be given as an input to the computation of
coefficient functions in the $n_f+1$ scheme. However, the very same derivation also allows us to construct the matching functions themselves.
This is true in general (see e.g.\ Ref.~\cite{Ball:2015dpa}), but we focus on the small-$x$ limit for simplicity.

The key observation is that, after scheme change, the coefficient functions in the $n_f+1$ scheme must not contain anymore
the collinear logarithms associated to the $(n_f+1)$-th flavour.
This is possible only if the matching functions subtract such collinear logarithms from the massive coefficients $\DCm_{a,i}(\mheavy)$,
such that the massless limit $\mheavy\to0$ of the coefficient functions in the new scheme $\DC_{a,i}(\mheavy)$ is finite.
If we further require that the $n_f$ and $n_f+1$ scheme are both of the same type, e.g.\ $\MSbar$-like,
we also need to impose that the massless limit of the massive coefficient $\DC_{a,i}(\mheavy)$
is just what we would have computed if the $(n_f+1)$-th flavour were massless:
\beq \label{eq:massless_limit}
\lim_{Q\gg \mheavy} \DC_{a,i}(\mheavy) = \DC_{a,i}(0), \qquad i=g,q.
\eeq
(Note that, at the logarithmic accuracy we are interested in,
$\DC_{a,i}(0)$ are the same in the $n_f$ and in the $n_f+1$ schemes.\footnote
{In the actual resummed expressions we compute, there is a non-zero $n_f$ dependence in the coefficient functions $\DC_{a,i}(0)$,
which is however subleading and can therefore be ignored.})
The massless limit Eq.~\eqref{eq:massless_limit} ensures that in the $n_f+1$ scheme the collinear logarithms are properly factorized into the PDFs
and resummed through DGLAP. It also fixes the ``constant'' (i.e., non mass dependent) part of the function $\DC_{a,i}(m_{q_{n_f+1}})$.
It does not tell us anything about the power corrections in $\mheavy/Q$ in the $n_f+1$ scheme, which have to be determined
by the matching procedure.

The results Eq.~\eqref{eq:CmatchingFL} show that, since $F_L$ has no collinear singularities at NLL, the massive coefficient in the $n_f+1$ scheme
smoothly approaches the massless one at large $Q$, without any scheme change to be applied.
On the other hand, in the case of $F_2$, Eq.~\eqref{eq:CmatchingF2}, which contains collinear singularities in the massless limit, the scheme-change effectively subtracts the matching function, and the difference will smoothly tend to the massless coefficient for $Q\gg \mheavy$.

We can exploit the last consideration to derive the desired resummed expressions for the matching functions $\Delta K_{hi}$, $i=g,q$.
Indeed, Eq.~\eqref{eq:CmatchingF2} can be inverted to give $\Delta K_{hi}$ in terms of the massive coefficient functions $\DCm_{2,i}(m)$,
which are known, and of the coefficient functions $\DC_{2,i}(m)$ in the $n_f+1$ scheme, the massless limits of which, $\DC_{2,i}(0)$, are known.
We can therefore consider the massless limit $\mheavy\to0$ of this expression, keeping in mind that $\Delta K_{hi}$ and $\DCm_{2,i}(m)$ are separately
logarithmically divergent, and therefore the massless limit has to be intended as setting to zero power-behaving contributions while keeping the logarithms finite.
Assuming, or better choosing, that $\Delta K_{hi}$ contain only logarithms of the mass $\mheavy$ or mass independent terms, we then find
\beq\label{eq:Kres}
\Delta K_{hi}(\mheavy) = \lim_{Q\gg \mheavy} \DCm_{2,i}(\mheavy) - \DC_{2,i}(0) , \qquad i=g,q,
\eeq
where the limit has to be intended as a formal expression as described above.
Note that including power-behaved mass dependent terms is possible and would simply change the form of $\DC_{2,i}(m)$ through Eq.~\eqref{eq:CmatchingF2},
as well as the PDFs in the $n_f+1$ scheme according to Eq.~\eqref{eq:matching}.
However, this ``factorization'' of power behaved contributions is beyond the control of the collinear factorization framework,
so the simplest choice Eq.~\eqref{eq:Kres} is perfectly acceptable and therefore universally adopted.

The definition Eq.~\eqref{eq:Kres} also shows that matching functions at NLL satisfy colour-charge relations.
Indeed, it is known that both massless and massive coefficient functions in the quark and gluon channels are related at NLL by\footnote
{These colour-charge relations are strictly speaking only valid when the resummed contributions refer to the pure-singlet.
If they refer to the singlet, in the massless case there is a fixed-order contribution which needs to be subtracted~\cite{Bonvini:2016wki}.
Alternatively, we can say that these relations are valid for $\DkC{k}_{a,i}$ with $k\geq1$.}
\beq\label{eq:Ccolourcharge}
\DC_{a,q}(0) = \frac{C_F}{C_A} \DC_{a,g}(0), \qquad
\DCm_{a,q}(m) = \frac{C_F}{C_A} \DCm_{a,g}(m).
\eeq
From Eq.~\eqref{eq:Kres} it then immediately follows
\beq\label{eq:Kcolourcharge}
\Delta K_{hq}(m) = \frac{C_F}{C_A} \Delta K_{hg}(m).
\eeq
Together, Eqs.~\eqref{eq:Ccolourcharge} and \eqref{eq:Kcolourcharge} imply through Eqs.~\eqref{eq:CmatchingF2} and \eqref{eq:CmatchingFL}
a colour-charge relation
\beq\label{eq:Ccolourcharge2}
\DC_{a,q}(m) = \frac{C_F}{C_A} \DC_{a,g}(m).
\eeq
for the coefficient functions in the $n_f+1$ scheme.

\subsubsection{Generalization to neutral and charged currents}
\label{sec:Vgen}

The previous results have been derived in the case of photon-exchange DIS.
We now comment on their generalization to the more general case of neutral current, where also the $Z$ boson contributes,
and the case of charged current, where the exchanged boson is a $W$.

Including a $Z$ boson in the discussion is rather trivial. What changes is just the coupling to the quark,
which is different to that of the photon both in the $Z$-only exchange and in the photon-$Z$ interference.
In $Z$-only exchange both vector and axial couplings contribute, which we denote generically as $g_V$ and $g_A$.
If we consider only massless quarks, the sum of the squares of each coupling, $g_V^2+g_A^2$, factorizes.
The only subtlety regarding $Z$ exchange is that in the massive case there is a contribution
which is not proportional to the sum of the squares of the vector and axial couplings.
However, in this case, one can still factor out $g_V^2+g_A^2$, but a contribution
proportional to $g_A^2/(g_V^2+g_A^2)$ appears, which is not present in the photon (or photon-$Z$ interference) case.
This term is not problematic, as it vanishes in the limit of massless quarks.
Therefore, the whole construction of the previous sections, and in particular Sect.~\ref{sec:211}, remains unchanged, provided the photon couplings
$e_k^2$ are replaced with the more general coupling $(g_V^2+g_A^2)_k$ for each quark $q_k$.

The charged-current case is less trivial.
In this case, the quark flavour changes after interacting with the $W$ and this results in two additional complications.
Firstly, the quark mass changes. 
In fact, for all practical applications, one of the quarks interacting with the $W$ can be considered as massless.
Indeed, ignoring the top quark which is too heavy to give a contribution at typical DIS energies,
the only two massive quarks are the charm and the bottom, but their interaction is suppressed by the $V_{cb}$ CKM entry
and can thus be neglected.
Hence, the only remaining combinations involve either a massless and a massive quarks, or two massless quarks.
The second complication arises due to the fact that the final state, composed by a quark $q$ and an anti-quark $\bar q'$,
is not self conjugate (as it is in the neutral-current case, where the final state contains $q\bar q$).
This means that the non-singlet coefficient in Eq.~\eqref{eq:Fscheme} is different for $q$ and $\bar q$, depending on the process.
In particular, this implies that when using Eq.~\eqref{eq:Cmatching} to derive Eq.~\eqref{eq:CmatchingNLL},
only either the heavy quark $q_{n_f+1}$ or the heavy anti-quark $\bar q_{n_f+1}$ contributes:
thus, the collinear subtraction, where present, contains only half of the matching function.
Additionally, for the computation of the parity-violating $F_3$ structure function (which contains a singlet contribution in the massive charged-current case),
the non-singlet LO coefficient has opposite sign depending on whether the initial state parton is a quark or an anti-quark,
specifically $C^{\rm CC, NS,(0)}_{3,q}=1$ and $C^{\rm CC, NS,(0)}_{3,\bar q}=-1$.\footnote
{Note that the opposite sign in these two contribution makes the collinear subtractions
in the fully massless case ineffective. Indeed, the collinear singularities in this case cancel automatically.
In fact, after cancellation, the non-singular part vanishes, thus making the massless singlet contribution to $F_3$ zero
(which remains true at higher orders, since the underlying reason is the antisymmetry of the $F_3$ contribution for the exchange of the two quark masses).
The same mechanism holds in the neutral-current massive case.}
Putting everything together, we have that the analogous of Eqs.~\eqref{eq:CmatchingF2} and \eqref{eq:CmatchingFL} are
\begin{subequations}\label{eq:CmatchingCC}
\begin{align}
  \DC_{2,i}^{\rm CC}(\mheavy) &= \DCm_{2,i}^{\rm CC}(\mheavy) - \Delta K_{hi}(\mheavy)/2, \qquad\qquad i=g,q, \label{eq:C2matchingCC}\\
  \DC_{L,i}^{\rm CC}(\mheavy) &= \DCm_{L,i}^{\rm CC}(\mheavy), \label{eq:CLmatchingCC}\\
  \DC_{3,i}^{\rm CC}(\mheavy) &= 
                                \begin{cases}
                                  \DCm_{3,i}^{\rm CC}(\mheavy) - \Delta K_{hi}(\mheavy)/2 & \text{if final state is $q \bar Q$},\\
                                  \DCm_{3,i}^{\rm CC}(\mheavy) + \Delta K_{hi}(\mheavy)/2 & \text{if final state is $Q \bar q$},
                                \end{cases}
                                                                                            \label{eq:C3matchingCC}
\end{align}
\end{subequations}
where we are denoting with $Q$ the heavy massive quark and with $q$ the companion massless quark
appearing in the final state. The massless limit of the latter is just zero, $\DC_{3,i}^{\rm CC}(0)=0$, since in the massless limit
$F_3$ is non-singlet and therefore it does not contain logarithmic enhancement at small $x$ (see App.~\ref{sec:DISoffshell} for further details).

\subsection{Small-$x$ resummation of coefficient functions and matching functions}
\label{sec:resum_coeff}

We are now ready to discuss the actual resummed expressions for coefficient functions both in the massless and massive case.
Additionally, using Eq.~\eqref{eq:Kres}, we also determine the all-order behaviour of the matching functions.
The all-order behaviour of partonic coefficient functions is obtained using the $\kt$-factorization theorem. In this framework one computes the gluon-initiated contribution to the process of interest, keeping the incoming gluon off its mass-shell. 
To the best of our knowledge, only the off-shell coefficient functions that are necessary to perform the resummation of the photon-induced DIS structure functions have been presented in the literature, both in the case of massless and massive quarks, see e.g.\ \cite{Catani:1990xk,Catani:1992zc,Catani:1990eg}. 
However, in this study we want to resum all DIS coefficient functions, both the neutral-current and charged-current contributions. Therefore, we perform a general calculations of DIS off-shell coefficient functions considering the coupling to the electro-weak bosons and, where relevant, the interference with the photon-induced contributions.
The calculation is detailed in Appendix~\ref{sec:DISoffshell}, where we collect all the relevant results for the off-shell DIS coefficient functions.
These results have been implemented in the public code \texttt{HELL}, version \texttt{2.0},
and are also accessible through the public code \texttt{APFEL}~\cite{Bertone:2013vaa}, to which \texttt{HELL} has been interfaced,
which directly constructs resummed structure functions.

\subsubsection{Resummed coefficient functions}

The resummation of the massless coefficient functions was originally performed to NLL in Ref.~\cite{Catani:1994sq}.
Following Ref.~\cite{Altarelli:2008aj}, in our recent work~\cite{Bonvini:2016wki} we have also included formally subleading,
but important, contributions such as the ones related to the running of the strong coupling.
In our approach, the partonic massless coefficient function ${\cal C}_a(N,\xi,0,0)$,
which is calculated with an incoming off-shell gluon (see App.~\ref{sec:DISoffshell} for details about our notation),
is convoluted with an evolution operator $U(N,\xi)$,
\begin{subequations}\label{eq:Cgmassless}
\begin{align}
\DkC0_{2,g}(0) &= -\int d\xi\, \frac{d}{d\xi}{\cal C}_2(0,\xi,0,0)\, U(N,\xi) - \frac{U_{qg}}{n_f}, \label{eq:C2gmassless} \\
\DkC0_{L,g}(0) &= -\int d\xi\, \frac{d}{d\xi}{\cal C}_L(0,\xi,0,0)\, U(N,\xi), \label{eq:CLgmassless}
\end{align}
\end{subequations}
where
\begin{equation}\label{eq:evol}
U(N,\xi)= \exp\[\int_1^\xi \frac{d \zeta}{\zeta} \gamma_+ \(N, \as(\zeta Q^2)\) \],
\end{equation}
and $\gamma_+$ is the small-$x$ resummed anomalous dimension.
Note that the above expressions hold in a factorization scheme denoted $Q_0\MSbar$~\cite{Catani:1993ww,Catani:1994sq,Ciafaloni:2005cg,Marzani:2007gk}, which differs from $\MSbar$ at relative $\Ord(\as^3)$. In the context of small-$x$ resummation this scheme is preferred because it leads to more stable results~\cite{Altarelli:2008aj}.
Furthermore, since the explicit $N$ dependence of the off-shell partonic coefficient function is subleading, we find advantageous to work at NLL with its $N=0$ moment.
In the resummed expression of the $F_2$ contribution, the subtraction term $U_{qg}$ appears.
Its role is to cancel the collinear singularity of ${\cal C}_2(0,\xi,0,0)$. Its expression reads~\cite{Bonvini:2016wki}
\begin{align}\label{eq:collsubRC}
U_{qg} = \int \frac{d\xi}{\xi} \gamma_{qg}(N,\as(\xi Q^2))\,\theta(1-\xi) \, U(N,\xi)
\end{align}
where $\gamma_{qg}$ is the resummed $qg$ anomalous dimension.
All $\xi$ integrals extend to $\infty$ and start from the position of the Landau pole, $\xi_0=\exp\frac{-1}{\as\beta_0}$.
In $\xi=\xi_0$ the evolution function $U$ is supposed to vanish
(this was e.g.\ a condition for neglecting a boundary term when integrating by parts in Ref.~\cite{Bonvini:2016wki}).
However, due to subleading contributions, this is not always true in our practical construction.
While the induced effect is subleading, this fact is undesirable: we discuss in Appendix~\ref{sec:ht} how we now deal with this issue.

We can apply the same procedure to the case of a heavy (non-active) flavour.
We start considering neutral currents.
Note that in this case the mass of the quark acts as a regulator and no subtraction term $U_{qg}$ appears,
\beq\label{eq:C2gmassive_tilde}
\DkCm0_{a,g}(m) = -\int d\xi\, \frac{d}{d\xi}{\cal C}_a\(0,\xi,\xi_m,\xi_m\)\, U(N,\xi), \quad a=2, L,
\eeq
where we have defined $\xi_m=m^2/Q^2$ (see again App.~\ref{sec:DISoffshell} for the precise definition of the off-shell coefficient and its arguments).
The massive coefficient functions entering the above formula have been computed a long time ago~\cite{Catani:1990eg,Catani:1996ny}
(see also Ref.~\cite{Bottazzi:1998rs}).
We report them in Appendix~\ref{sec:DISoffshell}, where we have also recomputed them in a more general set-up which covers
the full neutral-current case in which the mediator can be a $Z$ boson thus finding a new contribution proportional to the axial coupling,
and the charged-current scenario in which the mass of the quark changes after interacting with the $W$ boson.
We have already commented in Sect.~\ref{sec:Vgen} that for physically relevant processes only one of the quarks involved in charged current DIS is massive,
and the other can be treated as massless.
In this case the resummed coefficients read
\begin{subequations}\label{eq:CgmassiveCC_tilde}
\begin{align}
\DkCm0_{2,g}^{\rm CC}(m) &= -\int d\xi\, \frac{d}{d\xi}{\cal C}_2\(0,\xi,\xi_m,0\)\, U(N,\xi) - \frac{U_{qg}}{2n_f}, \\
\DkCm0_{L,g}^{\rm CC}(m) &= -\int d\xi\, \frac{d}{d\xi}{\cal C}_L\(0,\xi,\xi_m,0\)\, U(N,\xi) - \frac{\xi_m}{1+\xi_m} \,\frac{U_{qg}}{2n_f}, \\
\DkCm0_{3,g}^{\rm CC}(m) &= 
  \begin{cases}
    \displaystyle -\int d\xi\, \frac{d}{d\xi}{\cal C}_3\(0,\xi,\xi_m,0\)\, U(N,\xi) + \frac1{1+\xi_m} \,\frac{U_{qg}}{2n_f} & \;(q\bar Q), \\
    \displaystyle -\int d\xi\, \frac{d}{d\xi}{\cal C}_3\(0,\xi,0,\xi_m\)\, U(N,\xi) - \frac1{1+\xi_m} \,\frac{U_{qg}}{2n_f} & \;(Q\bar q).
  \end{cases}
\end{align}
\end{subequations}
Here, since one of the two quarks involved is massless, we need massless collinear subtractions, implemented through $U_{qg}$,
to take care of the collinear singularity.
Since only one out of two diagrams contains the singularity, there is a factor $1/2$ for each subtraction.
Each subtraction further multiplies the LO non-singlet diagram evaluated in $N=0$, corresponding to the process $q+W\to Q$
(or its conjugate) with massive $Q$, which has non-trivial mass dependence for $F_L$ and $F_3$ (and non-trivial sign for $F_3$).
For $F_L$ in particular, this term vanishes in the massless limit, consistently with Eq.~\eqref{eq:CLgmassless}.
In the last equation we are treating separately the cases in which the final state contains a massless quark plus a massive anti-quark and its conjugate process,
for the same reason discussed in Sect.~\ref{sec:Vgen}.
Note that, according to the notation defined in App.~\ref{sec:DISoffshell} where the third argument of the off-shell coefficient
is the mass (squared divided by $Q^2$) of the anti-quark and the fourth the mass of the quark in the final state,
the arguments are swapped in the two cases. Effectively, for $F_3$ swapping the arguments changes the sign, so the
difference between the two cases is just an overall sign.
For $F_2$ and $F_L$, instead, the coefficients are symmetric for final-state charge conjugation, and therefore the result does not
change when swapping the arguments.

We stress that in the massive case the partonic coefficients include non-trivial theta functions which restrict the available phase space.
This is originally encoded in the $N$ dependence of the off-shell coefficients, which we loose when setting $N=0$.
As these theta functions are very physical, it is important to restore them.
Details on how this is implemented in our resummed results are given in Appendix~\ref{sec:theta}.

The massless $\xi_m\to0$ limit of all the off-shell coefficients is finite, because the off-shellness $\xi$ regulates the collinear region,
and gives the massless off-shell coefficients entering Eqs.~\eqref{eq:Cgmassless} (and zero for $F_3$).
However, while the $\xi_m\to0$ limit for $F_L$ gives automatically the massless result, as it must according to Eqs.~\eqref{eq:CmatchingFL}, \eqref{eq:CLmatchingCC},
for $F_2$ and $F_3$ one further needs to subtract the matching condition, Eqs.~\eqref{eq:CmatchingF2}, \eqref{eq:C2matchingCC}, \eqref{eq:C3matchingCC}.
Accordingly, the resummed massive coefficient functions for the massive active flavour are given in neutral current by
\begin{subequations}
\begin{align}
\DkC0_{2,g}(m) &= -\int d\xi\, \frac{d}{d\xi}{\cal C}_2\(0,\xi,\xi_m,\xi_m\)\, U(N,\xi) - \Delta_0 K_{hg}(m),\label{eq:C2gmassive} \\
\DkC0_{L,g}(m) &= -\int d\xi\, \frac{d}{d\xi}{\cal C}_L\(0,\xi,\xi_m,\xi_m\)\, U(N,\xi), \label{eq:CLgmassive}
\end{align}
\end{subequations}
and in charged current by
\begin{subequations}\label{eq:CgmassiveCC}
\begin{align}
\DkC0_{2,g}^{\rm CC}(m) &= -\int d\xi\, \frac{d}{d\xi}{\cal C}_2\(0,\xi,\xi_m,0\)\, U(N,\xi) - \frac{U_{qg}}{2n_f} - \frac{\Delta_0 K_{hg}(m)}2,\label{eq:C2gmassiveCC} \\
\DkC0_{L,g}^{\rm CC}(m) &= -\int d\xi\, \frac{d}{d\xi}{\cal C}_L\(0,\xi,\xi_m,0\)\, U(N,\xi) - \frac{\xi_m}{1+\xi_m} \frac{U_{qg}}{2n_f}, \label{eq:CLgmassiveCC} \\
\DkC0_{3,g}^{\rm CC}(m) &= 
  \begin{cases}
    \displaystyle -\int d\xi\, \frac{d}{d\xi}{\cal C}_3\(0,\xi,\xi_m,0\)\, U(N,\xi) + \frac1{1+\xi_m} \frac{U_{qg}}{2n_f} - \frac{\Delta_0 K_{hg}(m)}2 & \;(q\bar Q), \\
    \displaystyle -\int d\xi\, \frac{d}{d\xi}{\cal C}_3\(0,\xi,0,\xi_m\)\, U(N,\xi) - \frac1{1+\xi_m} \frac{U_{qg}}{2n_f} + \frac{\Delta_0 K_{hg}(m)}2 & \;(Q\bar q)
  \end{cases}
\label{eq:C3gmassiveCC}
\end{align}
\end{subequations}
(again, the last equation is split in two depending on whether the final state is $q\bar Q$ or $Q\bar q$, the difference being an overall sign).
Comparison of Eq.~\eqref{eq:C2gmassive} (and equivalently Eq.~\eqref{eq:C2gmassiveCC}) with Eq.~\eqref{eq:C2gmassless} shows that the massless limit
does not commute with the on-shell limit in presence of collinear singularities.
In particular, the $\xi_m\to0$ limit applied to $\DkC0_{a,g}^{\rm (CC)}(m)$, $a=2,3$ does not commute with the $\xi$ integration.

\subsubsection{Resummed matching functions}

We can now use Eq.~\eqref{eq:Kres} to determine the resummed matching function $\Delta_0 K_{hg}(m)$.
Using Eqs.~\eqref{eq:C2gmassive_tilde} and \eqref{eq:C2gmassless} we can write\footnote
{Note that this equation can be written in two alternative forms by comparing Eq.~\eqref{eq:C2gmassiveCC} to Eq.~\eqref{eq:C2gmassless}
or by noting that the massless limit of Eq.~\eqref{eq:C3gmassiveCC} vanishes.
Using the results of App.~\ref{sec:appDISlimits} it is easy to verify that all these forms are equivalent and lead to the same result.}
\begin{align}\label{eq:Khgres_def}
  \Delta_0 K_{hg}(m) &= \frac{U_{qg}}{n_f}  
+\int d\xi\, \frac{d}{d\xi}{\cal C}_2(0,\xi,0,0)\, U(N,\xi)
  -\lim_{\xi_m\to0}\int d\xi\, \frac{d}{d\xi}{\cal C}_2\(0,\xi,\xi_m,\xi_m\)\, U(N,\xi),
\end{align}
where the last two terms are basically the commutator of $\xi$ integration and massless limit.
Computing this commutator in the general case is highly non-trivial.
Therefore, we consider first the limit in which the coupling is kept fixed.
In this limit the convolution over $\xi$ becomes a Mellin transformation with moment $M=\gamma_s\(\frac{\as}{N}\)$,
which is the the LL anomalous dimension, dual of the LO BFKL kernel.
This Mellin transform can be performed analytically, so the massless $\xi_m\to0$ limit can be safely taken afterwards.
We have (see Appendix~\ref{sec:appDISlimits})
\beq\label{eq:Khgres_fc}
\Delta_0 K_{hg}^\text{f.c.}(m) = \frac{U_{qg}}{n_f}-
\frac{\as}{\pi} \(\frac{m^2}{Q^2}\)^{\gamma_s} \frac{1-{\gamma_s}}{{\gamma_s}} \frac{\Gamma^3(1-{\gamma_s})\Gamma(1+{\gamma_s})}{(3-2{\gamma_s})\Gamma(2-2{\gamma_s})}.
\eeq
A non-trivial check of the above expression can be performed by considering its perturbative expansion,
\begin{align}\label{eq:Khg_exp}
  \Delta_0 K_{hg}^\text{f.c.} (m)
  &= \as\bigg[\frac{h_0/n_f}{\gamma_s}+h_1/n_f+(h_2/n_f)\gamma_s+\ldots \nonumber\\
  &\qquad\quad -\frac1{3\pi \gamma_s} - \frac{5+3\log(m^2/Q^2)}{9\pi} - \frac{56+30\log(m^2/Q^2)+9\log^2(m^2/Q^2)}{54\pi}\gamma_s +\ldots \bigg] \nonumber \\
  &= - \frac{\log(m^2/Q^2)}{3}\frac{\as}\pi - \frac{28+30\log(m^2/Q^2)+9\log^2(m^2/Q^2)}{18N}\(\frac{\as}{\pi}\)^2 +\Ord(\as^3),
\end{align}
where $h_i$ are the (known) coefficients of the perturbative expansion of $\gamma_s U_{qg}\overset{\text{f.c.}}{=} \gamma_{qg}$
in powers of $\gamma_s$~\cite{Catani:1994sq, Bonvini:2016wki}.
The first coefficients read $h_0=\frac{n_f}{3\pi}$, $h_1=\frac{n_f}{3\pi}\frac53$, $h_2=\frac{n_f}{3\pi}\frac{14}9$.
Note that the collinear pole $1/\gamma_s$ cancels out in the sum.
The second equality is then obtained replacing $\gamma_s=\frac{\as C_A}{\pi N}+\Ord(\as^2)$.
The above expression can be compared with the fixed-order results presented in Ref.~\cite{Buza:1996wv}.
The $\as$ term corresponds to the Mellin transform of the NLO result computed in $N=0$,
and the $\as^2$ term correctly reproduces the leading singularity of the NNLO contribution.
Checking the above result one order higher in perturbation theory is less trivial because at this order we start to become sensitive to the choice of factorization scheme. 
After taking into account the conversion from $Q_0 \MSbar$ to $\MSbar$, which affects the second term in Eq.~\eqref{eq:Khgres_fc}, we find full agreement with the high-energy limit of the $\as^3$ result~\cite{Ablinger:2014nga,Behring:2014eya}.

We can now restore the running-coupling effects in the resummation from the fixed-coupling result by computing
\beq\label{eq:Khgres}
\Delta_0 K_{hg}(m) = \frac{U_{qg}}{n_f} - \int d\xi\, \frac{d}{d\xi}{\cal K}_{hg}(\xi,\xi_m)\, U(N,\xi),
\eeq
where $\mathcal{K}_{hg}$ is obtained as the inverse Mellin transform of its fixed-coupling counterpart, second term in Eq.~\eqref{eq:Khgres_fc}.
The computation of this inverse Mellin transform is done in App.~\ref{sec:appDISlimits}, and its derivative is given by
\begin{align}\label{eq:DcalKhg}
\frac{d}{d\xi}\mathcal{K}_{hg}(\xi,\xi_m)
&= \frac{\as}{3\pi}\, \frac{6\xi_m}{\xi^2} \[1 - \frac{4\xi_m}{\xi}\sqrt{\frac{\xi}{\xi+4\xi_m}}\log\(\sqrt{\frac\xi{4\xi_m}}+\sqrt{1+\frac\xi{4\xi_m}}\)\].
\end{align}
Clearly, by construction, Eq.~\eqref{eq:Khgres} with Eq.~\eqref{eq:DcalKhg} reproduces the correct result
in the fixed-coupling limit, Eq.~\eqref{eq:Khgres_fc}.
Also, it clearly includes the correct resummation of the subleading running coupling effects, as the form Eq.~\eqref{eq:Khgres}
is the standard expression for such resummation~\cite{Bonvini:2016wki}.
Eq.~\eqref{eq:DcalKhg} is a new result. It allows to resum the matching conditions in $\MSbar$-like schemes
with full inclusion of running-coupling effects.

\subsubsection{Matching to fixed order and construction of the VFNS: FONLL as an example}
\label{sec:FONLL}

We conclude the section by giving some details on how the resummed results presented above are combined
with the fixed-order contributions to construct a VFNS.

First, we have to subtract from the resummed coefficient functions their expansion up to the order we want to match onto.
The $\Ord(\as)$ and $\Ord(\as^2)$ contributions to the resummed matching function are explicitly given in Eq.~\eqref{eq:Khg_exp}
in the fixed-coupling limit, but up to this order they are identical to their running coupling counterparts,
and so they can be used straight away to construct $\Delta_1 K_{hg}(m)$ and $\Delta_2 K_{hg}(m)$.
The same equation shows in the first line the expansion of $U_{qg}/n_f$, which is in turn needed for the massless collinear subtractions
of DIS coefficient functions in Eqs.~\eqref{eq:C2gmassless}, \eqref{eq:CgmassiveCC_tilde} and \eqref{eq:CgmassiveCC}.
To complete the list, one needs to expand the $\xi$-integrals of the off-shell coefficient functions with the evolution factor $U(N,\xi)$.
As for the matching function, up to $\Ord(\as^2)$ this expansion can be simply obtained by working in the fixed-coupling limit.
For this, we need the expansions in $M$ of the Mellin transforms of the off-shell coefficients, given in App.~\ref{sec:Mexpansion},
where $M$ should be replaced by $\gamma_s=\frac{\as C_A}{\pi N}+\Ord(\as^2)$ and expanded out.
By doing so, all the $\DkC1$'s and $\DkC2$'s can be constructed, making   the matching of each ingredient with the corresponding fixed order up to NNLO straightforward.

The actual construction of a VFNS is more delicate.
Indeed, there are at least two degrees of freedom that have been exploited in the literature to construct
different incarnation of VFNSs (at fixed order).
One degree of freedom is related to the inclusion of undetermined (by the matching conditions) power-behaving
mass dependent contributions in some coefficient functions, as already discussed in Sect.~\ref{sec:211}.
The second degree of freedom is related to how to combine the various ingredients at a given finite perturbative order.
The approach adopted so far in our construction can be identified with a plain (i.e., without any $\chi$-rescaling~\cite{Tung:2001mv})
S-ACOT construction, with a canonical perturbative counting based on explicit powers of $\as$
(at fixed $\as/N$ for resummed contributions).
This is equivalent to a plain (i.e., without damping~\cite{Forte:2010ta}) FONLL construction, even though
in FONLL the various ingredients are combined together with a different philosophy.
In the following we will briefly review the FONLL construction, which is implemented in the \texttt{APFEL+HELL} package,
with which our results can be directly used for resummed DIS phenomenology.

The FONLL approach~\cite{Cacciari:1998it} is a standard combination of fixed-order and resummed contributions,
in which these two ingredients are simply summed up and the double counting between them subtracted.
In the FONLL case, the distinction between ``fixed order (FO)'' and ``resummation (NLL)''
refers to collinear logarithms due to massive quarks.
In DIS, the FONLL construction~\cite{Forte:2010ta} of structure functions, assuming a single heavy quark $q_{n_f+1}$ with mass $m$, is performed as
\beq\label{eq:FONLL}
F_a^{\rm FONLL}(m) = F_a^{\nf}(m) + F_a^{\nfo}(0) - F_a^{\rm d.c.}(m), \qquad a=2,L,3,
\eeq
where $F_a^{\nf}(m)$ is the fixed-order (called massive) contribution,
in which the collinear logarithms are not resummed and which retains the full mass dependence of the heavy quark,
$F_a^{\nfo}(0)$ is the resummed (called massless) contribution,
computed assuming that the heavy quark is massless, and thus where the (singular) collinear logarithms are resummed,
and finally $F_a^{\rm d.c.}(m)$ is the double-counting term (called massive-zero),
which can be either seen as the fixed-order expansion of $F_a^{\nfo}(0)$ or the ``massless limit''
(in which divergent terms are kept finite) of $F_a^{\nf}(m)$.
The combination $F_a^{\nfo}(0) - F_a^{\rm d.c.}(m)$ can be seen as the resummed contribution
to be added to the fixed order to resum the collinear logarithms, or equivalently
$F_a^{\nf}(m) - F_a^{\rm d.c.}(m)$ can be interpreted as the power-behaving mass corrections
to the (resummed) massless calculation.

According to our nomenclature (extended to the structure functions) the FONLL result
is just the massive result in the $n_f+1$ scheme, i.e.\
\beq
F_a^{\rm FONLL}(m) = F_a^{\nfo}(m).
\eeq
Thus, the double counting term, which is the only non-trivial ingredient in Eq.~\eqref{eq:FONLL},
is given by
\beq\label{eq:FONLLdc}
F_a^{\rm d.c.}(m) = F_a^{\nf}(m) + F_a^{\nfo}(0) - F_a^{\nfo}(m)
\eeq
The corresponding small-$x$ resummed coefficient functions, analogous to $\DC_{a,i}$ and $\DCm_{a,i}$,
can be obtained from Eq.~\eqref{eq:FONLLdc} using Eq.~\eqref{eq:CmatchingNLL}
together with Eqs.~\eqref{eq:deltaCnf} and \eqref{eq:Cnf+1massive},
and are given by
\beq\label{eq:DeltaCdc}
\DC^{\rm d.c.}_{a,i}(m) = \DC_{a,i}(0) + C^{\rm NS,(0)}_{a,q} \Delta K_{hi}(m), \qquad i=g,q
\eeq
for NC, and similarly for CC (with an extra factor $1/2$ multiplying $\Delta K_{hi}$).
The discussion so far does not add anything to the results presented in the previous  sections. However,
having now the small-$x$ resummation for each individual ingredient appearing in Eq.~\eqref{eq:FONLL},
we can also consider the version of FONLL which includes a damping on the resummed contribution,
\beq\label{eq:FONLLdamp}
F_a^{\rm FONLL+damp}(m) = F_a^{\nf}(m) + \theta(1-\xi_m)\(1-\xi_m\)^2\[F_a^{\nfo}(0) - F_a^{\rm d.c.}(m)\],
\eeq
such that the resummation smoothly turns off at the scale $Q=m$.
This variant is often used in PDF fits, and effectively corresponds to damping the collinear subtraction term
$-C^{\rm NS,(0)}_{a,q} \Delta K_{hi}(m)$ in our resummed coefficients $\DC_{a,i}(m)$.

A word of caution is needed when discussing the perturbative counting. The canonical counting would consist in including
all contributions at $\Ord(\as)$ for NLO and all contributions at $\Ord(\as^2)$ at NNLO (and so on).
However, a non-standard counting is usually adopted at NLO (e.g.\ in NLO NNPDF fits),
where the massless contribution $F_a^{\nfo}(0)$ is retained at $\Ord(\as)$, as well as the matching functions,
but the massive contribution $F_a^{\nf}(m)$ is computed at one order higher, $\Ord(\as^2)$~\cite{Forte:2010ta}.
When this particular perturbative counting is adopted, the double-counting piece must be computed with care.
In particular, only the definition of $F_a^{\rm d.c.}(m)$ as the fixed-order expansion of $F_a^{\nfo}(0)$
to $\Ord(\as^2)$ gives the correct result.
As far as small-$x$ resummation is concerned, one has to use $\DkCm2_{a,i}(m)$ for the massive part
and $\DkC1_{a,i}(0)$ for the massless part,
while for the matching at the heavy quark threshold in DGLAP evolution $\Delta_1 K_{hi}(m)$ is to be used.
For the double-counting part, being it the expansion of the massless,
the use of $\DkC1_{a,i}(0)$ and $\Delta_1 K_{hi}(m)$ in Eq.~\eqref{eq:DeltaCdc} is needed.
It can be explicitly verified that in this way the ``resummed contribution'' $F_a^{\nfo}(0)-F_a^{\rm d.c.}(m)$
is indeed of $\Ord(\as^3)$ and subleading at small $x$.

Finally, we discuss a variant of the VFNS where the mass of the heavy quark is retained in all coefficient functions
in the $n_f+1$ scheme. This is the original ACOT~\cite{Aivazis:1993kh,Aivazis:1993pi,Collins:1998rz} construction, and it
also corresponds to the variant FONLL$_\text{IC}$ proposed in Refs.~\cite{Ball:2015tna,Ball:2015dpa}
to account for a possible intrinsic component of the charm PDF.
Following the latter references, we define
\begin{align}
\delta F_a
&= F_a^{{\rm FONLL}_{\rm IC}} - F_a^{\rm FONLL} \nonumber\\
&= F_a^{\rm ACOT} - F_a^\text{S-ACOT},
\end{align}
to be the term needed to upgrade S-ACOT/FONLL to ACOT/FONLL$_\text{IC}$ (ignoring damping, rescaling, etc.).
The small-$x$ resummation of this term can be simply obtained by computing the difference
between the resummation obtained with massive NS coefficients, Eq.~\eqref{eq:C0NSmassive},
and the one obtained with massless NS coefficients, Eq.~\eqref{eq:C0NSmassless}.
We thus have (we apologize with the Reader for the awkward notation)
\begin{subequations}\label{es:Deltadeltac}
\begin{align}
\Delta \delta c_{a,i}(m) &= \[C^{\rm NS,(0)}_{a,q}(0) - C^{\rm NS,(0)}_{a,q}(m) \] \Delta K_{hi}(m), &&a=2,L, &&i=g,q, \\
\Delta \delta c_{a,i}^{\rm CC}(m) &= \[C^{\rm CC,NS,(0)}_{a,q}(0) - C^{\rm CC,NS,(0)}_{a,q}(m) \] \Delta K_{hi}(m)/2, &&a=2,L,3, &&i=g,q,
\end{align}
\end{subequations}
which are the resummed contributions to the singlet coefficient functions $\delta c_{a,i}$ making up $\delta F_a$
for neutral current and charged current respectively.
Note that the massive NS coefficients, which have a non-trivial dependence on $N$,
can be computed in $N=0$ in Eqs.~\eqref{es:Deltadeltac}, as the $N$ dependence is a subleading effect as small $x$.

\section{A new approach to running-coupling resummation in DGLAP evolution}
\label{sec:RCnew}

In the original ABF construction~\cite{Ball:1995vc,Ball:1997vf,Altarelli:2001ji,Altarelli:2003hk,Altarelli:2005ni,Altarelli:2008aj},
which we followed with minor modifications in our previous work~\cite{Bonvini:2016wki},
the resummation of the anomalous dimension $\gamma_+$ (the largest eigenvalue of the singlet sector) is performed
through the exploitation of the duality relation between DGLAP and BFKL evolution kernels,
improved with symmetrization of the latter and the imposition of exact momentum conservation.
This result is usually referred to as double-leading (DL) resummation.

However, it was realized long ago~\cite{Altarelli:2001ji} that running coupling corrections to fixed-order duality
give rise to subleading terms which potentially spoil the perturbative stability of the result.
Therefore, despite their formally subleading nature, the resummation of these effects is of utmost importance 
in order to obtain stable and reliable resummed anomalous dimensions. Additionally, the resummation of these terms changes
the nature of the all-order small-$N$ singularity, converting a square-root branch-cut into a simple pole.
Therefore, the resummation of these contributions, known as running-coupling (RC) resummation,
is usually added to the DL result.

The RC resummation can be obtained by solving the BFKL equation with full running coupling dependence
(see e.g.~\cite{Thorne:1999sg,Thorne:1999rb,Thorne:2001nr}),
and then deriving from the solution (which is an eigenvector PDF) its anomalous dimension.
If we were able to perform this procedure with the full DL BFKL kernel,
the resulting anomalous dimension would just be the final result.
However, solving such equation analytically is not possible,
due to the complicated all-order $\as$-dependence and the non-trivial $M$-dependence of the DL kernel.
In some approaches, e.g.~\cite{Salam:1998tj,Ciafaloni:1999yw,Ciafaloni:2003kd,Ciafaloni:2003rd,Ciafaloni:2007gf},
the equation is thus solved numerically, and the resummed anomalous dimension derived in a numerical way.
Instead, in Refs.~\cite{Altarelli:2005ni,Bonvini:2016wki} an approximate analytic solution,
in which both $\as$- and $M$-dependencies of the kernel are simplified,
was constructed and added to the DL anomalous dimension, subtracting the appropriate double counting.

We find this second approach more convenient, and we keep adopting it here.
However, in this section we critically review the approximations used in Refs.~\cite{Altarelli:2005ni,Bonvini:2016wki}
and propose a new way of constructing and approximating the kernel from which the RC solution is computed.
Our approach has various advantages, from purely practical ones related to the numerical implementation
to most serious ones related to the physical nature of the solution and its $\as$ dependence.

\subsection{The choice of the kernel}
\label{sec:RCkernel}

The core of small-$x$ resummation of the largest eigenvalue $\gamma_+$ is encoded in the duality
between the DGLAP and BFKL equations. Imposing that the corresponding eigenvector PDF
is solution to both equations requires the duality relation
\beq\label{eq:duality0}
\chi_{\rm DL}\(\gamma_{\rm DL}(N,\as),\as\) = N,
\eeq
where $\chi_{\rm DL}\(M,\as\)$ is the BFKL kernel and $\gamma_{\rm DL}(N,\as)$ the DGLAP anomalous dimension.
The knowledge of the BFKL kernel at N$^k$LO provides by duality all the N$^k$LL contributions in the anomalous dimension,
and vice versa.
The name DL comes from the fact that both kernels are supposed to contain their fixed-order part at N$^k$LO,
thus implying that they also contain (by duality with each other) all N$^k$LL contributions.
Therefore, dual DL kernels obtained with fixed N$^k$LO in both, usually denoted DL-N$^k$LO, are matched results
of the form N$^k$LO+N$^k$LL, and so they are both double (next-to-$^k$) leading order and log.
The actual DL kernel and anomalous dimension further contain additional ingredients (from symmetrization and momentum conservation)
which are required to make the result perturbatively stable.

The duality relation Eq.~\eqref{eq:duality0} assumes that the strong coupling $\as$ does not run, namely it is $Q$-independent.
When the running of $\as$ is taken into account, the duality equation receives additional corrections.
If these corrections are included perturbatively,
new singularities appear which make the result perturbatively unstable.
For instance, at NLO+NLL one should include a purely NLL term of the form of a LL function of $\as/N$
times an overall factor $\as$~\cite{Altarelli:2001ji}
\beq\label{eq:gammass}
\gamma_{ss}^{\rm rc}(N,\as) = -\beta_0\as \left.\frac{\chi_0''(M)\chi_0(M)}{2{\chi_0'}^2(M)}\right|_{M=\gamma_s(\as/N)},
\eeq
where $\gamma_s(\as/N)$ is the dual of the LO BFKL kernel $\as\chi_0(M)$.
The new singularity is obtained when $\chi_0'(M)$ in the denominator vanishes, i.e.\ in $M=1/2$.
Higher-order corrections will have larger powers of $\chi_0'(M)$ in the denominator,
leading to a perturbatively unstable singularity for $N$ such that $\gamma_s(\as/N)=1/2$, i.e.\ $N=\as\chi_0(1/2)$.
This instability goes away if RC corrections are included to all orders in the running-coupling parameter $\beta_0$
(i.e., at NLO+NLL one should include a term of the form of a LL function of $\as/N$ times a function of $\as\beta_0$ to all orders),
and the various singularities sum up to a simple pole, whose position is perturbatively stable.
This is the main motivation for including the RC corrections to all orders.

RC corrections are resummed by solving the BFKL equation with the DL kernel $\chi_{\rm DL}$ in which $\as$ is not fixed but is running.
When the coupling runs, $\as(Q^2)$ becomes a differential operator $\ashat=\as/(1-\beta_0\as d/dM)$ in Mellin space,
with $\as=\as(Q_0^2)$, $Q_0$ being a fixed scale, and where we have assumed 1-loop running.
In principle, there can be $\as$ computed at different scales in the kernel, but one can always rewrite it as
$\as(Q^2)$ by evolving to that scale and expanding the relevant evolution factors.
In this way, the $\ashat$ operators are placed on the left of all the $M$-dependent terms of the kernel,
and act on everything to the right.
This ordering is the one chosen by ABF to derive their solution of the RC equation.
Two observations are now in order.
\begin{itemize}
\item While at DL-LO all the running coupling evolution factors are higher orders,
  and $\as$ can be set equal to $\as(Q^2)$ in all terms without
  modifying explicitly the kernel, at DL-NLO changing the argument of each $\as$ to $\as(Q^2)$
  in all LO contributions produces terms which are formally NLO and have to be included in the kernel.
  Because of the symmetry properties of the BFKL ladder, the DL-NLO kernel (see Eq.~\eqref{eq:chiDLNLO} later)
  does not correspond, by construction, to a kernel in which all $\as$'s are $\as(Q^2)$.
  Therefore, at NLO, the form of the kernel in which all $\as$'s are $\as(Q^2)$
  differs from that of the DL kernel by NLO terms.
  For this reason, a different kernel was used for DL and RC resummation at NLO
in Refs.~\cite{Altarelli:2005ni,Altarelli:2008aj,Bonvini:2016wki}.
\item Once all powers of $\as$ are computed at $Q^2$,
  which is equivalent to say that all powers of $\ashat$ have been commuted to the left,
  the running coupling equation cannot be solved directly, because it is in principle a differential equation of infinite order.
  Therefore, in ABF a linear approximation of the kernel, in which at maximum one power of $\ashat$ is retained
  and all others are frozen to $\as(Q_0^2)$, is used.
\end{itemize}
The fact that the complicated all-order $\ashat$ dependence of the kernel is approximated with a linear one may seem too crude.
However, the goal of the RC resummation is to resum to all orders a class of terms, behaving as powers of $\as\beta_0$ times a LL function of $\as/N$,
which originates from 1-loop running at lowest order. The NLO contribution to the BFKL kernel would produce corrections
which are of order $\as (\as\beta_0)^n (\as/N)^k$ for all $n,k$, and therefore beyond the formal accuracy we aim to.
This shows that the linear approximation suggested by ABF is in fact sufficient to the scope of this resummation.

In fact, this argument also suggests that using a NLO kernel for RC resummation which differs
from the DL one is unnecessary. Indeed, the ingredients which determine the leading RC corrections to all orders
are contained in the LO part of the kernel. Thus, correcting the kernel by NLO terms will change subleading RC contributions
which we do not aim to resum. Therefore, for the accuracy we are interested in, the NLO part of the RC kernel is immaterial,
and there is no reason for using a different kernel for RC resummation and DL resummation.

Using the same kernel for DL and RC resummations, as we now suggest, has an important consequence: we do not need any more
the infamous function $\gamma_{\rm match}$ introduced in Ref.~\cite{Altarelli:2005ni} and used later in Ref.~\cite{Altarelli:2008aj,Bonvini:2016wki}
to cure the mismatch of singularities between the DL part and the RC part of the result.
This function was needed to effectively remove the square-root branch-cut of the DL solution, since the
subtraction of the double counting term between DL and RC resummations does not cancel it, exactly because two different kernels were used.
What we have realized only with this work is that the function $\gamma_{\rm match}$ is not subleading as was originally claimed~\cite{Altarelli:2005ni},
but rather it is NLL and therefore ruins the formal accuracy of the NLO+NLL result.
We give more details about this function in Sect.~\ref{sec:gammamatch}.

Having understood that the linear $\ashat$ approximation is a good approximation,
and that as such the same BFKL kernel can be used for both DL and RC resummations,
in order to be able to solve the RC BFKL equation we further need to specify the $M$ functional form.
The approximation adopted by ABF, which we followed in Ref.~\cite{Bonvini:2016wki}, is a quadratic approximation
around the minimum of the kernel. Indeed, the minimum encodes, by duality, the information on the leading singularity,
and it is therefore sufficient to accurately describe the kernel and to perform the RC resummation.
However, we argue that this quadratic approximation has subtle undesired properties which makes it not ideal
for our purposes. For instance, the $\as$-expansion of the resulting anomalous dimension contains
half-integer powers of $\as$.
This is a direct consequence of the fact that a polynomial kernel, such as this quadratic approximation,
is non-physical.
Thus, here we propose a different approximation, which is physically motivated and which leads to an expansion
in integer powers of $\as$. We discuss this new approximation, denoted collinear approximation, in the next section.

\subsection{Solution of the RC differential equation in the collinear approximation}

We are now going to derive the solution of the RC BFKL equation in the linear $\ashat$ approximation and collinear $M$ approximation.
The starting point is the on-shell BFKL kernel in symmetric variables~\cite{Altarelli:2005ni} that we denote simply $\chi(M,\as)$,
whose $\ashat$ dependence is approximated as
\beq\label{eq:linearapprox}
\chi(M,\ashat)
= \chi(M,\as) + (\ashat-\as) \chi'(M,\as)
= \bar\chi(M,\as) + \ashat \chi'(M,\as)
\eeq
where prime denotes derivative with respect to $\as$.
It is important to observe that this approximation includes an $\Ord(\ashat^0)$ term,
$\bar\chi$, which is not physical and not present in the original kernel which is of lowest order $\Ord(\ashat)$.
For this reason the kernel Eq.~\eqref{eq:linearapprox} does not go to zero as $\ashat\to0$.
One could therefore consider another linear approximation,
\beq\label{eq:linearapprox2}
\chi(M,\ashat)
= \ashat \frac{\chi(M,\as)}{\as},
\eeq
which does go to zero as $\ashat\to0$, but does not reproduce the exact derivative in $\ashat=\as$.
In fact, both approximations are equally valid for our purposes, as they would be identical (and exact) in the case of a LO kernel $\chi(M,\ashat)=\ashat \chi_0(M)$.
In the following, we will use the first approximation, Eq.~\eqref{eq:linearapprox}, to derive our results,
which can then be easily translated into the second approximation, Eq.~\eqref{eq:linearapprox2}, by simply letting $\bar\chi=0$, $\chi' = \chi/\as$.
We stress that this simple translation rule could not be applied in the original ABF solution with quadratic kernel,
as the limit $\bar\chi\to0$ is not trivial in that case.

The (homogeneous) RC BFKL equation with kernel Eq.~\eqref{eq:linearapprox},
from which the resummed anomalous dimension can be derived,
is given by~\cite{Altarelli:2005ni}
\beq
\[N - \bar\chi(M,\as) \] f(N,M) = \ashat \chi'(M,\as) f(N,M),
\eeq
where $f(N,M)$ is the double Mellin transform of the eigenvector PDF.
Assuming 1-loop running, and taking the logarithmic derivative of the solution, we arrive at the anomalous dimension
\beq\label{eq:gamma_rc_linear_kernel}
\gamma_{\rm rc}(N,\as) = \[\frac{d}{dt}\log\int_{M_1-i\infty}^{M_1+i\infty}\frac{dM}{2\pi i}\,e^{Mt}
\exp\int_{M_0}^M dM'\,\frac1{\beta_0\as} \(1-\frac{\as\chi'(M',\as)}{N - \bar\chi(M',\as)}\)\]_{t=0},
\eeq
where $M_0$ and $M_1$ are free parameters which must be in the physical region $0<M<1$
(they can be conveniently chosen to be equal to each other, and equal to the position of the minimum).
In order to compute the integrals analytically, we need to specify the form of the kernels $\bar\chi$ and $\chi'$.
In the ABF construction, a quadratic approximation around the minimum of the actual kernel was considered,
\beq\label{eq:BFKL_quadratic_expansion}
\chi(M,\as) = c(\as) + \frac{\kappa(\as)}{2} \(M-M_{\rm min}(\as)\)^2  + \Ord\((M-M_{\rm min})^3\),
\eeq
where $M_{\rm min}$ differs in general from $1/2$ by terms of $\Ord(\as)$. The polynomial form of the quadratic kernel
is non-physical, as for instance the inverse Mellin transform of the $n$-th power of $M$ corresponds
to the $n$-th derivative of a $\delta$ function of $\kt$ in momentum space.
A better approximation, which we propose here, is inspired by a collinear plus anti-collinear approximation of the kernel~\cite{Salam:1999cn,Marzani:2007gk}
generalized to account for a minimum which is not in $1/2$:
\beq\label{eq:BFKL_coll_approx}
\chi_{\rm coll}(M,\as) = A(\as)\[\frac1{M}+\frac1{2M_{\rm min}(\as)-M}\] + B(\as).
\eeq
Expanding around its minimum $M=M_{\rm min}$ we find
\beq
\chi_{\rm coll}(M,\as) = \(B+\frac{2A}{M_{\rm min}}\) + \frac{2A}{M_{\rm min}^3} \(M-M_{\rm min}\)^2 + \Ord\((M-M_{\rm min})^3\),
\eeq
which leads to the identifications
\beq\label{eq:ABckappa}
A=\frac{M_{\rm min}^3\kappa}{4}, \qquad B=c - \frac{M_{\rm min}^2\kappa}{2},
\eeq
such that the collinear kernel incorporates exactly the same information as the quadratic kernel.
Therefore, from the point of view of the accuracy of the approximation, the new collinear kernel is as good as the old quadratic one.
However, as its form resembles the leading collinear and anticollinear poles of the actual kernel, it performs
better than the quadratic one, and leads to a solution with better features, as we shall now see.

To compute the solution of the BFKL equation we need to specify the $\ashat$ dependence of the collinear kernel Eq.~\eqref{eq:BFKL_coll_approx}.
For $A(\as)$ and $B(\as)$, we use the very same linear decomposition Eq.~\eqref{eq:linearapprox},
while the position of the minimum $M_{\rm min}(\as)$ will not be considered as an operator.\footnote
{Note that this assumption is less crude than the approach of previous works, where $M_{\rm min}$ was simply approximated to be $1/2$ in the RC equation.}
The integrals in Eq.~\eqref{eq:gamma_rc_linear_kernel} can be computed easily by noticing that the functional form
of the integrand is identical to that of the quadratic kernel used by ABF, for which the solution is already known.
Specifically, for quadratic kernel, the kernel-dependent part of the integrand of Eq.~\eqref{eq:gamma_rc_linear_kernel} is given by
\beq
\frac{\chi'(M,\as)}{N - \bar\chi(M,\as)}
= \frac{c'+\kappa'/2(M-M_{\rm min})^2}{N-\bar c-\bar\kappa/2(M-M_{\rm min})^2},
\eeq
while for collinear kernel we find
\begin{align}
\frac{\chi'(M,\as)}{N - \bar\chi(M,\as)}
&= \frac{2M_{\rm min}A'+B'M(2M_{\rm min}-M)}{(N-\bar B)M(2M_{\rm min}-M)-2M_{\rm min}\bar A} \nonumber\\
&= \frac{c'-B'/M_{\rm min}^2(M-M_{\rm min})^2}{N -\bar c - (N-\bar B)/M_{\rm min}^2(M-M_{\rm min})^2},
\end{align}
having used in the last step Eq.~\eqref{eq:ABckappa}. By direct comparison, we find the translation rules
\beq
\kappa'\to -2\frac{B'}{M_{\rm min}^2} = \kappa'-2\frac{c'}{M_{\rm min}^2},\qquad 
\bar\kappa\to 2\frac{N-\bar B}{M_{\rm min}^2} = \bar\kappa +2\frac{N-\bar c}{M_{\rm min}^2}.
\eeq
Hence, the final solution is given by~\cite{Altarelli:2005ni,Bonvini:2016wki}\footnote
{In order to account for a generalized position of the minimum, we have recomputed the solution analytically,
thus providing a useful cross-check of the result presented in the literature.}
\beq\label{eq:RCsol}
\gamma_{\rm rc}(N,\as) = M_{\rm min} + \beta_0\asb \[ z \frac{k_{\nu}'(z)}{k_{\nu} (z)} -1 \],
\eeq
where $k_\nu(z)$ is a Bateman function, with
\begin{subequations}\label{eq:BatemanFunctions}
\begin{align}
  \frac1{\asb} &= \frac1{\as} + \frac{\kappa'-2c'/M_{\rm min}^2}{\bar\kappa+2(N-\bar c)/M_{\rm min}^2}\\
  z &= \frac1{\beta_0\asb}\sqrt{\frac{N-\bar c}{\bar\kappa/2 +(N-\bar c)/M_{\rm min}^2}}\\
  \nu &= \(\frac{c'}{N-\bar c} + \frac{\kappa'-2c'/M_{\rm min}^2}{\bar\kappa+2(N-\bar c)/M_{\rm min}^2}\) \asb z.
\end{align}
\end{subequations}
We immediately observe that the limit $\bar c,\bar\kappa\to0$ of these expressions is finite and trivial,
so the solution in the approximation Eq.~\eqref{eq:linearapprox2} is a trivial limit of this solution.
This is in contrast with the analogous solution with a quadratic kernel, whose $\bar\chi\to0$ limit is not trivial
and leads to a solution in terms of Airy functions. This represents a first advantage of using the collinear kernel
with respect to the quadratic one.

Eq.~\eqref{eq:RCsol} with Eq.~\eqref{eq:BatemanFunctions} represents a new result with respect to the
old ``Bateman'' RC solution of Refs.~\cite{Altarelli:2005ni,Altarelli:2008aj,Bonvini:2016wki} and it generalizes it
to the case in which the minimum is not in $1/2$.
In order to study the properties of this result,
we start considering the saddle point expansion of Eq.~\eqref{eq:gamma_rc_linear_kernel},
which is equivalent to an expansion in powers of $\beta_0$ of Eq.~\eqref{eq:RCsol}.
This expansion is also needed to identify the proper double counting with the DL part.
We find
\begin{align}\label{eq:singular}
\gamma_{\rm rc}(N,\as) &= M_{\rm min} - \sqrt{\frac{N-c}{\kappa/2 +(N-c)/M_{\rm min}^2}} - \beta_0\as
+\frac14 \beta_0\as^2\[3\frac{\kappa'-2c'/M_{\rm min}^2}{\kappa+2(N-c)/M_{\rm min}^2}-\frac{c'}{N-c}\] \nonumber\\
&+\beta_0^2\as^3 \sqrt{\frac{\kappa/2 +(N-c)/M_{\rm min}^2}{N-c}} \frac{c'\kappa-\kappa'(N-c)}{32(N-c)^2[2(N-c)/M_{\rm min}^2+\kappa]^2}\nonumber\\
&\qquad\times\bigg[16c^2/M_{\rm min}^2 + 8N(\kappa+2N/M_{\rm min}^2) - c(16\as c' /M_{\rm min}^2+8\kappa-3\as\kappa'+32N /M_{\rm min}^2) \nonumber\\
 &\qquad\qquad +\as(5c'\kappa+16c'N /M_{\rm min}^2-3\kappa'N)\bigg]+ \Ord(\beta_0^3).
\end{align}
This result shows a number of interesting features, especially when compared with the analogous expansion in the case of the quadratic kernel~\cite{Altarelli:2005ni}.
First, we note that at large $N$ all terms go to zero except for the running coupling correction $-\beta_0\as$, which is finite. This is in agreement with
the fact that the starting kernel had a pole in $M=0$, which by fixed-coupling duality leads to an anomalous dimension that goes to zero at large $N$.
Note that this is in contrast with the case of the quadratic kernel, which diverges at large $N$ as $-\sqrt{2N/\kappa}$,
due to the absence in the first square root of the term $+(N-c)/M_{\rm min}^2$ in the denominator, in agreement with it being derived
from a BFKL kernel quadratic in $M$.
This term is crucial for another reason, as it makes the denominator of $\Ord(\as^0)$, while it would be of $\Ord(\as)$
if the denominator were just $\kappa/2$, as it happens with the quadratic kernel. Having a denominator of $\Ord(\as)$
produces an $\as$ expansion of this result which contains half-integer powers of $\as$.
Instead, the $\as$ expansion of Eq.~\eqref{eq:singular}, and hence of Eq.~\eqref{eq:RCsol}, is perfectly acceptable with only integer powers of $\as$.
These two differences represent two additional important benefits of using the collinear kernel rather than the quadratic one.

\subsection{Construction of matched results}

We recall that the solution Eq.~\eqref{eq:RCsol},
having been derived with an approximate $M$ dependence, cannot be regarded as the full solution.
Rather, it represents the all-order resummation of the $\beta_0$ terms which must be added to the DL result,
after subtracting those contributions which are already included (and not approximated) in the DL result.
In the following we will thus focus on the combination of our RC resummed anomalous dimension with the DL one,
also providing the $\as$ expansions of the RC contributions which will be needed in Sect.~\ref{sec:expansion} for matching
(N)LL resummation to (N)NLO.

When matching the RC resummation to the DL-LO result, the first three terms of the singular expansion
Eq.~\eqref{eq:singular} have to be subtracted
(the first two because they are LL, and the third because it is of $\Ord(\as)$, and hence already included in the DL-LO).
After subtraction we have the expansion
\begin{align}\label{eq:DeltaRCLO}
\Delta_\text{DL-LO}\gamma_{\rm rc}(N,\as)
&\equiv \gamma_{\rm rc}(N,\as) - \[M_{\rm min} - \sqrt{\frac{N-c}{\kappa/2 +(N-c)/M_{\rm min}^2}} - \beta_0\as\]
\nonumber\\
&= \beta_0\as^2 \frac{3\kappa_0/32-c_0}N + \Ord(\as^3),
\end{align}
where $\kappa_0$ and $c_0$ are the derivatives of $\kappa$ and $c$ computed in $\as=0$, and are given by
\beq\label{eq:c0kappa0}
c_0 = \frac{C_A}{\pi}4\log2,
\qquad
\kappa_0 = \frac{C_A}\pi 28\zeta_3.
\eeq
The careful Reader might wonder what are the consequences of starting with a BFKL kernel in symmetric variables.
Indeed, when combined with the DL result, the RC result must be translated to DIS (asymmetric) variables.
This amounts to adding $N/2$ to the RC anomalous dimension. However, such a term is automatically subtracted
in the construction of the RC contribution to the DL result, $\Delta_\text{DL-LO}\gamma_{\rm rc}(N,\as)$~\cite{Bonvini:2012sh}.
Therefore, the latter object is insensitive to the change of variables.

For NLL resummation the RC result must be matched to DL-NLO, so we further need to subtract the $\Ord(\as^2)$ of $\gamma_{\rm rc}$.
However, we observe that at $\Ord(\beta_0)$ the expansion of $\gamma_{\rm rc}$, Eq.~\eqref{eq:singular},
contains terms which are formally NLL, and specifically given by $\as\beta_0$ times a LL function.
These terms should be already included in the DL-NLO result. In fact, contributions of this form
originate from running coupling corrections to the duality relation~\cite{Altarelli:2001ji}, Eq.~\eqref{eq:gammass},
and are not automatically generated by fixed-coupling duality in the DL-NLO result.
Rather, they have to be supplied to the DL-NLO result as an additive correction~\cite{Altarelli:2005ni,Bonvini:2016wki},
\begin{align}\label{eq:A.16}
\Delta\gamma_{ss}^{\rm rc}(N,\as)
&= -\beta_0\as \[ \frac{\chi_0''(M)\chi_0(M)}{2{\chi_0'}^2(M)} - 1\]_{M=\gamma_s(\as/N)}
\nonumber\\ &
= \Ord(\as^4),
\end{align}
where $\as\chi_0(M)$ is the LO BFKL kernel and $\gamma_s(\as/N)$ its dual.
The $-1$ in square brackets represents the subtraction of the double counting with the fixed-order part of the DL-NLO;
after subtraction, this function starts at $\Ord(\as^4)$. 
Since the kernel used in the RC resummation is only approximate, the function $\gamma_{\rm rc}$ does not correctly
predict all the NLL contributions of Eq.~\eqref{eq:A.16}. Therefore, Eq.~\eqref{eq:A.16} must be still added to the DL-NLO
result, and the $\Ord(\beta_0)$ part of $\gamma_{\rm rc}$
has to be considered as a double counting term with respect to $\Delta\gamma_{ss}^{\rm rc}$, and hence subtracted.
Thus, for RC resummation matched to DL-NLO, we further need to subtract the fourth term of Eq.~\eqref{eq:singular},
\begin{align}\label{eq:DeltaRCNLO}
\Delta_\text{DL-NLO}\gamma_{\rm rc}(N,\as)
&\equiv \gamma_{\rm rc}(N,\as) - \Bigg[M_{\rm min} - \sqrt{\frac{N-c}{\kappa/2 +(N-c)/M_{\rm min}^2}} - \beta_0\as \nonumber\\
&\qquad\qquad\qquad\quad+\frac14 \beta_0\as^2\(3\frac{\kappa'-2c'/M_{\rm min}^2}{\kappa+2(N-c)/M_{\rm min}^2}-\frac{c'}{N-c}\)\Bigg]
\nonumber\\
&= \beta_0^2\as^3 \frac{\kappa_0}{16N} + \Ord(\as^4),
\end{align}
where we have also written its $\as$ expansion in the last line.
We observe that, formally, only the $\Ord(\as\beta_0)$ times a LL function is really doubly counted,
so in principle one could expand the last line of Eq.~\eqref{eq:DeltaRCNLO} at NLL, and remove the NNLL terms.
However, we think that it is most consistent to also remove these spurious higher logarithmic order contributions.
Interestingly, the expansions of both Eq.~\eqref{eq:DeltaRCLO} and \eqref{eq:DeltaRCNLO} at lowest order only involve lowest order derivatives of $c$ and $\kappa$,
Eq.~\eqref{eq:c0kappa0}, which in turn are determined from just the LO BFKL kernel,
and hence do not depend on the actual construction of the DL kernel.
Also, they are the same irrespectively of the kind of approximate dependence on $\ashat$ is used,
either Eq.~\eqref{eq:linearapprox} or Eq.~\eqref{eq:linearapprox2}.

\subsection{Singularity mismatch}
\label{sec:gammamatch}

The functions $\Delta_\text{DL-(N)LO}\gamma_{\rm rc}(N,\as)$, Eqs.~\eqref{eq:DeltaRCLO} and \eqref{eq:DeltaRCNLO},
contain the resummation of $\beta_0$ contributions which should be added to $\gamma_\text{DL-(N)LO}$, respectively.
The square root term in the subtractions of Eqs.~\eqref{eq:DeltaRCLO} and \eqref{eq:DeltaRCNLO}
contains the LL singularity which has to be removed from the DL result, and replaced with the pole singularity
contained in the $\gamma_{\rm rc}$ term.
This cancellation is exact if the parameters of the minimum, $c(\as)$ and $\kappa(\as)$, are computed
from the same DL kernel used for the fixed-coupling duality which defines $\gamma_\text{DL-(N)LO}$.

At LO+LL, the kernel is the one obtained putting on-shell Eq.~\eqref{eq:chiDLLO}, so the singularity automatically cancels.\footnote
{In fact, in Ref.~\cite{Bonvini:2016wki} two different expressions of $\chi_s$ were used for computing the DL kernel
and for the kernel used in RC resummation. Specifically, in the second case we used the dual of the exact LO anomalous dimension,
which however could not be used for the DL kernel as the exact LO anomalous dimension $\gamma_+$ has a square-root branch-cut,
due to the way the eigenvalue of the singlet anomalous dimension matrix is computed, which would produce a spurious
oscillating behaviour. For this reason, for the DL we used an approximate LO anomalous dimension, thereby creating a mismatch
in the singularities even at LO+LL.
Here, thanks to the approximation discussed in App.~\ref{sec:gammaapprox}, we use exactly the same kernel.}
In Ref.~\cite{Bonvini:2016wki} an intermediate result, denoted LO+LL$^\prime$, was introduced to
perform the resummation of quark entries of the anomalous dimension matrix and of coefficient functions.
This anomalous dimension is formally LO+LL, but uses the RC parameters of the NLO kernel
such that the position of the leading singularity is the same as that of the NLO+NLL result.
For this result, the cancellation of the branch-cut cannot take place.
In order to cure the mismatch in the singularities of the DL-LO result and the RC result with NLO parameters
we need a matching function $\gamma_{\rm match}$.
Its form must be
\beq\label{eq:gammamatch}
\gamma_{\rm match}^{\rm LO+LL'}(N,\as) = \gamma_{\rm m}\(N,c^{\rm NLO}(\as),\kappa^{\rm NLO}(\as),M_{\rm min}^{\rm NLO}(\as)\)
- \gamma_{\rm m}\(N,c^{\rm LO}(\as),\kappa^{\rm LO}(\as),\frac12\),
\eeq
where the function $\gamma_{\rm m}$ must reproduce the singular behaviour of the RC and DL parts, respectively,
and the parameters $c^{\rm (N)LO}$, $\kappa^{\rm (N)LO}$ and $M_{\rm min}^{\rm (N)LO}$ are those obtained from the (N)LO kernel.
For the case of collinear kernel, $\gamma_{\rm m}$ may be simply given by the first two terms of Eq.~\eqref{eq:singular}.
However, we have some latitude with the definition of the matching function as far as subleading corrections are concerned.
We can exploit this freedom to define a matching which numerically has a very moderate effect. We find that the choice
\beq\label{eq:gammam}
\gamma_{\rm m}(N,c,\kappa,M_{\rm min}) = M_{\rm min} - \sqrt{\frac{N-c}{\kappa/2+(N-c)/M_{\rm min}^2}} - \frac{M_{\rm min}^3\kappa}{4N},
\eeq
has the desired properties. We note that, in contrast with the case of the quadratic kernel~\cite{Bonvini:2016wki}, 
here we do not need any further contribution, as this function already vanishes as $1/N$ at large $N$.
Since the parameters in Eq.~\eqref{eq:gammamatch} start differing at $\Ord(\as^2)$, the function $\gamma_{\rm match}^{\rm LO+LL'}$
is formally NLL. This is not a problem here, as the formal accuracy of the LO+LL$^\prime$ result is just LL.
The expansion of the matching function Eq.~\eqref{eq:gammamatch} in powers of $\as$ is
\beq
\gamma_{\rm match}^{\rm LO+LL'} (N,\as) = \Ord(\as^3).
\eeq
Hence, the choice of subleading terms in Eq.~\eqref{eq:gammam} has the additional benefit that we do not need to keep
the matching function into account when expanding the LO+LL$^\prime$ result to $\Ord(\as^2)$.

At NLO+NLL, in Refs.~\cite{Altarelli:2005ni,Bonvini:2016wki} the kernel for RC resummation
was constructed with a different $\ashat$ ordering with respect to the DL one,
which had different minima and thus created a singularity mismatch between the DL and RC anomalous dimensions,
making it necessary the introduction of a matching function to cancel these singularities. 
We have already argued in Sect.~\ref{sec:RCkernel}
that using different kernels was not necessary, and we can actually use the same DL kernel also for RC resummation,
thereby ensuring automatic cancellation of the square-root branch-cut.
Therefore, in this work we no longer need to patch the NLO+NLL result with a matching function.

It is important to stress that, had we used two kernels for the DL and RC parts of the NLO+NLL resummation
differing by $\Ord(\as^2)$ terms,
the analogous matching function would have necessarily been NLL (as in LO+LL$^\prime$ case),
thus contaminating the result which could not be claimed to be NLL anymore.
This is indeed the case for the one used in Refs.~\cite{Altarelli:2005ni,Bonvini:2016wki}.
We have verified that it is not possible to modify the function $\gamma_{\rm m}$, Eq.~\eqref{eq:gammam},
to make the function $\gamma_{\rm match}$ NNLL without introducing new (uncanceled) singularities.
To prove this statement, we consider a generalization of Eq.~\eqref{eq:gammam}
\beq\label{eq:gammamgen}
\gamma_{\rm m}(N,c,\kappa,M_{\rm min}) = M_{\rm min} - \sqrt{\frac{N-c}{\kappa/2+(N-c)/M_{\rm min}^2}} + \eta(N,c,\kappa,M_{\rm min}),
\eeq
where $\eta(N,c,\kappa,M_{\rm min})$ is a function to be determined,
with the requirement that it must not introduce further leading singularities.
Expanding Eq.~\eqref{eq:gammamgen} to NLL, i.e.\ expanding in powers of $\as$ up to $\Ord(\as)$ at fixed $\as/N$, we find
\begin{align}
  \gamma_{\rm m}(N,c,\kappa,M_{\rm min})
  &= \frac12 - \sqrt{\frac{N/\as-c_0}{\kappa_0/2+4(N/\as-c_0)}} + \eta\(\frac{N}{\as},c_0,\kappa_0,\frac12\) \nonumber\\
  &+\as\Bigg[m_1+\frac{c_1 \kappa_0 - c_0 \kappa_1 - 32 c_0^2 m_1 + \kappa_1 N/\as + 64 c_0 m_1 N/\as - 32 m_1 (N/\as)^2}{\sqrt{2}\sqrt{N/\as-c_0}\(\kappa_0+8(N/\as-c_0)\)^{3/2}}\nonumber\\
&\qquad+ \(c_1\partial_2+ \kappa_1\partial_3+ m_1\partial_3\)\eta\(\frac{N}{\as},c_0,\kappa_0,\frac12\) \Bigg] + \text{NNLL},
\end{align}
where the index in the derivatives indicates with respect to which argument the derivative is computed,
and $m_1$ is the $\Ord(\as)$ contribution to $M_{\rm min}$.
The expansion of the square-root term at NLL depends on the NLO coefficients $c_1$, $\kappa_1$ and $m_1$.
If the two kernel used for RC and DL resummations differ at $\Ord(\as^2)$, then these coefficients
differ, and when building up the function $\gamma_{\rm match}$ these NLL terms do not cancel among the two $\gamma_{\rm m}$'s.
Thus, to only way to make the function $\gamma_{\rm match}$ a purely NNLL object is to choose the functions $\eta$
such that the NLL expansion of $\gamma_{\rm m}$ vanishes.
However, the expansion of the square-root term is singular at NLL in $N=\as c_0$,
so it is clear that in order to make $\gamma_{\rm match}$ vanishing at NLL the derivatives of the function $\eta$,
and thus the function itself, must be singular in $N=c$.
But if this is the case, then the function $\gamma_{\rm m}$ will contain additional singularities
with respect to those which it is suppose to cancel.
This violates our assumptions.
We must thus conclude that it is not possible to cancel a NLO singularity mismatch
and at the same time preserve NLL accuracy.
This conclusion remains true also for the matching function used with the quadratic kernel.
Therefore, the only way not to spoil the formal NLL accuracy of the NLO+NLL result is to use exactly
the same kernel for the DL and the RC parts of the result: this is the main motivation for this choice,
adopted in this work for the first time.

In the NLO+NLL result, there is a different singularity mismatch coming from the function $\Delta\gamma_{ss}^{\rm rc}$,
Eq.~\eqref{eq:A.16}, and the $\Delta_\text{DL-NLO}\gamma_{\rm rc}$, Eq.~\eqref{eq:DeltaRCNLO}.
The latter exhibits explicitly a pole in $N=c$, which is different from an analogous singularity
in the former,
\beq\label{eq:gammasssing}
\Delta\gamma_{ss}^{\rm rc}(N,\as) \sim -\frac14\beta_0\as^2\frac{c_0}{N-\as c_0},
\eeq
which is in $N=\as c_0$, as one can easily verify from the definition. The singularities would be identical if the parameters
of the RC kernel were those of the LO BFKL kernel, and would cancel in the sum.
However, due to the higher orders contained in the parameters used to construct the RC kernel,
the position of the singularity is shifted and the cancellation does no longer take place.
We can solve the problem by introducing a new matching function to be added to the final result,
which effectively replaces the singularity of Eq.~\eqref{eq:DeltaRCNLO} with Eq.~\eqref{eq:gammasssing}.
Being the singular contribution a NLL term, this matching function is formally NNLL, and therefore acceptable.
However, as in the LO+LL$^\prime$ case, it is convenient to subtract additional higher orders, such that
the effect of this function is as moderate as possible. Our choice is
\beq\label{eq:gammassmatch}
\gamma^{ss}_{\rm match}(N,\as) = \frac14\beta_0\as^2 \[ \frac{c_0}{N-\as c_0} - \frac{c'}{N-c} +\frac{c'-c_0}{N} \] = \Ord(\as^4),
\eeq
where the last term (which is formally NNLL) ensures cancellation of a number of subleading contributions
from the difference between the first two terms which could potentially spoil the accuracy of the result.
Additionally, because of the last term, the function $\gamma^{ss}_{\rm match}$ starts at $\Ord(\as^4)$ and therefore it does
not contribute to the $\as$-expansion of the NLO+NLL result to $\Ord(\as^3)$.
Note that this singularity mismatch was present also in the original works using a quadratic kernel~\cite{Altarelli:2005ni,Bonvini:2016wki};
the problem there was solved by replacing by hand the singularity in $\Delta_\text{DL-NLO}\gamma_{\rm rc}$
with that of $\Delta\gamma_{ss}^{\rm rc}$, which effectively corresponds to using the same matching function Eq.~\eqref{eq:gammassmatch} but without the last term.

\section{Resummed DGLAP evolution matched to NNLO}
\label{sec:expansion}

As already discussed in Sect.~\ref{sec:RCnew}, the ABF construction of the resummation of the anomalous dimension $\gamma_+$
relies on a double-leading (DL) part, which is based on the duality between the DGLAP and BFKL kernels (at the core of the resummation),
and on a running-coupling (RC) part, which includes a class of subleading but very important effects
which change the nature of the small-$x$ singularity.
The DL resummation is naturally performed at LO+LL or at NLO+NLL, which are obtained by combining together the (N)LO DGLAP
anomalous dimension and the (N)LO BFKL kernel.
Therefore, previous results on small-$x$ resummation have always been presented at these orders.

As already mentioned in the introduction, it is of great importance being able to match the resummed result to fixed NNLO in order to obtain state-of-the art theoretical predictions.
Matching the resummation to NNLO is in principle straightforward: starting from the NLO+NLL resummed result,
one just needs to subtract its $\as$ expansion up to $\Ord(\as^3)$, and replace it with the exact NNLO expression.
While subtracting the NLO from the NLO+NLL is trivial, further subtracting the $\Ord(\as^3)$ term is not,
due to the fact that the DL resummation is expressed in terms of implicit equations, which are usually solved numerically.
One could think of different alternatives.
One possibility is to expand the resummed result numerically, which however does not seem to be a reasonable option, as the numerical
solution of the implicit equations is already challenging and slow, and one cannot hope in general to obtain sufficient precision
in a reasonable amount of time from numerical techniques (unless further numerical developments are made\footnote
{Some developments with respect to our previous implementations have been performed, which make the code faster and more reliable.
  See App.~\ref{sec:gammaapprox} for further details.}).
A second option is to construct a DL result starting from NNLO DGLAP and NLO BFKL, so that the result would be naturally
NNLO+NLL. This option is itself non-trivial, as it requires the computation of a new class of double-counting terms
between the two kernels, and has the undesirable disadvantage that the resummed result one obtains would differ by subleading NNLL
terms from the original NLO+NLL.

In this work we have opted for a third, and perhaps more natural, option, namely expanding the resummed result analytically.
Despite the rather technical nature of this computation, we find it illustrative to give its details in the following Sect.~\ref{sec:DL}.
Indeed, for instance, this exercise allowed us to find a small mistake in the original ABF construction
of the DL part~\cite{Altarelli:2005ni}, which we also have
inherited in our previous work~\cite{Bonvini:2016wki}, and which we correct here.
Then, in Sect.~\ref{sec:final}, we present all the final expressions for the resummed splitting functions,
providing a detailed explanation of the implementation of small-$x$ resummation
that constitutes the backbone of \texttt{HELL}, version \texttt{2.0}.

\subsection{Expansion of the Double Leading anomalous dimension}
\label{sec:DL}

In the ABF construction, the DL resummed anomalous dimension $\gamma_{\rm DL}$,
Eq.~\eqref{eq:duality0}, is obtained from an implicit equation of the form
\beq\label{eq:duality}
\chi_{\Sigma}\(\gamma_{\rm DL}(N,\as),N,\as\) = N,
\eeq
where the function $\chi_{\Sigma}\(M,N,\as\)$ is a so-called off-shell BFKL kernel~\cite{Altarelli:2005ni,Bonvini:2016wki}.
The DL anomalous dimension obtained through Eq.~\eqref{eq:duality} assumes fixed coupling $\as$,
and it thus receives a correction, Eq.~\eqref{eq:A.16}, due to running coupling effects.
This correction starts at $\Ord(\as^4)$ and it is therefore of no interest for the expansion of the DL result to $\Ord(\as^3)$.
In the following, we explain how to construct a perturbative expansion of $\gamma_{\rm DL}(N,\as)$
defined by the implicit equation \eqref{eq:duality}, focussing first on the simpler LL case, and moving next to the NLL case.

\subsubsection{Expansion of the LL resummed result}

We start from LL resummation for simplicity. We seek its expansion to $\Ord(\as^2)$, which would be needed to match LL resummation to NLO.
The off-shell kernel at LO, needed for LL resummation, is given by~\cite{Altarelli:2005ni}
\beq\label{eq:chiDLLO}
\chi_{\Sigma}^{\rm LO}(M,N,\as)
= \chi_{s}\(\frac{\as}M\) + \chi_{s}\(\frac{\as}{1-M+N}\) + \as \tilde\chi_0(M,N) + \chi_{\rm mom}^{\rm LO}(N,\as)
\eeq
where the function $\chi_s(\as/M)$ is defined as the dual to the LO anomalous dimension $\gamma_0$,
\beq\label{eq:chis}
\as\gamma_0\(\chi_s\(\frac{\as}{M}\)\) = M \qquad \Leftrightarrow\qquad \chi_s\(\frac1{\gamma_0(N)}\)=N,
\eeq
the function
\beq\label{eq:chi0tilde}
\tilde\chi_0(M,N) = \frac{C_A}\pi \Big[\psi(1) + \psi(1+N) - \psi(1+M) - \psi(2-M+N)\Big]
\eeq
is the off-shell extension of the LO BFKL kernel with double counting with $\chi_s$ subtracted, and
the function
\beq\label{eq:chimom}
\chi_{\rm mom}(N,\as) = c_{\rm mom}(\as) f_{\rm mom}(N), \qquad
f_{\rm mom}(N) = \frac{4N}{(N+1)^2}
\eeq
restores momentum conservation, i.e.\ the constraint $\gamma_{\rm DL}(N=1,\as)=0$ which translates into $\chi_\Sigma(0,1,\as)=1$,
through a suitable coefficient $c_{\rm mom}$. Because of the definition Eq.~\eqref{eq:chis}, $\chi_s(\as/M)$ in $M=0$ equals $1$,
so we have
\beq
c_{\rm mom}^{\rm LO}(\as) = -\chi_{s}\(\frac{\as}{2}\) - \as \tilde\chi_0(0,1).
\eeq
Note that the LO anomalous dimension $\gamma_0(N)$ that we use for the definition of $\chi_s$
does not necessarily need to be the exact LO anomalous dimension.
In fact, it can be replaced with an approximate expression with the same qualitative features and which preserves its small-$x$ behaviour.
This was already done in both Refs.~\cite{Altarelli:2008aj,Bonvini:2016wki}, in slightly different ways,
to cure a problem due to a branch-cut present in the $n_f\neq0$ case.
Here, we adopt another, simpler, approximation, which circumvents the same problem and also solves another issue.
Additionally, it allows us to exploit duality relations analytically,
which is a great advantage from the numerical implementation point of view.
Further details are given in App.~\ref{sec:gammaapprox}.

In order to obtain the coefficient of the $\as$-expansion of the DL anomalous dimension,
we substitute the formal expansion
\beq\label{eq:gammaDLLOexp}
\gamma_\text{DL-LO}(N,\as) = \as\gamma_0(N)+\as^2\tilde\gamma_1(N)+\ldots
\eeq
into Eq.~\eqref{eq:duality} with $\chi_\Sigma$ given in Eq.~\eqref{eq:chiDLLO}, and expand the equation in powers of $\as$.
We stress that we are implicitly assuming that $\gamma_0(N)$ in Eq.~\eqref{eq:gammaDLLOexp} is the same used in the definition
of $\chi_s$, Eq.~\eqref{eq:chis}; this has to be correct because of the way $\chi_\Sigma^{\rm LO}$ Eq.~\eqref{eq:chiDLLO} is constructed,
as we shall verify shortly.
On the other hand, $\tilde\gamma_1$ is a prediction of the resummation (hence the tilde).
The tricky part for performing the expansion is the first (collinear) $\chi_s$ in Eq.~\eqref{eq:chiDLLO},
as its argument $\as/M$ is of $\Ord(\as^0)$. For this term we then compute the expansion as
\begin{align}\label{eq:chisexpansion}
\chi_s\(\frac{\as}{\gamma_\text{DL-LO}(N,\as)}\)
&= \chi_s\(\frac1{\gamma_0}\[1-\as\frac{\tilde\gamma_1}{\gamma_0}+\Ord(\as^2)\]\) \nonumber\\
&= \chi_s\(\frac{1}{\gamma_0}\) - \as\frac{\tilde\gamma_1}{\gamma_0^2}\chi_s'\(\frac{1}{\gamma_0}\) + \Ord(\as^2)\nonumber\\
&= N + \as\frac{\tilde\gamma_1}{\gamma_0'} + \Ord(\as^2),
\end{align}
where in the last equality we have used the definition Eq.~\eqref{eq:chis}, assuming that $\gamma_0$ is the one appearing in Eq.~\eqref{eq:chis},
and the formula for the derivative
\beq
\chi_s'\(\frac{1}{\gamma_0}\) = -\frac{\gamma_0^2}{\gamma_0'},
\eeq
which can be obtained by deriving both sides of the first of Eq.~\eqref{eq:chis} with respect to $\as/M$.
Here $\gamma_0'$ denotes a derivative with respect to $N$.
All the other terms can be simply expanded in powers of $\as$. Up to the first non-trivial order we get
\beq
N = N +\as\frac{\tilde\gamma_1}{\gamma_0'} + \as\frac{\chi_{01}}{1+N} + \as \tilde\chi_0(0,N) - \as\[\frac{\chi_{01}}{2} + \tilde\chi_0(0,1)\]f_{\rm mom}(N) + \Ord(\as^2),
\eeq
where $\chi_{01}=\chi_s'(0)=C_A/\pi$, from which it immediately follows
\beq
\tilde\gamma_1(N) = -\gamma_0'(N) \[ \frac{\chi_{01}}{1+N} + \tilde\chi_0(0,N)
 -\(\frac{\chi_{01}}2 + \tilde\chi_0(0,1)\)f_{\rm mom}(N) \].
\eeq
Note that the $\Ord(\as^0)$ term cancels automatically, which confirms that indeed the LO part of $\gamma_\text{DL-LO}$
is given by the same $\gamma_0$ appearing in Eq.~\eqref{eq:chis}.
Now, it happens that, due to the explicit form of $\tilde\chi_0(M,N)$, Eq.~\eqref{eq:chi0tilde},
\beq
\tilde\chi_0(0,N) = \frac{C_A}{\pi} \[\psi(1+N)-\psi(2+N)\] = -\frac{C_A}{\pi}\frac1{1+N},
\eeq
and hence we find
\beq
\tilde\gamma_1(N) = 0.
\eeq
This might come as a surprise, however it does not. Indeed, the LL pole of the exact NLO $\gamma_1$ is accidentally zero,
so the only part which is supposed to be predicted correctly by this kernel had to be zero.
In principle there could be non-zero subleading corrections, which in practice are absent (at DL level --- RC contributions do produce extra terms, see Eq.~\eqref{eq:DeltaRCLO}).
If we wish to match LL resummation to NNLO, we should expand to one extra order, but we are not interested in doing so, thus we move to the next logarithmic order.

\subsubsection{Expansion of the NLL resummed result}
\label{sec:DLNLL}

For NLL resummation we need the NLO off-shell kernel
\begin{align}\label{eq:chiDLNLO}
\chi_\Sigma^{\rm NLO}(M,N,\as)
  &= \chi_{s,\rm NLO}(M,\as) + \chi_{s,\rm NLO}(1-M+N,\as) \nonumber\\
  &+ \as \tilde\chi_0(M,N)  + \as^2 \tilde\chi_1(M,N) + \as^2\chi_{1}^{\rm corr}(M,N) \nonumber\\
  &+ \chi_{\rm mom}^{\rm NLO}(N,\as).
\end{align}
Here, $\chi_{s,\rm NLO}(M,\as)$ is the generalization of $\chi_s$, constructed as the exact dual of the
NLO anomalous dimension\footnote{As before, we use an approximate NLO anomalous dimension, see App.~\ref{sec:gammaapprox}.}
$\as\gamma_0(N)+\as^2\gamma_1(N)$, which is an input at this order. This kernel satisfies the formal expansion
\beq\label{eq:chisNLO}
\chi_{s,\rm NLO}(M,\as) = \sum_{j=0}^\infty \as^j \sum_{k=1}^\infty \chi_{jk} \(\frac{\as}M\)^k;
\eeq
the $j=0$ term corresponds to $\chi_s(\as/M)$, Eq.~\eqref{eq:chis}.
The kernel $\tilde\chi_1(M,N)$ was given in Eqs.~(A.23)--(A.29) of Ref.~\cite{Bonvini:2016wki}, and we do not report it here.
The extra term $\chi_{1}^{\rm corr}(M,N)$ takes into account running coupling corrections; its correct expression is
\beq\label{eq:chi1corr}
\chi_1^{\rm corr}(M,N) = \beta_0\[-\frac{C_A}\pi \psi_1(2-M+N) - \frac{1}{(1-M+N)^2}\chi_s'\(\frac{\as}{1-M+N}\) + \frac{\chi_0(M,N)}M-\frac{C_A}{\pi M^2}\].
\eeq
This equation corrects Eq.~(A.18) of Ref.~\cite{Bonvini:2016wki} (i.e.\ Eq.~(6.19) of Ref.~\cite{Altarelli:2005ni}), which did not contain the second term.
In fact, the second term was not really necessary, as it is subleading, but then the argument of $\psi_1$ in the first term should be $1-M+N$,
as the $2$ comes from the subtraction of double-counting with the second term.
In practice, however, we have verified that neglecting the second term and correcting the argument of the first leads to a kernel
which is unstable close to the anticollinear pole $M=1$, instability which is cured (resummed) by including the second term.
We verified that the overall effect of this correction is mild, but not negligible.
Finally, $\chi_{\rm mom}^{\rm NLO}(N,\as)$ restores momentum conservation, in the same form as Eq.~\eqref{eq:chimom}, with
\beq
c_{\rm mom}^{\rm NLO} = -\chi_{s,\rm NLO}(2,\as) - \as \tilde\chi_0(0,1) - \as^2 \tilde\chi_1(0,1) - \as^2\chi_{1}^{\rm corr}(0,1).
\eeq
Note that since $\chi_{s,\rm NLO}(M,\as)$ is the exact dual of the NLO anomalous dimension, it equals $1$ in $M=0$.
Now we consider the expansion of the DL-NLO anomalous dimension
\beq\label{eq:gammaDLNLOexp}
\gamma_\text{DL-NLO}(N,\as) = \as\gamma_0(N)+\as^2\gamma_1(N)+\as^3\tilde\gamma_2(N) + \ldots,
\eeq
where both $\gamma_0(N)$ and $\gamma_1(N)$ are assumed to be those used in the definition of $\chi_{s,\rm NLO}$
(as before, this will be confirmed by the explicit computation),
and $\tilde\gamma_2(N)$ is what we aim to find.
The expansion of $\chi_{s,\rm NLO}$ is obtained by using the same technique used in Eq.~\eqref{eq:chisexpansion}, and leads to
\beq\label{eq:chisNLOexpa}
\chi_{s,\rm NLO}(\gamma_\text{DL-NLO}(N,\as),\as) = N + \as^2\frac{\tilde\gamma_2}{\gamma_0'} + \Ord(\as^3).
\eeq
Note that up to this order the expanded kernel $\chi_s(\as/M)+\as\chi_{ss}(\as/M)$,
corresponding to the first two terms in the $j$-sum of Eq.~\eqref{eq:chisNLO} and originally used in ABF~\cite{Altarelli:2005ni},
gives identical results.
Substituting Eq.~\eqref{eq:gammaDLNLOexp} into Eq.~\eqref{eq:duality} with the NLO kernel Eq.~\eqref{eq:chiDLNLO} and expanding
in powers of $\as$ using Eq.~\eqref{eq:chisNLOexpa} we find the following expression
\begin{align}\label{eq:gamma2}
  \tilde\gamma_2(N) = -\gamma_0'(N)
  &\bigg[ \frac{\chi_{11}}{1+N} + \frac{\chi_{02}}{(1+N)^2} +\tilde\chi_1(0,N) + \chi_{1}^{\rm corr}(0,N) \nonumber\\
  & -\(\frac{\chi_{11}}2 + \frac{\chi_{02}}4 + \tilde\chi_1(0,1) +\chi_1^{\rm corr}(0,1)\)f_{\rm mom}(N) \nonumber\\
  &+ \frac{C_A}{\pi}\[\psi_1(1+N)-\psi_1(1)\]\gamma_0(N) \bigg].
\end{align}
The coefficients of the expansion of $\chi_{s,\rm NLO}$ are given by (see App.~\ref{sec:gammaapprox})
\begin{subequations}\label{eq:chikj}
\begin{align}
  \chi_{02} &= -\frac{11C_A^2}{12\pi^2} + \frac{n_f}{6\pi^2}(2C_F-C_A)\\
  \chi_{11} &= -\frac{n_f}{36\pi^2}\(23C_A-26C_F\).
\end{align}
\end{subequations}
Now, from the definition of $\tilde\chi_1$ (see Ref.~\cite{Bonvini:2016wki})
\beq
\tilde\chi_1(M,N) = \tilde\chi_1^{\rm u}(M,N) - \tilde\chi_1^{\rm u}(0,N) + \tilde\chi_1^{\rm u}(0,0),
\eeq
we immediately find
\beq
\tilde\chi_1(0,N) = \tilde\chi_1^{\rm u}(0,0)
\eeq
which is $N$-independent, and thus also equal to the momentum conservation subtraction $\tilde\chi_1(0,1)$.
Its value is (from Eq.~(A.29) of Ref.~\cite{Bonvini:2016wki})
\beq
\tilde\chi_1(0,N) = \frac{1}{\pi^2} \[ -\frac{74}{27}C_A^2+\frac{11}6C_A^2\zeta_2+\frac52C_A^2\zeta_3 + n_f\(\frac4{27}C_A+\frac7{27}C_F-\frac13C_F\zeta_2\) \].
\eeq
On the other hand we have
\beq\label{eq:chi1correxp}
\chi_1^{\rm corr}(0,N) = -\beta_0\frac{C_A}\pi \zeta_2,
\eeq
which is again $N$-independent.
We can then rewrite Eq.~\eqref{eq:gamma2} as
\begin{align}\label{eq:gamma2v2}
  \tilde\gamma_2(N) = -\gamma_0'(N)
  &\bigg[ \rho + \frac{\chi_{11}}{1+N} + \frac{\chi_{02}}{(1+N)^2} -\(\rho + \frac{\chi_{11}}2 + \frac{\chi_{02}}4 \)f_{\rm mom}(N) \nonumber\\
  &+ \frac{C_A}{\pi}\[\psi_1(1+N)-\psi_1(1)\]\gamma_0(N) \bigg]
\end{align}
with
\begin{align}\label{eq:A}
  \rho &= \frac{1}{\pi^2} \[ -\frac{74}{27}C_A^2+\frac{11}{12}C_A^2\zeta_2+\frac52C_A^2\zeta_3 + n_f\(\frac4{27}C_A+\frac7{27}C_F+\frac16C_A\zeta_2-\frac13C_F\zeta_2\) \] .
\end{align}
This represents the final result for our expansion of the DL anomalous dimension.
As a cross-check, we can now expand $\tilde\gamma_2$ about $N=0$. Given that
\beq
\gamma_0(N) = \frac{C_A}{\pi N} + \Ord(N^0), \qquad
\gamma_0'(N) = -\frac{C_A}{\pi N^2} + \Ord(N^{-1}), \qquad
\eeq
we have that up to NLL the singular behaviour of $\tilde\gamma_2$ in $N=0$ is given by (using $f_{\rm mom}(0)=0$)
\begin{align}
\tilde\gamma_2(N) &= \frac{C_A}{\pi N^2} \[\rho+\chi_{11}+\chi_{02}-2\zeta_3\frac{C_A^2}{\pi^2}\] \nonumber\\
&=\frac{C_A}{\pi^3N^2} \frac{C_A^2 (54 \zeta_3 + 99 \zeta_2 - 395) + n_f (C_A - 2 C_F) (18 \zeta_2 - 71)}{108},
\end{align}
which indeed reproduces the correct NLL pole of the known three-loop anomalous dimension $\gamma_2(N)$~\cite{Vogt:2004mw},
while the LL $1/N^3$ pole is accidentally zero.
We stress that without the correction of the error in Eq.~\eqref{eq:chi1corr} the constant in Eq.~\eqref{eq:chi1correxp}
would change, and thus the NLL singularity of $\gamma_2(N)$ would not be reproduced.

\subsection{Resummed splitting functions matched to NNLO}
\label{sec:final}

In the previous sections we have computed the expansion of the DL result,
and in Sect.~\ref{sec:RCnew} we have presented a new running coupling resummation and provided its $\as$ expansion.
We are now ready to construct the final expressions for the resummed anomalous dimension $\gamma_+$, and write their expansions.
With those, we can then also construct the resummed expressions of all the entries of the singlet anomalous dimension matrix
in the physical basis~\cite{Bonvini:2016wki}, which by Mellin inversion give the singlet splitting functions.

\subsubsection{Anomalous dimensions}

As a first step, we need to add the running coupling contribution to the DL result, with the proper matching functions to cure
the singularity mismatches.
Note that adding the RC resummed functions $\Delta_\text{DL-(N)LO}\gamma_{\rm rc}(N,\as)$ to the DL results violates momentum,
which is further violated in the LO+LL$^\prime$ by the matching function
and in the NLO+NLL by $\Delta\gamma_{ss}^{\rm rc}$ and its matching function.
Momentum conservation can be restored by simply adding a function proportional to $f_{\rm mom}(N)$, Eq.~\eqref{eq:chimom}.
In summary, we have\footnote
{Note that we are here using a notation for the resummed results, ``res (N)LL$^{(\prime)}$'',
which differs from the one used in Ref.~\cite{Bonvini:2016wki}, ``(N)LO+(N)LL$^{(\prime)}$''.
The reason is that we will now use the latter name for the actual resummed results matched to any fixed order,
while in these resummed results the fixed order is only approximate.}
\begin{subequations}
\begin{align}
\gamma_+^{\rm res\,LL }(N,\as)
&= \gamma_\text{DL-LO }(N,\as) + \Delta_\text{DL-LO}\gamma_{\rm rc}^{\rm  LL}(N,\as) - \Delta_\text{DL-LO}\gamma_{\rm rc}^{\rm  LL}(1,\as) f_{\rm mom}(N), \\
\gamma_+^{\rm res\,LL'}(N,\as)
&= \gamma_\text{DL-LO }(N,\as) + \Delta_\text{DL-LO}\gamma_{\rm rc}^{\rm NLL}(N,\as) + \gamma_{\rm match}^{\rm LO+LL'} (N,\as) \nonumber\\
&\quad - \[\Delta_\text{DL-LO}\gamma_{\rm rc}^{\rm NLL}(1,\as) + \gamma_{\rm match}^{\rm LO+LL'} (1,\as) \] f_{\rm mom}(N), \\
\gamma_+^{\rm res\,NLL}(N,\as)
&= \gamma_\text{DL-NLO}(N,\as) + \Delta_\text{DL-NLO}\gamma_{\rm rc}^{\rm NLL}(N,\as) + \Delta\gamma_{ss}^{\rm rc}(N,\as) + \gamma^{ss}_{\rm match}(N,\as) \nonumber\\
&\quad - \[\Delta_\text{DL-LO}\gamma_{\rm rc}^{\rm NLL}(1,\as) + \Delta\gamma_{ss}^{\rm rc}(1,\as) + \gamma^{ss}_{\rm match}(1,\as)\] f_{\rm mom}(N),
\end{align}
\end{subequations}
where the various functions have been introduced in Sect.~\ref{sec:RCnew} and Sect.~\ref{sec:DL}.
Using the results in there, these expressions admit the following $\as$ expansions
\begin{subequations}\label{eq:(N)LLexp}
\begin{align}
\gamma_+^{\rm res\,LL}(N,\as)  \label{eq:LLexp}
&= \as \gamma_0(N) + \as^2 \beta_0 \(\frac{3}{32}\kappa_0-c_0\) \(\frac1N - f_{\rm mom}(N)\) + \Ord(\as^3),\\
\gamma_+^{\rm res\,LL'}(N,\as) \label{eq:LL'exp}
&= \as \gamma_0(N) + \as^2 \beta_0 \(\frac{3}{32}\kappa_0-c_0\) \(\frac1N - f_{\rm mom}(N)\) + \Ord(\as^3),\\
\gamma_+^{\rm res\,NLL}(N,\as)  \label{eq:NLLexp}
&= \as \gamma_0(N) + \as^2 \gamma_1(N) \nonumber\\
&\quad+ \as^3 \Bigg\{ \beta_0^2 \frac{\kappa_0}{16} \(\frac1N - f_{\rm mom}(N)\) \nonumber\\
  &\qquad\qquad -\gamma_0'(N)
  \bigg[ \rho + \frac{\chi_{11}}{1+N} + \frac{\chi_{02}}{(1+N)^2} -\(\rho + \frac{\chi_{11}}2 + \frac{\chi_{02}}4 \)f_{\rm mom}(N) \nonumber\\
  &\qquad\qquad\qquad\qquad\quad+ \frac{C_A}{\pi}\[\psi_1(1+N)-\psi_1(1)\]\gamma_0(N) \bigg] \Bigg\}
+ \Ord(\as^4),
\end{align}
\end{subequations}
with coefficients defined in Eqs.~\eqref{eq:c0kappa0}, \eqref{eq:chikj} and \eqref{eq:A}.
Note that the functions $\gamma_0$ and $\gamma_1$ are those entering the definition of $\chi_s$ and $\chi_{s,\rm NLO}$
in the DL kernel, which are not the exact fixed-order results (see App.~\ref{sec:gammaapprox}).
Thus, it is convenient to introduce the pure resummed contributions, defined by the difference
between the above expressions and their expansion up to a given order, e.g.\
\beq
\Delta_n\gamma_+^{\rm NLL}(N,\as) = \gamma_+^{\rm res\,NLL}(N,\as) - \sum_{k=1}^n \as^k \[\gamma_+^{\rm res\,NLL}(N,\as)\]_{\Ord(\as^k)},
\eeq
and similarly for the LL result. In this way, the resummed anomalous dimension matched to the (exact) fixed order is given by
\beq
\gamma_+^\text{N$^n$LO+N$^k$LL}(N,\as) = \gamma_+^\text{N$^n$LO}(N,\as) + \Delta_n\gamma_+^\text{N$^k$LL}(N,\as),
\eeq
where $\gamma_+^\text{N$^n$LO}(N,\as)$ is the exact N$^n$LO anomalous dimension.
In Ref.~\cite{Bonvini:2016wki} we only considered the ``natural'' contributions
\begin{subequations}
\begin{align}
\Delta_1\gamma_+^{\rm LL^{(\prime)}}(N,\as)  \label{eq:deltaLL}
&= \gamma_+^{\rm res\,LL^{(\prime)}}(N,\as) - \as \gamma_0(N), \\
\Delta_2\gamma_+^{\rm NLL}(N,\as)  \label{eq:deltaNLL}
&= \gamma_+^{\rm res\,NLL}(N,\as) - \as \gamma_0(N) - \as^2 \gamma_1(N)  .
\end{align}
\end{subequations}
With the results of Eqs.~\eqref{eq:(N)LLexp}
we can now also compute
\begin{subequations}
\begin{align}
\Delta_2\gamma_+^{\rm LL^{(\prime)}}(N,\as)  \label{eq:delta2LL}
&= \gamma_+^{\rm res\,LL^{(\prime)}}(N,\as) - \as \gamma_0(N) - \as^2 \beta_0 \(\frac{3}{32}\kappa_0-c_0\) \(\frac1N - f_{\rm mom}(N)\), \\
\Delta_3\gamma_+^{\rm NLL}(N,\as)
&= \gamma_+^{\rm res\,NLL}(N,\as) - \as \gamma_0(N) - \as^2 \gamma_1(N)  \nonumber\\
&\quad-\as^3 \Bigg\{ \beta_0^2 \frac{\kappa_0}{16} \(\frac1N - f_{\rm mom}(N)\) \nonumber\\
&\qquad\qquad -\gamma_0'(N)
  \bigg[ \rho + \frac{\chi_{11}}{1+N} + \frac{\chi_{02}}{(1+N)^2} -\(\rho + \frac{\chi_{11}}2 + \frac{\chi_{02}}4 \)f_{\rm mom}(N) \nonumber\\
  &\qquad\qquad\qquad\qquad\quad+ \frac{C_A}{\pi}\[\psi_1(1+N)-\psi_1(1)\]\gamma_0(N) \bigg] \Bigg\},
  \label{eq:delta3NLL}
\end{align}
\end{subequations}
which are the resummed contributions needed for NLO+LL and NNLO+NLL.

The previous equations are the primary ingredients which allow us to match NLL resummation to NNLO.
Having the expansion of the eigenvalue anomalous dimension $\gamma_+$, we can now construct
the expansions for all the entries of the anomalous dimension matrix in the physical basis.
For achieving this, the secondary ingredient is the expansion of the resummed $\gamma_{qg}$ entry of the evolution matrix.
We refer the Reader to Ref.~\cite{Bonvini:2016wki} for all the details on its resummation, and we report here its
expansion in powers of $\as$,
\beq\label{eq:gammaqg}
\gamma_{qg}^{\rm NLL}(N,\as) = \as h_0 + \as^2 h_1 \gamma_0^{\rm LL'}(N)
+ \as^3 \[h_2 \gamma_0^{\rm LL'}(N)\(\gamma_0^{\rm LL'}(N) -\beta_0\) + h_1 \gamma_1^{\rm LL'}(N)\] + \Ord(\as^4),
\eeq
where $h_0=\frac{n_f}{3\pi}$, $h_1=\frac{n_f}{3\pi}\frac53$ and $h_2=\frac{n_f}{3\pi}\frac{14}9$ are numerical coefficients
(already introduced in Sect.~\ref{sec:resum_coeff}),
$\gamma_1^{\rm LL'}(N)$ is the $\Ord(\as^2)$ term of $\gamma^{\rm LO+LL'}(N,\as)$, which can be read off Eq.~\eqref{eq:LL'exp},
and $\gamma_0^{\rm LL'}(N)$ is given (up to a factor $\as$) by Eq.~(B.10) of Ref.~\cite{Bonvini:2016wki} which we report here for convenience
(with the notations of App.~\ref{sec:gammaapprox})
\beq\label{eq:gamma0LLp}
\gamma_0^{\rm LL'}(N) = \frac{a_{11}}{N} + \frac{a_{10}}{N+1}.
\eeq
Note that in Ref.~\cite{Bonvini:2016wki} two forms for $\gamma_{qg}^{\rm NLL}$, differing by subleading terms, were considered
and used to estimate an uncertainty of the resummation. Up to the order of Eq.~\eqref{eq:gammaqg},
both expressions give identical results, the difference starting at $\Ord(\as^4)$.
Using Eq.~\eqref{eq:gammaqg} it is possible to construct the resummed contributions
\begin{align}
\Delta_2\gamma_{qg}^{\rm NLL}(N,\as) &= \gamma_{qg}^{\rm NLL}(N,\as)- \as h_0 - \as^2 h_1 \gamma_0^{\rm LL'}(N), \\
\Delta_3\gamma_{qg}^{\rm NLL}(N,\as) &= \gamma_{qg}^{\rm NLL}(N,\as)- \as h_0 - \as^2 h_1 \gamma_0^{\rm LL'}(N) \nonumber\\
&\quad- \as^3 \[h_2 \gamma_0^{\rm LL'}(N)\(\gamma_0^{\rm LL'}(N) -\beta_0\) + h_1 \gamma_1^{\rm LL'}(N)\],
\end{align}
which are needed for NLO+NLL and NNLO+NLL resummations, respectively.
The resummed contributions for other entries of the anomalous dimension matrix can be constructed in terms of
$\Delta_n\gamma_+^{\rm NLL}$ and $\Delta_n\gamma_{qg}^{\rm NLL}$, as described in Ref.~\cite{Bonvini:2016wki}.

\subsubsection{Splitting functions}

From the resummed anomalous dimension we can obtain resummed contributions for the splitting function matrix by Mellin inversion.
Additionally, in order to ensure a smooth matching onto the fixed-order at large $x$, a damping is applied. 
Furthermore, we enforce exact momentum conservation on our final results by requiring the first moments of $P_{gg}+P_{qg}$ and $P_{gq}+P_{qq}$ to vanish. 
The final expressions are given by
\beq
P_{ij}^\text{N$^n$LO+N$^k$LL}(x,\as) = P_{ij}^\text{N$^n$LO}(x,\as) + \Delta_n P_{ij}^\text{N$^k$LL}(x,\as)
\eeq
with
\begin{subequations}\label{eq:DeltanPfinal}
\begin{align}
\Delta_n P_{gg}^{\rm NLL}(x,\as) &= (1-x)^2 \(1-\sqrt{x}\)^4 \[\Delta_n P_+^{\rm NLL}(x,\as) - \frac{C_F}{C_A}\Delta_n P_{qg}^{\rm NLL,nodamp}(x,\as) - D\] \\
\Delta_n P_{qg}^{\rm NLL}(x,\as) &= (1-x)^2 \(1-\sqrt{x}\)^4 \Delta_n P_{qg}^{\rm NLL,nodamp}(x,\as) \\
\Delta_n P_{gq}^{\rm NLL}(x,\as) &= \frac{C_F}{C_A} \Delta_n P_{gg}^{\rm NLL}(x,\as) \\
\Delta_n P_{qq}^{\rm NLL}(x,\as) &= \frac{C_F}{C_A} \Delta_n P_{qg}^{\rm NLL}(x,\as)
\end{align}
\end{subequations}
(and similarly at LL)
where $\Delta_n P_+^{\rm NLL}$ and $\Delta_n P_{qg}^{\rm NLL,nodamp}$ are the inverse Mellin of $\Delta_n\gamma_+^{\rm NLL}$ and $\Delta_n\gamma_{qg}^{\rm NLL}$, respectively.
The constant $D$ is given by
\beq\label{eq:momconsc}
D = \frac1{d(1)} \int_0^1 dx\, x (1-x)^2 (1-\sqrt{x})^4 \[\Delta_n P_+^{\rm NLL}(x,\as) +\(1- \frac{C_F}{C_A}\)\Delta_n P_{qg}^{\rm NLL,nodamp}(x,\as) \],
\eeq
where $d(N)$ is the Mellin transform of $(1-x)^2 (1-\sqrt{x})^4$.
Note that, with respect to Ref.~\cite{Bonvini:2016wki}, we have introduced a further damping function, $(1-\sqrt{x})^4$,
to ensure a smoother matching with the fixed-order at large $x$.

From a numerical point of view, we proceed as follows:
\begin{itemize}
\item With a new private code (available upon request) we produce the resummed anomalous dimension $\gamma_+$.
  Specifically, we output $\Delta_2\gamma_+^{\rm LL^{(\prime)}}$ and $\Delta_3\gamma_+^{\rm NLL}$,
  i.e.\ the ones with the expansion terms computed in this work subtracted,
  along the inverse Mellin integration contour, for a grid of values of $\as$ and $n_f$.
  This ensures that these contributions start at $\Ord(\as^3)$ and $\Ord(\as^4)$ respectively,
  as the subtraction is performed within the same code
  (subtracting later in a different code is of course possible,
  but subject to a higher chance of introducing bugs or numerical instabilities).
\item The output of the first code (in the form of publicly available tables) is then read by the public code \texttt{HELL}, version \texttt{2.0} onwards.
  This code essentially computes the resummation of coefficient functions (and equally of the $qg$ anomalous dimension) as described in Ref.~\cite{Bonvini:2016wki},
  and partly summarized in Sect.~\ref{sec:resum_coeff}. The splitting function matrix is then constructed and momentum conservation imposed.
  The objects which are computed are $\Delta_2P_{ig}^{\rm LL}$, $\Delta_3P_{ig}^{\rm NLL}$ ($i=g,q$),
  $\Delta_2K_{hg}$, $\DkC{2}_{a,g}$, $\DkCm{2}_{a,g}(m)$ ($a=2,L$) and $\DkCm{2}^{\rm CC}_{a,g}(m)$ ($a=2,L,3$)\footnote
  {Massive DIS coefficient functions are further sampled for various values of the quark massess.}
  on a $n_f, \as, x$ grid.
  Coefficient functions for additional processes will be also added in the future in the same form.
\item These grids (again publicly available) are then read by the public code \texttt{HELL-x}, version \texttt{2.0} onwards, where
  the $\as,x$ grid is interpolated (cubicly), the quark components of the splitting, matching and coefficient functions are computed by colour-charge relations,
  and $\Delta_1P_{ij}^{\rm LL}$, $\Delta_2P_{ij}^{\rm NLL}$, $\Delta_1K_{hg}$, $\DkC1_{a,i}$ and $\DkCm1^{\rm (CC)}_{a,i}(m)$
  are also constructed by adding the respective lower orders directly in $x$-space.
  For this, we need the analytic inverse Mellin transforms of the non-trivial expansion terms in
  Eq.~\eqref{eq:LLexp}, \eqref{eq:LL'exp} and \eqref{eq:NLLexp}, as well as the analogous for the coefficient functions.
  The latter was already done in the massless case in Ref.~\cite{Bonvini:2016wki}; explicit results for the massive case are presented in App.~\ref{sec:appDISlimits}.
  Explicit results for the splitting functions are presented in App.~\ref{sec:invMellin}.
\end{itemize}
We underline that the second step is the slowest, as the \texttt{HELL} code needs to compute several integrals for the varius grid points and the various
functions to be resummed.
The last step, performed by the \texttt{HELL-x} code, is instead extremely fast, as it simply amounts to an interpolation and the evaluation of simple functions.
For practical applications of our results, it is sufficient to use the \texttt{HELL-x} code, making the inclusion of small-$x$ resummation very handy.
As far as PDF evolution and construction of DIS observables is concerned, the code \texttt{HELL-x} has been included in the \texttt{APFEL} code~\cite{Bertone:2013vaa},
which can be used to access all our results and construct resummed predictions for physical observables.

\section{Numerical results}
\label{sec:results}

In order to illustrate the capabilities of \texttt{HELL 2.0}, we present here some representative results for the small-$x$ resummation of splitting functions and DIS coefficient functions obtained with the techniques described in Ref.~\cite{Bonvini:2016wki} and improved as described in the previous sections.
Moreover, we also show new results for the coefficient functions with massive quarks.

\subsection{Splitting functions}

Let us start with DGLAP evolution. With respect to our previous work~\cite{Bonvini:2016wki} we have made substantial
changes in the resummation of the anomalous dimensions, mostly due to the treatment of running coupling effects,
as described in Sect.~\ref{sec:RCnew}.
Additionally, we are now able to match the NLL resummation of the splitting functions to their fixed-order expressions up to NNLO,
as presented in Sect.~\ref{sec:expansion}.

\begin{figure}[t]
  \centering
  \includegraphics[width=0.495\textwidth,page=1]{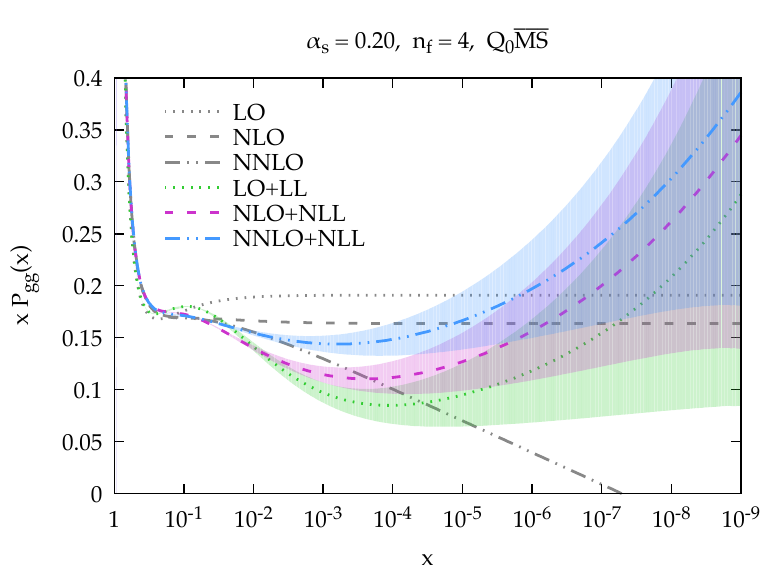}
  \includegraphics[width=0.495\textwidth,page=2]{images/plot_P_nf4_paper.pdf}\\
  \includegraphics[width=0.495\textwidth,page=3]{images/plot_P_nf4_paper.pdf}
  \includegraphics[width=0.495\textwidth,page=4]{images/plot_P_nf4_paper.pdf}
  \caption{The resummed and matched splitting functions at LO+LL (dotted green), NLO+NLL (dashed purple) and NNLO+NLL (dot-dot-dashed blue) accuracy:
    $P_{gg}$ (upper left), $P_{gq}$ (upper right), $P_{qg}$ (lower left) and $P_{qq}$ (lower right).
    The fixed-order results at LO (dotted) NLO (dashed) and NNLO (dot-dot-dashed) are also shown (in black).
    The results also include an uncertainty band, as described in the text.
    The plots are for $\as=0.2$ and $n_f=4$ in the $Q_0\MSbar$ scheme. We note that difference between $Q_0\MSbar$ and $\MSbar$ for the fixed-order results is immaterial at this accuracy.}
  \label{fig:Pres}
\end{figure}
In Fig.~\ref{fig:Pres} we show the fixed-order splitting functions at LO (black dotted), NLO (black dashed) and NNLO (black dot-dot-dashed)
compared to resummed results at LO+LL (green dotted), NLO+NLL (purple dashed) and NNLO+NLL (blue dot-dot-dashed).
In principle, we also have the technology for matching LL resummation to NLO, but this is of very limited interest,
so we do not show these results here (they can be obtained from the \texttt{HELL-x} code).
The gluon splitting functions $P_{gg}$ and $P_{gq}$ are shown in the upper plots,
and the quark ones $P_{qg}$ and $P_{qq}$ are shown in the lower plots
(the latter two start at NLL so the LO+LL curve is absent there).
All splitting functions are multiplied by $x$ for a clearer visualization.
The scheme of the resummed splitting functions is $Q_0\MSbar$
(the fixed-order ones are the same in both $\MSbar$ and $Q_0\MSbar$ at these orders).
The number of active flavours is $n_f=4$, and the value of the strong coupling is $\as=0.2$, corresponding to $Q\sim6$~GeV.
Note that for such value of $Q$ in a VFNS one usually has $n_f=5$ active flavours;
however, the difference between the results in the $n_f=4$ and $n_f=5$ schemes at the same value of $\alpha_s$ is modest,
and our choice allows to directly compare with previous results presented in the literature.

The results of Fig.~\ref{fig:Pres} can be compared directly to the ones presented in our previous paper~\cite{Bonvini:2016wki}.
It can be noticed that the LO+LL result is rather different: in our new implementation this
curve is lower than in the previous version. This is entirely due to the new treatment of running coupling effects, which clearly differs by subleading logarithmic terms.
At the next order, NLO+NLL, there is again a difference with respect to our previous work.  
As before, this is due to subleading terms, which are now NNLL, and so, as expected, they lead to smaller discrepancies. Indeed, NLO+NLL results are much closer to the ones of our previous implementation.
We recall that the new version of the NLO+NLL results also includes the correction of an error in the original expressions~\cite{Altarelli:2005ni}, as detailed in Sect.~\ref{sec:DLNLL},
which has a non-negligible impact on the result, even though the effect is not as large as the one induced by the
new treatment of running coupling effects.

The notable novelty is the presence of the NNLO+NLL curve. The asymptotic small-$x$ behaviour is identical to the NLO+NLL
curve, except for a constant shift, which represents a term of the form $\as^3/x$ in the splitting functions.
Indeed, this term is NNLL, and it was therefore not correctly captured by the NLO+NLL result.
Its impact is larger in the gluon splitting functions $P_{gg}$ and $P_{gq}$, while it is rather small for the quark splitting functions $P_{qg}$ and $P_{qq}$.
At larger $x$, the NNLO+NLL curves smoothly match onto the NNLO result. For $P_{gg}$ and $P_{gq}$ this happens already at $x\sim10^{-2}$.
This is due to the fact that the dip at $x\sim 10^{-3}\div10^{-4}$, which is a known feature of the resummed result at moderate $x$
(see e.g.~\cite{Altarelli:1999vw,Altarelli:2005ni,Ciafaloni:2003kd,Ciafaloni:2003rd}),
is determined by the NNLO logarithmic term, which goes down and dominates at moderate values of $x$,
before the onset of the smaller $x$ asymptotic behaviour, which goes up.
Hence, when matching to NNLO, the initial descent of the splitting functions is automatically described,
and the resummed result deviates only when the rise due to the asymptotic small-$x$ behaviour sets in.
Note that, because of this difference between NLO+NLL and NNLO+NLL, we expect the latter to be much more accurate,
especially in the region $x\gtrsim 10^{-5}$, where the majority of DIS data lie.

The resummed curves are supplemented with an uncertainty, which aims to estimate the size of subleading logarithmic effects.
As far as $P_{qg}$ and $P_{qq}$ are concerned, it is defined exactly as in Ref.~\cite{Bonvini:2016wki},
namely symmetrizing the difference between our default construction of the $\gamma_{qg}$ resummation which uses the evolution function in Eq.~\eqref{eq:UABF},
and what we obtain by switching to a simpler and formally equally valid version, Eq.~\eqref{eq:RCmode}.
The uncertainty band is bigger than in our previous work just because the LL$^\prime$ anomalous dimension used to resum $\gamma_{qg}$
differs in the treatment of running coupling effects.
Because of the way $P_{gg}$ and $P_{gq}$ are constructed, Eq.~\eqref{eq:DeltanPfinal}, these splitting functions inherit the uncertainty band of $P_{qg}$ at NLL.
However, because the bulk of resummed contribution to these entries comes from $\gamma_+$, we have decided to also account for the uncertainty due to subleading contributions to $\gamma_+$. This was not considered in our previous work.
In order to construct this band, we use the same kind of variation used for $\gamma_{qg}$. Specifically, we symmetrize the difference
between the result obtained using Eq.~\eqref{eq:linearapprox} for the resummation of running coupling effects (our default)
and the variant obtained using the simpler, yet equally valid Eq.~\eqref{eq:linearapprox2}.
The uncertainty bands from both sources are then combined in quadrature.
At LL, there is no contribution from $\gamma_{qg}$, and the whole resummed curve is given by $\gamma_+$:
in this case the uncertainty band is just determined from the variation in the construction of the latter.

We note that there is nice overlapping between NLO+NLL and NNLO+NLL bands for the $P_{qg}$ and $P_{qq}$ splitting functions, giving us a good confidence that
they appropriately represent the uncertainty from missing subleading logarithmic orders.
In contrast, the uncertainty band on $P_{gg}$ and $P_{gq}$ does not fully cover
the effects of higher orders in the initial small-$x$ region,
$10^{-4}\lesssim x\lesssim10^{-2}$, as demonstrated by the fact that NLO+NLL and NNLO+NLL do not overlap there.
However, this effect is mostly driven by the largish NNLL effects at $\Ord(\as^3)$, which are those that are included
in the NNLO+NLL but not in the NLO+NLL results. At higher orders the effects of subleading logs in this region is likely to be smaller.
In support of this hypothesis, we can note that the distance between NNLO+NLL and NNLO for $x\sim10^{-2}$  is significantly smaller than the distance
between NLO+NLL and NLO, in the same region.
Thus, we believe that, while the uncertainty on the NLO+NLL result is not satisfactory in the intermediate $x$ region,
the uncertainty on the NNLO+NLL should be reliable.

These plots are also instructive to study the stability of the perturbative expansion.
By looking at the fixed-order splitting functions, we see that small-$x$ logarithms start being dominant already at $x\lesssim10^{-2}$, where the
logarithmic term of the NNLO contribution sets in. 
We note that the small-$x$ growth could have been in principle much stronger.
Indeed, the leading logarithmic contributions have vanishing coefficients both at NLO and NNLO
and the sharp rise of the NNLO splitting function is driven by its NLL contribution $\as^3 x^{-1} \log x$.
These accidental zeros are not present beyond NNLO and so we expect the yet-unknown N$^3$LO splitting functions
to significantly deteriorate the stability of the perturbative expansion because of their $\as^4 x^{-1} \log^3 x$ growth at small $x$.
Therefore, we anticipate that the inclusion of the resummation to stabilize the small-$x$ region will be even more crucial at N$^3$LO.

\subsection{DIS coefficient functions}

\begin{figure}[t]
\centering
  \includegraphics[width=0.495\textwidth,page=3]{images/plot_C_nf4_paper.pdf}
  \includegraphics[width=0.495\textwidth,page=1]{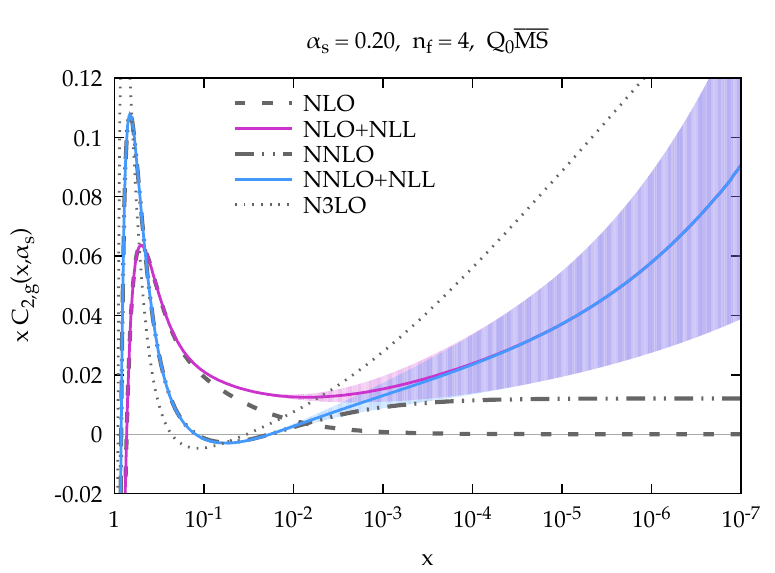}
  \caption{The resummed and matched massless coefficient functions $C_{L,g}$ (left) and $C_{2,g}$ (right)
    at NLO+NLL accuracy (solid purple) and at NNLO+NLL accuracy (solid blue).
    Fixed-order results are also shown in black: NLO in dashed, NNLO in dot-dot-dashed and N$^3$LO in dotted.
    The plots are for $\as=0.2$ and $n_f=4$ in the $Q_0\MSbar$ scheme.
    We note that difference between $Q_0\MSbar$ and $\MSbar$ for the fixed-order results is immaterial at this accuracy,
    except for the N$^3$LO contribution to $F_2$, which is shown in $\MSbar$.}
  \label{fig:c2cL}
\end{figure}
We now move to DIS coefficient functions and we first present updated predictions for the massless coefficients,
i.e.\ assuming that there are only $n_f$ massless quarks and no heavy quarks.
The construction did not change compared to our previous work~\cite{Bonvini:2016wki}, but the input
LL$^\prime$ anomalous dimension used for computing the resummed coefficients did, as explained in Sect.~\ref{sec:expansion}.\footnote
{Additionally, we changed the overall large-$x$ damping, which is now uniformly chosen to be $(1-x)^2(1-\sqrt x)^4$,
as for the splitting functions, Eq.~\eqref{eq:DeltanPfinal}.}
The updated results are shown in Fig.~\ref{fig:c2cL} for $C_{L,g}$ (left) and $C_{2,g}$ (right).
The quark contributions at small $x$ are very similar (due to colour-charge relation) and are not shown.
We observe some differences with respect our previous work, although within uncertainties, for the coefficient function $C_{2,g}$ due to the modified running-coupling resummation. These changes appear to have instead a small numerical effect on $C_{L,g}$.
The other noticeable difference with respect to our previous results is the size of the theoretical uncertainty,
which is now larger: this effect is entirely due to the different LL$^\prime$ used in the construction,
and is therefore ultimately due to the treatment of running-coupling effects.

\begin{figure}[t]
\centering
  \includegraphics[width=0.495\textwidth,page=2]{images/plot_Cm_c_nf3_as028_paper.pdf}
  \includegraphics[width=0.495\textwidth,page=1]{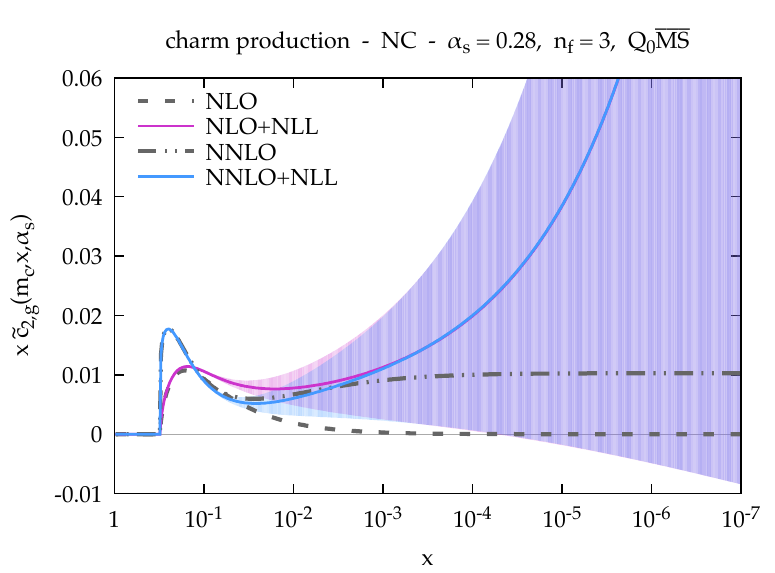}
  \includegraphics[width=0.495\textwidth,page=2]{images/plot_Cm_b_nf4_as020_paper.pdf}
  \includegraphics[width=0.495\textwidth,page=1]{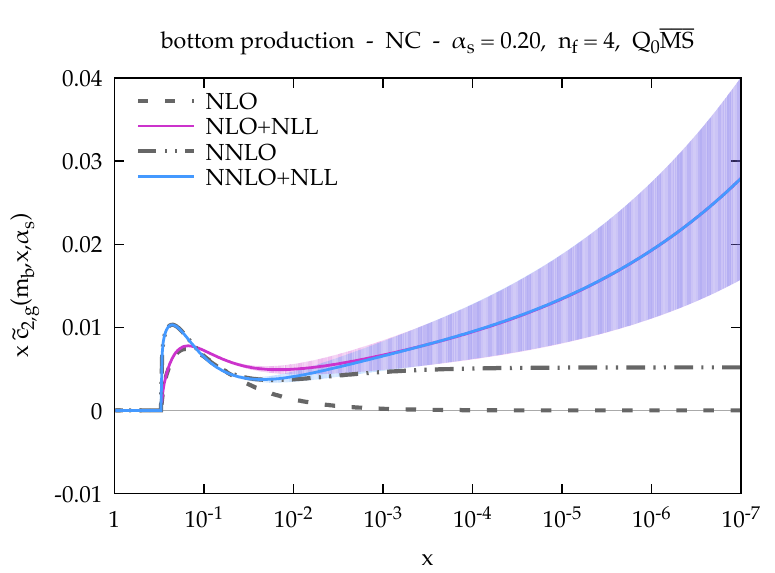}
  \caption{Same as Fig.~\ref{fig:c2cL}, but for the massive coefficient functions $\Cm_{L,g}$ (left) and $\Cm_{2,g}$ (right),
    for both charm production (upper plots) and bottom production (lower plots) in NC.
    The charm production plots are for $n_f=3$ and $\as=0.28$, corresponding to $Q\sim2$~GeV,
    slightly above the charm mass $m_c=1.5$~GeV.
    The bottom production plots are for $n_f=4$ and $\as=0.2$, corresponding to $Q\sim6$~GeV,
    slightly above the bottom mass $m_b=4.5$~GeV.}
  \label{fig:Cmass}
\end{figure}

We now move to the new results which include mass dependence.
We first show in Fig.~\ref{fig:Cmass} the analogous of Fig.~\ref{fig:c2cL} for the massive \emph{unsubtracted}
coefficient functions, both for charm production and for bottom production close to the production threshold.
As usual in theory papers, we define these contributions as the ones for which the heavy quark is struck by the photon
(at these energies, the $Z$ contribution in NC and the CC production mechanism are negligible).
We call generically these contributions $\Cm_{a,i}$, with $a=2,L$ and $i=g,q$, of which the functions $\DCm_{a,i}$
defined in Sect.~\ref{sec:211} are the resummed contributions.
For charm production (upper plots) we use $\as=0.28$, corresponding to $Q\sim2$~GeV, which is a scale right above
the charm mass assumed to be $m_c=1.5$~GeV, while for bottom production we use $\as=0.20$, corresponding to $Q\sim6$~GeV,
right above the bottom mass assumed to be $m_b=4.5$~GeV.
The number of active flavours is set to be $n_f=3$ for charm production and $n_f=4$ for bottom production,
i.e.\ the massive quark is treated as heavy and its collinear logarithms are not factorized.
In particular, the massive coefficients for bottom production are those contributions which should be added to the corresponding
massless coefficients in the same $n_f=4$ scheme, Fig.~\ref{fig:c2cL}, which instead assumed only coupling to light quarks,
to obtain a complete prediction (see e.g.\ the decomposition Eq.~\eqref{eq:deltaCnf} at resummed level).
We observe that the effect of adding the bottom production contribution to the purely massless contributions
is a rather small correction for $F_L$, while it is comparable in size to each individual massless contribution for $F_2$.

The pattern observed in Fig.~\ref{fig:Cmass} between fixed-order and resummed contributions
is very similar to that of the massless results in Fig.~\ref{fig:c2cL}.
The most notable difference is the visibly larger effect of resummation for charm production,
accompanied by a larger uncertainty band.
This effect is entirely due to the smaller scale, i.e.\ the larger value of $\as$, used in the charm production plots.
Another interesting feature of these massive coefficient functions is the very visible presence
of the physical threshold for heavy quark production, which lies at $x=x_{\rm th}\equiv 1/(1+4m^2/Q^2)$.
Because of our choice of scales, $x_{\rm th}\sim0.3$ for both processes.

\begin{figure}[t]
  \centering
  \includegraphics[width=0.495\textwidth,page=2]{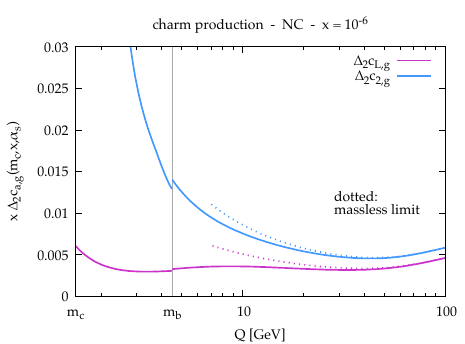}
  \includegraphics[width=0.495\textwidth,page=5]{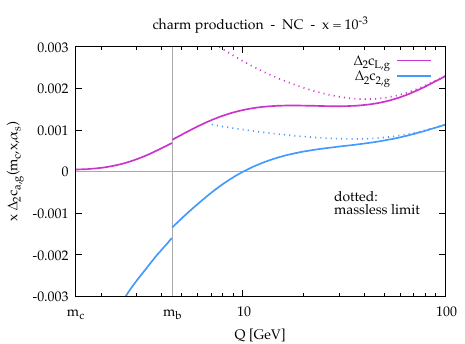}\\
  \includegraphics[width=0.495\textwidth,page=2]{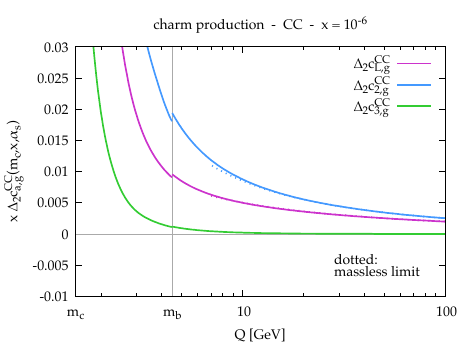}
  \includegraphics[width=0.495\textwidth,page=5]{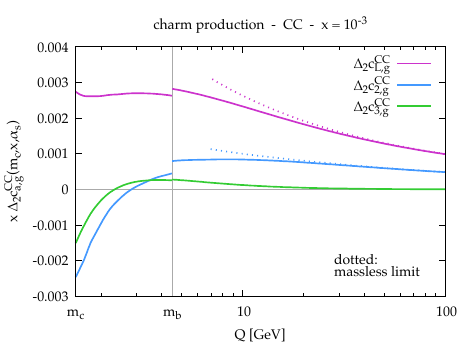}
  \caption{Resummed contributions to the coefficient functions including mass effects in neutral current,
    $\DkC{2}_{L,g}(m_c)$ and $\DkC{2}_{2,g}(m_c)$ (upper plots), and in charged current,
    $\DkC{2}^{\rm CC}_{L,g}(m_c)$, $\DkC{2}^{\rm CC}_{2,g}(m_c)$ and $\DkC{2}^{\rm CC}_{3,g}(m_c)$ (lower plots),
    shown as a function of $Q$ from $m_c$ to $100$~GeV. The left plots show the results for $x=10^{-6}$, the right plots for $x=10^{-3}$.
    The value of $n_f$ changes from 4 to 5 at the bottom matching scale taken to be equal to $m_b$.
    The massless result, to which the massive coefficients tend at large $Q$, is also shown as dotted curves
    (except for $\DkC{2}^{\rm CC}_{3,g}(m_c)$, which tends to zero).}
  \label{fig:scanC}
\end{figure}

The results presented so far do not include the resummation of collinear logarithms due to massive quarks.
For the scales considered, which are in both cases larger than the heavy quark mass, these collinear logarithms
are already usually resummed in most implementations of VFNSs.
We now thus consider the scenario in which a VFNS is used and heavy-quark collinear logarithms are resummed.
Since at fixed order there are various incarnations of VFNSs,
differing just by subleading effects but nonetheless being practically different (see e.g.\ discussion in Sect.~\ref{sec:FONLL}),
we prefer to focus on the resummed contributions only.
We focus on the charm-production case, with $m_c=1.5$~GeV.
For the sake of this study, we find more interesting to show a plot as a function of the momentum transfer $Q$,
in order to emphasize the importance of mass effects at different scales.
Therefore, in Fig.~\ref{fig:scanC} we  plot the resummed contributions $\DkC{2}_{2,g}(m_c)$ and $\DkC{2}_{L,g}(m_c)$ in NC
and $\DkC{2}^{\rm CC}_{L,g}(m_c)$, $\DkC{2}^{\rm CC}_{2,g}(m_c)$ and $\DkC{2}^{\rm CC}_{3,g}(m_c)$ in CC\footnote
{For CC, we assume the production of a charm quark together with a massless anti-quark. This fixes the sign of the $F_3$ contribution.}
as a function of $Q$ for two representative values of $x$, namely $x=10^{-6}$ (small) and $x=10^{-3}$ (moderate).
The plot starts from $Q=m_c$, where $n_f=4$, and then it transitions to $n_f=5$ when crossing the bottom threshold
(assumed to be at $m_b=4.5$~GeV). At the transition point a small discontinuity appears, due to the different
value of $n_f$ used in the computation of the LL$^\prime$ anomalous dimension.
This discontinuity is a standard consequence of the scheme change, and does not constitute any practical problem in the computation of physical observables.

At large $Q$, the massive resummed coefficient functions (which are the collinear subtracted ones) tend to the massless results,
shown in dotted style.
It is clearly visible that charm mass effects are significant for small $Q\lesssim(10\div30)$~GeV, and are more pronounced
at larger $x$, where however the effect of resummation is smaller.
Charm mass effects are also stronger in the NC case than in the CC case.
In practice, massive corrections on the resummed coefficient functions are a small effect on the full structure function,
especially when resummation is matched to NNLO.
Still, keeping into account these mass effects is important for an accurate description of the low-$Q$ data, and in particular
for the charm structure function $F_{a}^c$, $a=2,L$, which is entirely determined by the charm coefficient function.

\begin{figure}[t]
  \centering
  \includegraphics[width=0.5\textwidth]{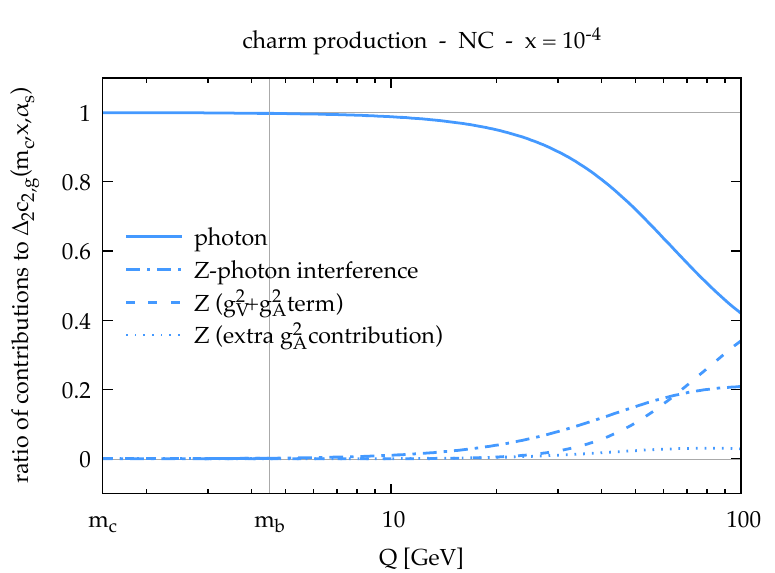}
  \caption{Ratio of the photon, $Z$ and photon-$Z$ interference contributions to the resummed massive
    coefficient function $\DkC{2}_{2,g}(m_c)$, shown for fixed $x=10^{-4}$ as a function of $Q$ from $m_c$ to $100$~GeV.}
  \label{fig:scanCcontributions}
\end{figure}
In the upper plots of Fig.~\ref{fig:scanC} we are showing the full NC coefficients, namely the sum of the contributions
from photon, $Z$ and photon-$Z$ interference to the structure functions, normalized to the photon couplings.
It is interesting to investigate how much the various terms contribute to the full result.
To do so, we show in Fig.~\ref{fig:scanCcontributions} the ratio of the individual contributions to the resummed contribution
to the structure function $F_2$, $\DkC{2}_{2,g}(m_c)$.
We stress that if the axial contribution proportional to $g_A^2$ which remains after factoring out the $g_V^2+g_A^2$ coupling to the $Z$
were absent, then the resummed coefficient function for the various contributions would be identical, up to the overall coupling,
and the ratio of the various contributions would be independent of $x$ and of the observable.
However, since this axial contribution is non-zero, a small dependence remain. However, for $x\lesssim10^{-3}$,
the differences between $F_2$ and $F_L$ are very small, and reducing with lowering $x$.
Thus, the plot in Fig.~\ref{fig:scanCcontributions}, obtained from $F_2$ for $x=10^{-4}$, remains pretty much unchanged
for lower $x$ and for $F_L$.
From the plot we see that the contribution from the $Z$ boson is basically negligible for all $Q\lesssim10$~GeV,
and becomes of the same size of the photon contribution at the $Z$ peak.
The photon-$Z$ interference dominates over the pure $Z$-exchange contribution below the $Z$ peak.
The axial contribution proportional to $g_A^2$, which is a new result of our computation,
turns out to be mostly insignificant, as it gives a small contribution (a few percent at most) for scales $Q$
where small-$x$ resummation is further suppressed by a smaller strong coupling.

\section{Conclusions} \label{sec:conclusions}
In this paper we have performed a comprehensive study of the high-energy, i.e.\ small-$x$,
resummation in deep-inelastic scattering of a lepton off a proton.
In particular, we have collected all the ingredients to perform NLL resummation matched to fixed-order up to NNLO. 

In order to achieve this we have considered the resummation of splitting functions, which govern DGLAP evolution of parton densities.
With respect to our previous work, we have modified the way running coupling corrections are treated and we have managed to match the resummation to NNLO,
thus obtaining state-of-the art NNLO+NLL results for the splitting kernels.
While fixed-order predictions at NNLO exhibit instabilities at small $x$ due to large logarithms,
the resummed results are stable and at small $x$ appear to be much closer to NLO than NNLO.
Furthermore, our results can be easily extended to match NLL to fixed N$^3$LO, when it becomes available. 
In this case we expect the resummation to have an even more substantial effect because of the larger fixed-order instabilities at small $x$ appearing at this order.

We have also considered the resummation of DIS partonic coefficient functions.
In order to obtain reliable results in a wide range of $x$ and $Q^2$ we have studied small-$x$ resummation in the context
of a variable flavour number scheme in which heavy and light quark coefficient functions are matched together.
In this context, we have considered mass effects originating from both the charm and the bottom quarks.
We have produced NNLO+NLL results for the coefficient functions relevant for $F_2$ and $F_L$ for neutral-current DIS,
considering the effect of both a virtual photon or a $Z$ boson exchange, as well as charged-current processes.  
If all quarks are massless, the structure function $F_3$ is purely non-singlet and therefore is not enhanced at small-$x$.
However, we have found that in the charged-current case with $W$ boson exchange, if at least one of the quarks interacting with the $W$
is massive there is a non-zero contribution at small $x$ to the parity violating structure function $F_3$.
We have also noted that in neutral-current DIS with massive quarks there is a difference between the $\gamma$ exchange or $Z\gamma$ interference
and the pure $Z$ exchange, other than the overall coupling.

We have implemented all these new results in a new version of our code, \texttt{HELL} version \texttt{2.0}.
A fast interface to these results is available through a new version of its companion code, \texttt{HELL-x} version \texttt{2.0}.
Both codes are publicly available at the webpage \href{https://www.ge.infn.it/~bonvini/hell/}{\texttt{www.ge.infn.it/$\sim$bonvini/hell}}.
The fast \texttt{HELL-x 2.0} code can be directly used to compute PDF evolution and DIS cross sections through the public code \texttt{APFEL}~\cite{Bertone:2013vaa}.

The main motivation behind this work was to compute and implement all the ingredients that are necessary to perform a state-of-the-art fit of parton distribution functions which consistently includes small-$x$ resummation both in the evolution of the parton densities and in the coefficient functions.
This task is being now pursued by the NNPDF collaboration.
Preliminary and very encouraging results have been presented in~\cite{Rottoli:2017ifw} in the case of a PDF fit that includes DIS-only data.
For the near future, we look forward to implementing other processes in \texttt{HELL}, the most relevant of which in the context of PDF extractions is the production of lepton pair via the Drell-Yan mechanism. 
We conclude by noting that the resummed results produced by \texttt{HELL}, both splitting functions and coefficient functions, are supplemented by a band representing the theoretical uncertainty due to missing higher-logarithmic orders. This information can (and should) be used in phenomenological studies and it could be also be fed into PDF fits together with other sources of theoretical uncertainty. However, the debate about how to best include theoretical uncertainties into PDF fits is not settled yet, see for instance~\cite{Forte:2010dt,Forte:2013wc,Ball:2017nwa,Ball:2015oha}.

\acknowledgments
{
We are always indebted to Richard Ball and Stefano Forte for many discussions and guidance on this topic. 
We are particularly grateful to Tiziano Peraro and Luca Rottoli for discussions and collaboration in the early stages of this project.
We also thank Giovanni Ridolfi and Juan Rojo for discussions and comments on our manuscript.
The work of MB was partly supported by the Marie Sk\l{}odowska Curie grant HiPPiE@LHC.
}

\appendix

\section{Off-shell DIS coefficient functions}\label{sec:DISoffshell}

\begin{figure}[t]
  \centering
  \includegraphics[trim=5.5cm 19.6cm 5.5cm 4.4cm,clip,width=0.7\textwidth]{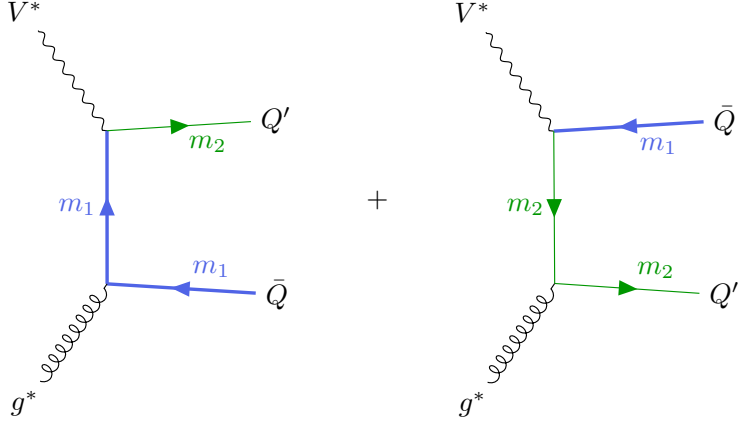}
  \caption{Diagrams entering the computation of the DIS coefficient function with off-shell incoming gluons at lowest order.
    The (off-shell) vector boson can either be a photon, a $Z$ or a $W$. In the latter case, the quark changes flavour after interacting with it,
    and thus does the mass.}
  \label{fig:diagramsDIS}
\end{figure}

In this section we report the computation and the results for the off-shell coefficient functions needed for the small-$x$ resummation of DIS observables.
We focus on the contributions with an incoming gluon, which is off its mass-shell, as this is what enters in the resummation formula.
The relevant diagrams are shown in Fig.~\ref{fig:diagramsDIS}.
The off-shellness of the incoming gluon regulates the collinear divergence of the produced quark pair,
so the case with massless quarks is just a finite limit of the case with massive quarks.
Therefore, we start considering the most general case, in which the gluon converts to a
massive quark (pair), with mass $m_1$, and then after interacting with the vector boson the quark changes mass, $m_2$
(and of course the opposite setup, as in the right diagram of Fig.~\ref{fig:diagramsDIS}).
This accounts for all possible type of interaction:
\begin{itemize}
\item $m_1=m_2$ is the case in which the boson is either a photon or a $Z$ (neutral current)
\item $m_1\neq m_2$ is the case in which the boson is a $W$ (charged current).
\end{itemize}
Additionally, we consider a generic coupling which includes vector and axial currents,
even though we will see that the couplings will mostly factor out.

Note that, in practice, the charged-current case is relevant only if one of the two quark is massive and the other massless.
Indeed only charm, bottom and top masses cannot be neglected, but the top contribution to DIS is always negligible.
Therefore, the only processes for which two different non-zero masses would be needed are $c+W\to b$ and $b+W\to c$.
But these processes are suppressed by the CKM matrix element $V_{cb}\sim 4\, 10^{-2}$,
and by the phase-space restrictions due to the quark masses.
Therefore, the only significant combinations in charged current will involve at most a single massive quark.
These combinations are $c+W\to s,d$ and $s,d+W\to c$,
on top of the fully massless contributions $u+W\to s,d$ and $s,d+W\to u$.

\subsection{Calculation of DIS off-shell partonic cross section}

In this section we summarize the computation of the DIS off-shell partonic cross section in the most general case
in which the quarks in the final state have different masses and their coupling to the boson contains vector and axial components.
We consider the process
\beq
\label{eq:DISprocess}
V^*\(q\) + g^*\(k\) \to \bar{Q}\(p_3\)+Q'\(p_4\),
\eeq
where $V^*$ is the generic off-shell vector boson, $g^*$ is the off-shell gluon and $Q$ and $Q'$ are quarks.
The final state quarks are on-shell, and their flavour is in general different (to cover the charged-current case), so their masses are $p_3^2=m_1^2$,  $p_4^2=m_2^2$.
The two diagrams contributing to this process are depicted in Fig.~\ref{fig:diagramsDIS}.
The invariant matrix element for the two diagrams is given by
\beq
\label{eq:MDIS}
i\mathcal{M}^{\mu\rho}=-i g_s t^c \bar{u}\(p_4\)\left[\gamma^{\mu}\(g_V+g_A\gamma^5\)\frac{\slashed{k}\gamma^\rho- 2 p_3^\rho}{t-m_1^2}
  -\frac{\gamma^{\rho}\slashed{k}-2p_4^\rho}{u-m_2^2}\gamma^\mu\(g_V+g_A\gamma^5\)\right]v\(p_3\),
\eeq
where $g_s, g_V, g_A$ are the strong, vector and axial couplings respectively, and $t=\(k-p_3\)^2=\(q-p_4\)^2$ and $u=\(k-p_4\)^2=\(q-p_3\)^2$.
For photon-induced DIS, the vector coupling is just the quark electric charge $g_V^\gamma=e_Q$
and the axial coupling is zero.
For $Z$-induced DIS, the vector and axial couplings depend on the quark isospin and are given by
\begin{align}
  g_V^Z &=
  \begin{cases}
    +\frac12-2e_Q\sin^2\theta_{\rm w} &Q=u,c,t\\
    -\frac12-2e_Q\sin^2\theta_{\rm w} &Q=d,s,b
  \end{cases},
&
  g_A^Z &=
  \begin{cases}
    +\frac12 &Q=u,c,t\\
    -\frac12 &Q=d,s,b
  \end{cases},
\end{align}
where $\theta_{\rm w}$ is the weak mixing angle.
For $W$-induced DIS the vector and axial couplings depend on both quark flavours and are given by
\begin{align}
  g_V^W &= \frac1{\sqrt{2}}V_{QQ'}, &
  g_A^W &= -\frac1{\sqrt{2}}V_{QQ'},
\end{align}
being $V_{ij}$ the CKM matrix.
Note that we are assuming that the vector boson is just $V$, and so the computation will not cover explicitly
the photon-$Z$ interference: however, this case is easily obtained in the final results by simply replacing
$g_V^2\to 2g_V^Zg_V^\gamma$ and $g_A^2\to0$
(the combination $g_Vg_A$ appears only in $F_3$, the gluonic contribution of which is zero in neutral current).

In the high-energy regime we are interested in, we decide to parametrize the kinematics of the process
in terms of dimensionless variables $z_1$, $z_2$, $\tau$, $\bar\tau$ and transverse vectors\footnote
{A transverse vector $p_\perp$ is defined to have components only along directions orthogonal to the time and $z$ directions,
i.e.\ in components $p_\perp=(0,p_x,p_y,0)=(0,\mathbf{p},0)$, where we have also defined the two-vector $\mathbf{p}$.
Note that $p_\perp^2=-\mathbf{p}^2$.}
$q_\perp$, $k_\perp$ and $\Delta_\perp$ defined by
\begin{subequations}
\label{eq:kinematics}
\begin{align}
q^\mu&= z_1 p_1^\mu+ q_\perp^\mu, \\
k^\mu&= z_2 p_2^\mu- k_\perp^\mu, \\
p_3^\mu&=\(1-\tau\) z_1 p_1^\mu + \bar{\tau} z_2 p_2^\mu + \(q_\perp^\mu-\Delta_\perp^\mu\),\\
p_4^\mu&=\tau z_1 p_1^\mu + \(1-\bar{\tau}\) z_2 p_2^\mu - \(k_\perp^\mu-\Delta_\perp^\mu\),
\end{align}
\end{subequations}
where $p_1=\frac{\sqrt{s}}{2}\(1,0,0,1\)$ and $ p_2=\frac{\sqrt{s}}{2}\(1,0,0,-1\)$ are light-cone vectors.
We also define the momentum transferred $Q$ by $q^2=-Q^2$. It is important to note that this definition is valid only in the high-energy limit, where $s \gg M_V^2$.

We define the parton-level \emph{off-shell} hadronic tensor as the squared of the matrix element Eq.~\eqref{eq:MDIS}
averaged over the off-shell gluon polarizations~\cite{Catani:1990eg,Caola:2010kv,Caola:2016upw}
\beq
\label{eq:offshellW}
{\cal W}_{\mu\nu}=-\frac{k_\perp^\rho k_\perp^\sigma}{k_{\perp}^2} \(\mathcal{M}_{\nu\sigma}\)^\dagger \mathcal{M}_{\mu\rho}.
\eeq
This object contains all the information on the DIS cross-section from the hadronic side.
It can be decomposed into different contributions with a given tensor structure as\footnote
{Note that in the photon-mediated DIS the contributions ${\cal W}_4$ and ${\cal W}_5$ can be related to ${\cal W}_1$ and ${\cal W}_2$ through Ward identities.
However, since we want to cover also the more general $Z$- and $W$-mediated DIS processes, we must keep them separate.
Nevertheless, their contribution to the DIS cross section, as well as the one from ${\cal W}_6$,
is of the order of the lepton mass and thus negligible, and will not be further considered.}
\beq
{\cal W}_{\mu\nu}=
-g_{\mu\nu}{\cal W}_1
+ k_\mu k_\nu {\cal W}_2
-i\epsilon_{\mu\nu\rho\sigma}k^\rho q^\sigma {\cal W}_3
+q_\mu q_\nu {\cal W}_4
+\(k_\mu q_\nu +q_\mu k_\nu\){\cal W}_5
+i \(k_\mu q_\nu -q_\mu k_\nu\){\cal W}_6.
\eeq
The various contributions to the structure functions
$F_1$, $F_2$ and $F_3$ (and $F_L=F_2-2xF_1$) are obtained from the respective counterparts in the hadronic tensor ${\cal W}_1$, ${\cal W}_2$ and ${\cal W}_3$.
These, in turn, can be obtained from the full tensor ${\cal W}_{\mu\nu}$ using suitable projector operators.
To this end, we define
\begin{subequations}\label{eq:squaredamplitudes}
\begin{align}
\abs{A}_2^2&=\frac{q_\perp^\mu q_\perp^\nu}{Q^2}{\cal W}_{\mu\nu}, \\
2\abs{A}_2^2 -3\abs{A}_L^2 = \abs{A}_g^2&=\(-g^{\mu\nu}+\frac{q^{\mu} q^{\nu}}{q^2}\) {\cal W}_{\mu\nu}, \\
\abs{A}_3^2&=-i\epsilon_{\mu\nu\alpha\beta}\frac{p_2^\alpha q^\beta}{4\(p_2 \cdot q\)} {\cal W}^{\mu\nu},
\end{align}
\end{subequations}
where $\abs{A}_2^2$, $\abs{A}_L^2$ and $\abs{A}_3^2$ are directly related to $F_2$, $F_L$ and $F_3$, respectively.
Using the definitions of the kinematic variables in the high-energy limit, Eqs.~\eqref{eq:kinematics},
the squared amplitudes Eqs.~\eqref{eq:squaredamplitudes} can be rewritten as
\begin{subequations}
\label{eq:amplitudes}
\begin{align}
\abs{A}_2^2&=8g_s^2\left[\frac{\(p_1 \cdot k\)\(p_2 \cdot q\)}{\(p_1 \cdot p_2\)}\right]^2 \notag\\
&\times \Bigg\{\frac{\(g_V^2+g_A^2\)}{\(m_1^2-t\)\(m_2^2-u\)}\left[1-\frac{1}{q^2 k^2}\(1-2\frac{\(p_1\cdot p_3\)\(p_2 \cdot p_4\)}{m_2^2-u}-2\frac{\(p_1 \cdot p_4\)\(p_2 \cdot p_3\)}{m_1^2-t}\)^2\right]\notag\\
&\qquad -g_A^2\frac{\(m_1+m_2\)^2}{q^2}-g_V^2\frac{\(m_1-m_2\)^2}{q^2}\frac{1}{\(m_1^2-t\)\(m_2^2-u\)}\Bigg\},\\
\abs{A}_g^2&=8g_s^2 \frac{\(p_1\cdot k\)^2}{\(p_1 \cdot p_2\)^2}\Bigg\{\(g_V^2+g_A^2\)\Bigg[\frac{\(p_2 \cdot p_3\)^2+\(p_2 \cdot p_4\)^2+\frac{\(m_1^2+m_2^2\)}{2q^2} \(p_2 \cdot q\)^2}{\(m_1^2-t\)\(m_2^2-u\)}\notag\\
&\qquad\qquad\qquad\qquad\qquad\qquad+\(\frac{m_1^2+m_2^2-2q^2}{2k^2}+\frac{\(m_1^2-m_2^2\)^2}{2k^2 q^2}\)\(\frac{\(p_2 \cdot p_3\)}{m_1^2-t}-\frac{\(p_2 \cdot p_4\)}{m_2^2-u}\)^2\Bigg]\notag\\
&\qquad -\(g_V^2-g_A^2\)\Bigg[\frac{3 m_1 m_2}{k^2}\(\frac{\(p_2 \cdot p_3\)}{m_1^2-t}-\frac{\(p_2 \cdot p_4\)}{m_2^2-u}\)^2+\frac{m_1 m_2}{q^2}\frac{\(p_2 \cdot q\)^2}{\(m_1^2-t\)\(m_2^2-u\)}\Bigg]\Bigg\},\\
\abs{A}_3^2&= 4 g_s^2 \(2 g_V g_A\) \frac{\(p_1 \cdot k\)^2}{\(p_1 \cdot p_2\)^2}\notag\\
&\times\Bigg\{\frac{\(p_2 \cdot p_3\)-\(p_2 \cdot p_4\)}{\(p_2 \cdot q\)}\left[\frac{q^2}{k^2}\(\frac{\(p_2\cdot p_3\)}{m_1^2-t}-\frac{\(p_2 \cdot p_4\)}{m_2^2-u}\)^2-\frac{\(p_2\cdot q\)^3}{\(m_1^2-t\)\(m_2^2-u\)}\right]\notag\\
&\qquad -\frac{m_1^2-m_2^2}{k^2}\(\frac{\(p_2 \cdot p_3\)}{m_1^2-t}-\frac{\(p_2 \cdot p_4\)}{m_2^2-u}\)^2\Bigg\}.
\end{align}
\end{subequations}
These squared amplitudes must be integrated over the final state two-body phase space.
In terms of the kinematic variables Eqs.~\eqref{eq:kinematics} we can express it as
\begin{align}
\label{eq:Phi}
d\Phi&=\frac{\nu}{2\pi^2} d\tau\,d\bar{\tau}\,d^2\mathbf{\Delta}\,\delta\(\bar{\tau}\(1-\tau\)\nu-\(\mathbf{q}-\mathbf{\Delta}\)^2-m_1^2\)
\,\delta\(\tau\(1-\bar{\tau}\)\nu-\(\mathbf{k}-\mathbf{\Delta}\)^2-m_2^2\) \nonumber\\
&=\frac{1}{8\pi^2}\frac{d\tau}{\tau\(1-\tau\)}d^2\tilde{\mathbf{\Delta}}\,\delta\(\nu-\frac{\tilde{\mathbf{\Delta}}^2+\(1-\tau\)m_2^2+\tau m_1^2}{\tau\(1-\tau\)}-\(\mathbf{q}-\mathbf{k}\)^2\),
\end{align}
where we have further defined
\begin{align}
\nu&=2z_1 z_2\(p_1\cdot p_2\),\\
\tilde{\mathbf{\Delta}}&=\mathbf{\Delta}-\tau \mathbf{q}-\(1-\tau\) \mathbf{k}.
\end{align}
We recall that bold symbols represent the two-dimensional components of a transverse vector.

The partonic off-shell coefficient functions (in Mellin space) for the three structure functions we are interested in are given
in terms of the squared amplitudes of Eqs.~\eqref{eq:squaredamplitudes} by
\begin{subequations}\label{eq:calCdefs}
\begin{align}
{\cal C}_2(N,\xi,\xi_{m_1},\xi_{m_2})&=\frac{1}{4\pi^2 \(g_V^2+g_A^2\)}\int_0^1 d\eta\,\eta^N\int d\Phi\, \abs{A}_2^2,\\
{\cal C}_L(N,\xi,\xi_{m_1},\xi_{m_2})&=\frac{1}{4\pi^2 \(g_V^2+g_A^2\)}\int_0^1 d\eta\,\eta^N\int d\Phi\, \frac{1}{3}\left[2\abs{A}_2^2- \abs{A}_g^2\right],\\
{\cal C}_3(N,\xi,\xi_{m_1},\xi_{m_2})&=\frac{1}{4\pi^2 \(2 g_V g_A\)}\int_0^1 d\eta\,\eta^N\int d\Phi\, \abs{A}_3^2,
\end{align}
\end{subequations}
where we have introduced the variable
\beq
\eta=\frac{Q^2}{\nu}
\eeq
and we are expressing the result in terms of dimensionless ratios
\begin{align}
\xi&=\frac{\mathbf{k}^2}{Q^2}, \qquad \xi_m=\frac{m^2}{Q^2}. 
\end{align}
Note that, as explained in the text, we are only interested in the $N=0$ Mellin moment.

To carry out the integrations in Eqs.~\eqref{eq:calCdefs} it is useful to express the following combinations
\begin{subequations}
\label{eq:Man}
\begin{align}
m_1^2-t&=\frac{1}{\tau}\left[\(1-\tau\)m_2^2+\tau m_1^2+\(\tilde{\mathbf{\Delta}}-\tau \mathbf{k}\)^2+\tau\(1-\tau\)Q^2\right],\\
m_2^2-u&=\frac{1}{1-\tau}\left[\(1-\tau\)m_2^2+\tau m_1^2+\(\tilde{\mathbf{\Delta}}+\(1-\tau\)\mathbf{k}\)^2+\tau\(1-\tau\)Q^2\right],\\
\bar{\tau}&=\frac{\left[\(1-\tau\)\(\mathbf{q}-\mathbf{k}\)-\tilde{\mathbf{\Delta}}\right]^2+m_1^2}{\(1-\tau\)\nu},\\
\(1-\bar{\tau}\)&=\frac{\left[\tilde{\mathbf{\Delta}}+\tau \(\mathbf{q}-\mathbf{k}\)\right]^2+m_2^2}{\tau \nu},
\end{align}
\end{subequations}
in terms of the phase-space variables.
In addition, it is convenient to use the following Feynman parametrizations
\begin{align*}
\frac{1}{\(m_1^2-t\)\(m_2^2-u\)}&=\int_0^1 dy\frac{\tau\(1-\tau\)}{\left[\(1-\tau\)m_2^2 +\tau\, m_1^2+\(\tilde{\mathbf{\Delta}}+\(y-\tau\)\mathbf{k}\)^2+\tau\(1-\tau\)\mathbf{q}+y\(1-y\) \mathbf{k}^2\right]^2},\\
\scriptsize
\frac{1}{\(m_1^2-t\)^2\(m_2^2-u\)^2}&=\int_0^1 dy\frac{6\tau^2\(1-\tau\)^2 y\(1-y\)}{\left[\(1-\tau\)m_2^2 +\tau\,m_1^2+\(\tilde{\mathbf{\Delta}}+\(y-\tau\)\mathbf{k}\)^2+\tau\(1-\tau\)\mathbf{q}+y\(1-y\) \mathbf{k}^2\right]^4}.
\end{align*}
In this way, most integrations are easy to perform,\footnote
{Note that using the $\delta$ function in Eq.~\eqref{eq:Phi} to perform the $\eta$ integration imposes restrictions on the remaining integrations.
  In particular, the restriction can be cast as a lower integration limit for $\tilde{\mathbf{\Delta}}^2$.
  However, in the high-energy regime, this lower bound is immaterial, and the integral can be extended down to zero.}
leaving the results in the form of a double integral in $y$ and $\tau$.
In order to simplify this integration (and to make contact with previous literature) it is also convenient
to perform the change of variables
\begin{align}
x_1&=4y\(1-y\), \qquad x_2=4\tau\(1-\tau\),
\end{align}
and express the results as integrals over $x_1$ and $x_2$.
General results for ${\cal C}_a\(0,\xi,\xi_{m_1},\xi_{m_2}\)$, $a=2,L,3$ are rather long and will not be reported here;
in the next section we focus on the physical combinations that are relevant for neutral and charged currents,
where the masses are either equal or at least one of them is vanishing.

\subsection{Results}

In this section we collect the results of the off-shell coefficient functions for neutral and charged currents,
contributing to the three structure functions $F_2$, $F_L$ and $F_3$.
The fully massless case $m_1=m_2=0$, which is common to neutral and charged currents, is of course the simplest limit and yields
\begin{subequations}
\label{eq:Cgmasslessapp}
\begin{align} 
{\cal C}_2\(0,\xi,0,0\)&=\frac{\as}{3\pi}\int_0^1\frac{dx_1}{\sqrt{1-x_1}}\int_0^1\frac{dx_2}{\sqrt{1-x_2}}\frac{3}{8\(\xi x_1+x_2\)^3}\notag\\
&\qquad\qquad\times\[(2-x_1) x_2^2+ x_1 x_2 \xi (3 x_1+3 x_2-4 x_1 x_2)+(2-x_2)x_1^2 \xi^2 \]\\
{\cal C}_L\(0,\xi,0,0\)&=\frac{\as}{3\pi}\int_0^1 \frac{dx_1}{\sqrt{1-x_1}}\int_0^1\frac{dx_2}{\sqrt{1-x_2}}\frac{x_2^2\(x_2+x_1\xi\(5-4x_1\)\)}{4\(\xi x_1+x_2\)^3}\\
{\cal C}_3\(0,\xi,0,0\)&=0.
\end{align}
\end{subequations}
These results coincide with those presented in Ref.~\cite{Catani:1994sq}, even though the longitudinal coefficient function
was not written explicitly there. An equivalent (but simpler) integral form for ${\cal C}_L(0,\xi,0,0)$ was also given in Ref.~\cite{Bonvini:2016wki}.

We now move to the case in which both masses are equal, $m_1=m_2\equiv m$, which is relevant for neutral current.
The results read
\begin{subequations}
\label{eq:Cgmassiveapp}
\begin{align}
{\cal C}_2\(0,\xi,\xi_m,\xi_m\)
&=\frac{\as}{3\pi}\int_0^1\frac{dx_1}{\sqrt{1-x_1}}\int_0^1 \frac{dx_2}{\sqrt{1-x_2}}\frac{3}{8\({4\xi_m}+\xi x_1+x_2\)^3}\\
&\times\Bigg\{\Big[2\xi^2 x_1^2-4\xi x_1^2x_2^2-\xi^2 x_1^2 x_2+3\xi x_1^2 x_2+3\xi x_1 x_2^2-x_1 x_2^2+2x_2^2\notag\\
&\qquad+{4\xi_m}\(-\xi x_1^2 x_2+4\xi x_1-x_1 x_2^2-\xi x_1 x_2 -x_1 x_2+4x_2\)+16\xi_m^2\(2-x_1 x_2\)\Big]\notag\\
&\qquad+\frac{8 g_A^2 \xi_m }{\(g_V^2+g_A^2\)}\({4\xi_m}+\xi x_1+ x_2\)^2\Bigg\},\notag\\
{\cal C}_L\(0,\xi,\xi_m,\xi_m\)
&=\frac{\as}{3\pi}\int_0^1\frac{dx_1}{\sqrt{1-x_1}}\int_0^1 \frac{dx_2}{\sqrt{1-x_2}}\frac{1}{4\({4\xi_m}+\xi x_1+x_2\)^3}\\
&\times\Bigg\{\Big[x_2^2\(x_2+x_1\xi\(5-4 x_1\)\)+2\xi_m x_2\(\xi x_1^2-2\xi x_1-3 x_1 x_2 +4 x_2\)+8\xi_m^2 x_2\(2-3 x_1\)\Big]\notag\\
&\qquad+\frac{ 6 g_A^2 \xi_m}{\(g_V^2+g_A^2\)}\Big[2\(4\xi_m+\xi x_1\)^2+\(2+x_1\) x_2^2+x_2\(\xi\(8-3x_1\)x_1+4\xi_m\(4+x_1\)\)\Big]\Bigg\},\notag\\
{\cal C}_3\(0,\xi,\xi_m,\xi_m\)&=0.
\end{align}
\end{subequations}
Note the presence in the above expressions of a term proportional to $g_A^2/(g_V^2+g_A^2)$. This contribution
is present only when the vector boson is a $Z$, while for photon and photon-$Z$ interference this term is zero.
This axial contribution is a new result. The remaining of the expressions were already known~\cite{Catani:1990xk,Catani:1992zc,Catani:1990eg},
however an explicit integral form of this kind for ${\cal C}_L$ is presented here for the first time.
Of course, the massless $\xi_m\to0$ limit of Eqs.~\eqref{eq:Cgmassiveapp} reduces to Eqs.~\eqref{eq:Cgmasslessapp}.

Finally, we move to the case in which one quark is massless (say, $m_2=0$) and the other is massive ($m_1\equiv m$), which is relevant for charged current.
According to the definition of the process, Eq.~\eqref{eq:DISprocess}, this choice corresponds to the production of a heavy anti-quark.
Here, it is most convenient to leave integration over $\tau$ untouched and to change variable only from $y$ to $x_1$.
We thus obtain
\begin{subequations}
\label{eq:Cgsinglemassiveapp}
\begin{align}
{\cal C}_2\(0,\xi,\xi_m,0\)
&=\frac{\as}{3\pi}\int_0^1 d\tau\int_0^1 \frac{dx_1}{\sqrt{1-x_1}}\frac{3}{2\(4\tau \xi_m+\xi x_1 +4\tau\(1-\tau\)\)^3}\\
&\times\Bigg\{16\(\xi_m+1\)\tau^2\(\xi_m+1-\tau\)^2+\xi^2 x_1^2\(\xi_m+\(1-\tau\)^2+\tau^2\)\notag\\
&\qquad+2\xi x_1^2 \(1-\tau\)\tau\(3\xi_m^2+\xi_m\(6-16\tau\)+16\tau^2-16\tau+3\)\notag\\
&\qquad+8\tau^2 x_1\[\xi\(\xi_m^2+\xi_m\(4-3\tau\)+3\(1-\tau\)^2\)-\(\xi_m+1\)^2\(1-\tau\)\(\xi_m+1-\tau\)\]\Bigg\},\notag\\
{\cal C}_L\(0,\xi,\xi_m,0\)
&=\frac{\as}{3\pi}\int_0^1 d\tau\int_0^1\frac{dx_1}{\sqrt{1-x_1}}\frac{1}{\(4\tau \xi_m+\xi x_1 +4\tau\(1-\tau\)\)^3}\\
&\times\Bigg\{16\tau^2\(\xi_m+1-\tau\)^2\(3\xi_m+4\(1-\tau\)\tau\)+3\xi^2 x_1^2\xi_m\notag\\
&\qquad+2\xi x_1^2 \(1-\tau\)\tau\(9\xi_m^2+\xi_m\(9-32 \tau\)-32\tau\(1-\tau\)\)\notag\\
&\qquad+8\tau^2 x_1\[\xi\(3\xi_m^2+10\xi_m\(1-\tau\)+10\(1-\tau\)^2\)-3\xi_m\(\xi_m+1\)\(1-\tau\)\(\xi_m+1-\tau\)\]\Bigg\},\notag\\
{\cal C}_3\(0,\xi,\xi_m,0\)
&=\frac{\as}{3\pi}\int_0^1 d\tau\int_0^1 \frac{dx_1}{\sqrt{1-x_1}}\frac{3}{2\(4\tau \xi_m+\xi x_1 +4\tau\(1-\tau\)\)^3}\\
&\times\Bigg\{16\tau^2\(2\tau-1\)\(\xi_m+1-\tau\)^2+\xi x_1^2\(\xi\(2\tau-1\)-6\(1-\tau\)\tau\(\xi_m+1-2\tau\)\)\notag\\
&\qquad+8\tau^2 x_1\[\(1-\tau\)\(\xi_m^2+\xi_m\(2-3\tau\)+2\tau^2-3\tau+1\)+\xi \xi_m\]\Bigg\}.\notag
\end{align}
\end{subequations}
To the best of our knowledge, these results are all new. Most notably, ${\cal C}_3$ does not vanish in this case.
Note that choosing $m_2=m$, $m_1=0$ corresponds to charge-conjugating the final state, thus producing a heavy quark.
Therefore, ${\cal C}_a\(0,\xi,0,\xi_m\) = {\cal C}_a\(0,\xi,\xi_m,0\)$ for $a=2,L$,
while there is a sign change in the parity-violating coefficient, ${\cal C}_3\(0,\xi,0,\xi_m\) = -{\cal C}_3\(0,\xi,\xi_m,0\)$
(see also Ref.~\cite{Buza:1997mg,Blumlein:2014fqa}).
As expected, the massless limit of Eqs.~\eqref{eq:Cgsinglemassiveapp} reduces to Eqs.~\eqref{eq:Cgmasslessapp}.

We now consider the Mellin transform with respect to $\xi$ of these results. This is particularly useful
for studying the massless limit of the resummed result, and for asymptotic expansions.
We denote the Mellin transform with a tilde, and replace the argument $\xi$ with the Mellin moment $M$:
\beq
\tilde{\cal C}_a\(N,M,\xi_{m_1},\xi_{m_2}\) = M \int_0^\infty d\xi\, \xi^{M-1} {\cal C}_a\(N,\xi,\xi_{m_1},\xi_{m_2}\), \qquad a=2,L,3.
\eeq
In the massless case we find
\begin{subequations}
\label{eq:Cgtildemasslessapp}
\begin{align}
\tilde{\cal C}_2\(0,M,0,0\) &= \frac{\as}{3\pi}\, \frac{\Gamma^3(1-M)\Gamma^3(1+M)}{\Gamma(2-2M)\Gamma(2+2M)} \, \frac{3(2+3 M-3 M^2)}{2M(3-2M)} ,\\
\tilde{\cal C}_L\(0,M,0,0\) &= \frac{\as}{3\pi}\, \frac{\Gamma^3(1-M)\Gamma^3(1+M)}{\Gamma(2-2M)\Gamma(2+2M)} \, \frac{3(1-M)}{3-2M} , \\
\tilde{\cal C}_3\(0,M,0,0\) &= 0,
\end{align}
\end{subequations}
which reproduce the results of Ref.~\cite{Catani:1994sq}.
In the massive case in neutral current we obtain
\begin{subequations}
\label{eq:Cgtildemassiveapp}
\begin{align}
\tilde{\cal C}_2\(0,M,\xi_m,\xi_m\)
&=\frac{\as}{3\pi}\, \frac32\xi_m^M\frac{\Gamma^3\left(1-M \right) \Gamma\left(1+M \right)}{(3-2M)(1+2M)\Gamma\left(2-2M \right)}\\
&\times \Bigg\{ 1+M-\[1+M-\frac{2+ 3 M -3 M^2}{2\xi_m} \] \,{}_2F_1 \left(1-M,1,\frac{3}{2}; -\frac1{{4\xi_m}} \right)\nonumber\\
&\qquad-\frac{2g_A^2}{g_A^2+g_V^2}\frac{1+2M}{1-2M}\bigg[8\xi_m\(M-2\xi_m-2\)\nonumber\\
&\qquad\qquad\qquad\qquad\qquad\quad+\({4\xi_m}+1\)^2 \,{}_2F_1\(1-M,1,-\frac{1}{2},-\frac{1}{{4\xi_m}}\)\bigg]\Bigg\},\nonumber\\
\tilde{\cal C}_L\(0,M,\xi_m,\xi_m\)
&=\frac{\as}{3\pi}\, \frac32\xi_m^M\frac{\Gamma^3\left(1-M \right) \Gamma\left(1+M \right)}{(3-2M)(1+2M)\Gamma\left(2-2M \right)} \frac{{4\xi_m}}{1+{4\xi_m}} \\
&\times \Bigg\{3+\frac{1-M}{2\xi_m} -\[3+\frac{1-M}{\xi_m} \left(1- \frac{M}{{4\xi_m}} \right)  \]\,{}_2F_1 \left(1-M,1,\frac{3}{2}; -\frac1{{4\xi_m}} \right)\nonumber\\
&\qquad-\frac{g_A^2}{g_A^2+g_V^2}\frac{\(1+4\xi_m\)\(1+2M\)}{12\xi_m^2}  \bigg[6\xi_m\(2M-3\) \,{}_2F_1\(2-M,1,\frac{3}{2},-\frac1{{4\xi_m}}\)\nonumber\\
&\qquad\qquad\qquad\qquad\qquad\qquad\qquad\qquad+\(3M-4\)\,{}_2F_1\(2-M,2,\frac{5}{2},-\frac1{{4\xi_m}}\)\bigg]\Bigg\},\nonumber\\
\tilde{\cal C}_3\(0,M,\xi_m,\xi_m\) &=0.
\end{align}
\end{subequations}
Apart from the contribution proportional to the axial coupling $g_A^2$, which is new,
the other terms reproduce the results of Refs.~\cite{Catani:1990xk,Catani:1992zc,Catani:1990eg}.
Finally, the massive case in charged current is given by
\begin{subequations}
\label{eq:Cgtildesinglemassiveapp}
\begin{align}
\tilde{\cal C}_2\(0,M,\xi_m,0\)
&= \frac{\as}{3\pi}\frac{3}{4}\xi_m^M \frac{\Gamma\(1-M\)^3 \Gamma\(1+M\)}{ M\(3-2M\)\(4M^2-1\)\Gamma\(2-2M\)}\frac{1}{\xi_m}\\
&\times\Bigg\{M\(M^2\(\xi_m^2-2\xi_m-3\)-M\(\xi_m^2-4\xi_m-3\)+\xi_m+2\)\notag\\
&\qquad+\(1-M\)\(1+\xi_m\)\(M^2\(\xi_m+3\)-3M-2\)\,{}_2F_1\(2M-1,1,M+1,\frac{1}{1+\xi_m}\)\Bigg\},\notag\\
\tilde{\cal C}_L\(0,M,\xi_m,0\)
&= \frac{\as}{3\pi} \frac{3}{4}\xi_m^M \frac{\Gamma\(1-M\)^3\Gamma\(1+M\)}{M\(3-2M\)\(1+2M\)\(1-2M\)\Gamma\(2-2M\)}\frac{1}{\xi_m}\\
&\times\Bigg\{2M^3\(3\xi_m+1\)-M^2\(\xi_m^2+11 \xi_m+2\)-M \xi_m\(\xi_m-2\)+2\xi_m^2 \notag\\
&\qquad +\(M-1\)\(2M^2\(\xi_m+1\)+M\(\xi_m^2-5\xi_m-2\)+2\xi_m^2\)\notag\\
&\qquad\qquad\qquad\times{}_2F_1\(2M-1,1,M+1,\frac{1}{1+\xi_m}\)\Bigg\},\notag\\
\tilde{\cal C}_3\(0,M,\xi_m,0\)
&=\frac{\as}{3\pi}\frac{3}{2}\xi_m^M\frac{\(1-M\)\Gamma\(1-M\)^3\Gamma\(1+M\)}{M\(3-2M\)\Gamma\(2-2M\)}\frac{\xi_m}{1+\xi_m}\\
&\times \Bigg[{}_2F_1\(2M,1,1+M,\frac{1}{1+\xi_m}\)-1-\frac1{\xi_m}\Bigg].\notag
\end{align}
\end{subequations}

\subsection{Special limits}
\label{sec:appDISlimits}

We find instructive to study the above expressions in two particular limits, namely the massless limit and the limit $M\to 0$.
The latter is useful to construct the fixed-order expansion of the resummed result.

\subsubsection{Massless limit}
We can use the Mellin forms to study the massless limit $\xi_m\to0$ of the resummed expressions.
We have for neutral current
\begin{align}
\label{eq:masslesslimit2massive}
\lim_{\xi_m \to 0} \tilde{\cal C}_2\(0,M,\xi_m,\xi_m\)
&= \tilde {\cal C}_2(0,M,0,0) + \tilde{\cal K}_{hg}(M,\xi_m), \\
\label{eq:masslesslimitLmassive}
\lim_{\xi_m \to 0} \tilde{\cal C}_L\(0,M,\xi_m,\xi_m\)
&= \tilde {\cal C}_L(0,M,0,0),
\end{align}
and for charged current
\begin{align}
\label{eq:masslesslimit2CC}
\lim_{\xi_m \to 0}   \tilde{\cal C}_2\(0,M,\xi_m,0\)
&= \tilde {\cal C}_2(0,M,0,0)  + \frac12\tilde{\cal K}_{hg}(M,\xi_m), \\
\label{eq:masslesslimitLCC}
\lim_{\xi_m \to 0}   \tilde{\cal C}_L\(0,M,\xi_m,0\)
&= \tilde {\cal C}_L(0,M,0,0),\\
\label{eq:masslesslimit3CC1}
\lim_{\xi_m \to 0}   \tilde{\cal C}_3\(0,M,\xi_m,0\)
&= \cancel{\tilde {\cal C}_3(0,M,0,0)}  + \frac12\tilde{\cal K}_{hg}(M,\xi_m), \\
\label{eq:masslesslimit3CC2}
\lim_{\xi_m \to 0}   \tilde{\cal C}_3\(0,M,0,\xi_m\)
&= \cancel{\tilde {\cal C}_3(0,M,0,0)} - \frac12\tilde{\cal K}_{hg}(M,\xi_m),
\end{align}
where we have defined
\beq\label{eq:calKhgMell}
\tilde{\cal K}_{hg}(M,\xi_m) = - \frac{\as}{\pi} \xi_m^M\frac{1-M}{M} \frac{\Gamma^3(1-M)\Gamma(1+M)}{(3-2M)\Gamma(2-2M)}.
\eeq
The function $\tilde{\cal K}_{hg}(M,\xi_m)$ contains the collinear singularity, appearing as a $M=0$ pole,
and produces the logarithmic mass contributions when expanded in powers of $M$. In a sense, it represents the conversion
from the collinear singularity regularized by the off-shellness and the very same singularity regularized by the mass,
see Eq.~\eqref{eq:Khgres_fc}.

The inverse Mellin of Eq.~\eqref{eq:calKhgMell}, needed for the running coupling resummation Eq.~\eqref{eq:Khgres},
can be obtained in the following way.
We first split the function as the product of three different factors which we write in integral form:
\begin{align}
  \Gamma(1-M)\Gamma(1+M) &= M\int_0^\infty dt\, \frac{t^{M-1}}{1+t}, \nonumber\\
  4^{1-M}\frac{(1-M)\Gamma^2(1-M)}{(3-2M)\Gamma(2-2M)} &= \int_0^1 dx\, \frac{x^{1-M}}{\sqrt{1-x}}, \nonumber\\
  \frac{(4\xi_m)^M}M &= \int_0^1 \frac{dy}y\, (4y\xi_m)^M.
\end{align}
Then we change variable from $t$ to $\xi=4\xi_mty/x$ to write the product in the form of a Mellin transfrom:
\begin{align}
\tilde{\cal K}_{hg}(M,\xi_m)
&= - \frac{\as}{4\pi} M\int_0^\infty dt\, \frac{t^{M-1}}{1+t}  \int_0^1 dx\, \frac{x^{1-M}}{\sqrt{1-x}}   \int_0^1 \frac{dy}y\, (4y\xi_m)^M \nonumber\\
&= - \frac{\as}{4\pi} M\int_0^\infty d\xi\, \xi^{M-1}  \int_0^1\frac{dx}{\sqrt{1-x}}\int_0^1\frac{dy}y \frac{xy{4\xi_m}}{x\xi+y{4\xi_m}}.
\end{align}
At this point we can simply read off the inverse Mellin transform in integral form
\begin{align}\label{eq:calKhg}
\mathcal{K}_{hg}(\xi,\xi_m)
&= - \frac{\as}{4\pi} \int_0^1\frac{dx}{\sqrt{1-x}}\int_0^1\frac{dy}y \frac{xy{4\xi_m}}{x\xi+y{4\xi_m}} \nonumber\\
&= - \frac{\as}{3\pi} \[\frac{{4\xi_m}}{\xi}+\log \frac{\xi_m}{\xi}+ \(2-\frac{{4\xi_m}}{\xi}\)\sqrt{1+\frac{{4\xi_m}}\xi}\log\(\sqrt{\frac\xi{{4\xi_m}}}+\sqrt{1+\frac\xi{{4\xi_m}}}\)\].
\end{align}
which has been computed explicitly in the second line. The $\xi$-derivative of this expression is Eq.~\eqref{eq:DcalKhg}.

\subsubsection{Fixed-order expansion and collinear singularities}
\label{sec:Mexpansion}

The Mellin form of the off-shell coefficients can be also used to compute expansions in $M=0$,
which are needed for computing the $\as$ expansion of the resummed results.
For matching up to NNLO, i.e.\ $\Ord(\as^2)$, we need the expansions up to $\Ord(M)$.
In the massless case we have
\begin{subequations}\label{eq:expCmassless}
\begin{align}
\tilde{\cal C}_2\(0,M,0,0\)&=\frac{\as}{3\pi}\left[\frac{1}{M}+\frac{13}{6}+\(\frac{71}{18}-\zeta_2\)M+\Ord\(M^2\)\right],\\
\tilde{\cal C}_L\(0,M,0,0\)&=\frac{\as}{3\pi}\left[1-\frac{M}{3}+\Ord\(M^2\)\right],
\end{align}
\end{subequations}
while in the massive case in neutral current we obtain
\begin{subequations}\label{eq:expCmassive}
\begin{align}
\tilde{\cal C}_2\(0,M,\xi_m,\xi_m\)&=\frac{\as}{3\pi}\Bigg[\frac{1+4\sqrt{1+4\xi_m}\csch^{-1}\(2\sqrt{\xi_m}\)}{2}+\frac{g_A^2}{g_A^2+g_V^2}\frac{12\xi_m 	\csch^{-1}\(2\sqrt{\xi_m}\)}{\sqrt{1+4\xi_m}}\notag\\
&\qquad\quad+\frac{M}{6\sqrt{1+4\xi_m}}\Bigg\{\sqrt{1+4\xi_m}\(3\ln\xi_m+5\)+2\csch^{-1}\(2\sqrt{\xi_m}\)\notag\\
&\qquad\qquad\times\(13-10\xi_m+6\(1-\xi_m\)\ln\xi_m\)-6\(1-\xi_m\) H_{-,+}\(-\frac{1}{\sqrt{1+4\xi_m}}\)\notag\\
&\qquad\qquad+\frac{12\xi_mg_A^2}{g_A^2+g_V^2}\Bigg(8\ln(4\xi_m)-16\ln\(1+\sqrt{1+4\xi_m}\)\notag\\
&\qquad\qquad\quad+\csch^{-1}\(2\sqrt{\xi_m}\)\(6\ln\xi_m+28\)-3H_{-,+}\(-\frac{1}{\sqrt{1+4\xi_m}}\)\Bigg)\Bigg\}  \notag\\
&\qquad\quad+\Ord\(M^2\)\Bigg],\\
\tilde{\cal C}_L\(0,M,\xi_m,\xi_m\)&=\frac{\as}{3\pi}\Bigg[\frac{\sqrt{1+4\xi_m}\(1+6\xi_m\)-8\xi_m\(1+3\xi_m\)\csch^{-1}\(2\sqrt{\xi_m}\)}{\(1+4\xi_m\)\sqrt{1+4\xi_m}}\notag\\
&\qquad\quad+\frac{g_A^2}{g_A^2+g_V^2}\frac{2\xi_m\(4\(7\xi_m+2\)\csch^{-1}\(2\sqrt{\xi_m}\)-\sqrt{1+4\xi_m}\)}{3\(1+4\xi_m\)\sqrt{1+4\xi_m}}\notag\\
&\qquad\quad+\frac{M}{3\(1+4\xi_m\)\sqrt{1+4\xi_m}}\Bigg\{\sqrt{1+4\xi_m}\(12\xi_m-1+3\(1+6\xi_m\)\ln\xi_m\)\notag\\
&\qquad\qquad+\csch^{-1}\(2\sqrt{\xi_m}\)\(6+8\(1-6\xi_m\)\xi_m-24\xi_m\(1+3\xi_m\)\ln\xi_m\)\notag\\
&\qquad\qquad+12\xi_m\(1+3\xi_m\) H_{-,+}\(-\frac{1}{\sqrt{1+4\xi_m}}\)\notag\\
&\qquad\qquad+\frac{2\xi_m g_A^2}{g_V^2+g_A^2}\Bigg(2\csch^{-1}\(2\sqrt{\xi_m}\)\(23+82\xi_m+6\(7\xi_m+2\)\ln\xi_m\)\notag\\
&\qquad\qquad\qquad\qquad-\sqrt{1+4\xi_m}\(8+3\ln\xi_m\)-6\xi_m\ln\frac{\sqrt{1+4\xi_m}-1}{\sqrt{1+4\xi_m}+1}\notag\\
&\qquad\qquad\qquad\qquad-6\(2+7\xi_m\)H_{-,+}\(-\frac{1}{\sqrt{1+4\xi_m}}\)\Bigg)\Bigg\} \notag\\
&\qquad\quad+\Ord\(M^2\)\Bigg],
\end{align}
\end{subequations}
having defined the harmonic polylogarithm
\beq
\label{eq:Hminusplus}
H_{-,+}\(z\)=\Li_2\(\frac{1-z}{2}\)-\Li_2\(\frac{1+z}{2}\)+\frac{1}{2}\ln\(\frac{1-z^2}{4}\)\ln\(\frac{1-z}{1+z}\).
\eeq
In the charged-current case, with only one massive quark, we have instead the following expansion:
\begin{subequations}\label{eq:expCchargecurrent}
\begin{align}
\tilde{\cal C}_2\(0,M,\xi_m,0\)&=\frac{\as}{3\pi}\Bigg[\frac{1}{2M}+\frac{8-3\ln\xi_m+6\ln\(1+\xi_m\)}6\notag\\
&\qquad\quad+\frac{M}{36}\Bigg(86-3\ln\xi_m\(10+3\ln\xi_m\)+6\ln\(1+\xi_m\)\(13+3\ln\(1+\xi_m\)\)\notag\\
&\qquad\qquad\qquad-36\Li_2\(\frac{1}{1+\xi_m}\)\Bigg)+\Ord\(M^2\)\Bigg],\\
\tilde{\cal C}_L\(0,M,\xi_m,0\)&=\frac{\as}{3\pi}\Bigg[\frac{\xi_m}{1+\xi_m}\frac{1}{2M}
+\frac{6+14\xi_m+3\xi_m\(1+2\xi_m\)\ln\xi_m-6\xi_m^2\ln\(1+\xi_m\)}{6\(1+\xi_m\)}\notag\\
&\qquad\quad+\frac{M}{36\(1+\xi_m\)}\Bigg(-12+92\xi_m-18\xi_m^2\ln^2\(1+\xi_m\)+36\xi_m^2\Li_2\(\frac{1}{1+\xi_m}\)\notag\\
&\qquad\qquad+9\xi_m\(1+2\xi_m\)\ln^2\xi_m+6\xi_m\(7\xi_m-1\)\ln\xi_m\notag\\
&\qquad\qquad+6\(6+\xi_m\(15-7\xi_m\)\)\ln\(1+\xi_m\)\Bigg)+\Ord\(M^2\)\Bigg],\\
\tilde{\cal C}_3\(0,M,\xi_m,0\)&=\frac{\as}{3\pi}\Bigg[-\frac{1}{1+\xi_m}\frac{1}{2M}-\frac{5+\(3+6\xi_m\)\ln\xi_m-6\xi_m\ln\(1+\xi_m\)}{6\(1+\xi_m\)}\notag\\
&\qquad\quad-\frac{M}{36\(1+\xi_m\)}\Bigg(56+30\(1+2\xi_m\)\ln\xi_m+9\(1+2\xi_m\)\ln^2\xi_m\notag\\
&\qquad\qquad-60\xi_m\ln\(1+\xi_m\)-18\xi_m\ln^2\(1+\xi_m\)+36\xi_m\Li_2\(\frac{1}{1+\xi_m}\)\Bigg)\notag\\
&\qquad\quad+\Ord\(M^2\)\Bigg].
\end{align}
\end{subequations}
In each of the above equation, the pole in $M=0$, where present, identifies the collinear singularity
and is given by the LO $P_{qg}$ in Mellin space times the LO non-singlet process $q+W\to q'$.
The latter has non-trivial mass dependence in charged current, Eqs.~\eqref{eq:expCchargecurrent},
since in this case the final state quark $q'$ is massive.
Note that in this case even $F_L$ has a non-vanishing contribution at LO, which is proportional to the mass and thus vanishes in the massless limit.
The $\Ord(M^0)$ terms in these expansions, after subtraction of massless collinear singularities as in Eqs.~\eqref{eq:C2gmassless} and \eqref{eq:CgmassiveCC_tilde},
reproduce the known $\Ord(\as)$ contributions~\cite{Gluck:1987uk,Gluck:1996ve}.
Higher-order corrections exist both for neutral and charged currents (see e.g.\ \cite{Laenen:1992zk,Berger:2016inr}), however not in a form we could compare our expansion to.

\section{Details on numerical implementation}
\label{sec:appRes}

\subsection{The evolution function $U$}
\label{sec:ht}

A key ingredient of the formalism for resumming coefficient functions is the evolution function $U(N,\xi)$, defined in Eq.~\eqref{eq:evol}.
As discussed in Ref.~\cite{Bonvini:2016wki}, computing it exactly with the resummed anomalous dimension $\gamma_+$
(specifically, with the LL$^\prime$ anomalous dimension) requires integrating $\gamma_+$ over all values of $\as$ from 0 to $\infty$,
which is numerically inconvenient.
Therefore, following Refs.~\cite{Altarelli:2008aj,Ball:2007ra}, we use an approximate expression for the evolution factor,
where the anomalous dimension is assumed to depend on $\as$ only at LO, with 1-loop running.
This leads to the ABF evolution factor~\cite{Bonvini:2016wki}
\beq\label{eq:UABF}
U_{\rm ABF}(N,\xi) = \Big(1+r(N,\as)\log\xi\Big)^{\gamma_+(N,\as)/r(N,\as)}
\eeq
with
\beq
r(N,\as) = -\frac{\frac{Q^2 d}{dQ^2}\gamma_+(N,\as)}{\gamma_+(N,\as)}
= \as^2 \beta_0\, \frac{\frac{d}{d\as}\gamma_+(N,\as)}{\gamma_+(N,\as)}.
\eeq
Note that the ratio $r(N,\as)$ is such that the approximation reproduces the correct derivative of $\gamma_+$ in $\as$.
However, this effect is strictly speaking beyond the formal accuracy we work with, so one could ignore
it and replace
\beq\label{eq:RCmode}
r(N,\as)\to\as\beta_0.
\eeq
This variant is used to construct the uncertainty band on our resummed predictions.
As a simpler approximation we could also consider the fixed-coupling limit, in which all the scale dependence in $\as$ is ignored
and the evolution factor becomes simply
\beq\label{eq:Ufc}
U_{\rm f.c.}(N,\as) = \xi^{\gamma_+(N,\as)}.
\eeq
In this case our formula for computing the resummation of coefficient functions simply reduces to a Mellin transformation
with moment $\gamma_+$.

The integration range in the off-shellness $\xi$ in the resummation formula extends to all accessible values between $0$ and $\infty$.
In the running-coupling case, $\as$ is computed at $\xi Q^2$ in $U$, Eq.~\eqref{eq:evol}, so at some small value of $\xi$ the Landau pole is hit,
and the integration must stop there.
With 1-loop running (and also at higher loops if the expanded solution for the running coupling is used) the position
of the Landau pole is given by
\beq
\xi_0=\exp\frac{-1}{\as\beta_0},
\eeq
and $\xi$-integration must be limited to the region $\xi\geq\xi_0$.
In the fixed-coupling limit, $\as$ is frozen at its value in $Q^2$, so all values of $\xi$ are in principle accessible, i.e.\ $\xi_0\overset{\text{f.c.}}{=}0$.

In Ref.~\cite{Bonvini:2016wki} we derived the resummation formula which had originally the form (schematically)
\beq\label{eq:Cres1}
C_{\rm res}(N) = \int_{\xi_0}^\infty d\xi \, {\cal C}(0,\xi) \,\frac{d}{d\xi}U(N,\xi).
\eeq
Then, for convenience in the numerical implementation, we integrated by parts to get
\beq\label{eq:Cres2}
C_{\rm res}(N) = -\int_{\xi_0}^\infty d\xi \, \frac{d}{d\xi}{\cal C}(0,\xi) \,U(N,\xi),
\eeq
where the boundary term at infinity vanishes thanks to ${\cal C}$, and the boundary term in $\xi_0$
is assumed to vanish because
\beq\label{eq:U=0}
U(N,\xi_0) = 0.
\eeq
The latter assumption is not always true.
However, at the leading logarithmic accuracy, which is the only accuracy on which we have control at the moment,
the resummed result is only governed by the fixed-coupling anomalous dimension $\gamma_s$, dual of the LO BFKL kernel.
Thus, the formula Eq.~\eqref{eq:Ufc} applies, with $\gamma_+=\gamma_s$.
To obtain the resummed coefficient function in momentum space, an inverse Mellin transform has to be computed.
This amounts to integrating in $N$ over an imaginary contour with abscissa to the right of the small-$x$ singularity,
which in the case of $\gamma_s$ is placed in $N=\as c_0$, with $c_0$ given in Eq.~\eqref{eq:c0kappa0}.
Along such contour, the real part of $\gamma_s$ is always positive, and therefore $U_{\rm f.c.}(N,\xi_0=0)=0$.
Therefore, at the accuracy we are working with, the boundary term indeed vanishes.

In practice, however, we include in our resummation subleading contributions which spoil the condition Eq.~\eqref{eq:U=0}.
Indeed the anomalous dimension $\gamma_+$ that we use has a more complex structure than $\gamma_s$. Additionally,
we use the approximation Eq.~\eqref{eq:UABF}, which is typically finite in $\xi=\xi_0$:
\beq\label{eq:UABF0}
U_{\rm ABF}(N,\xi_0) = \Big(1-\frac{r(N,\as)}{\as\beta_0}\Big)^{\gamma_+(N,\as)/r(N,\as)}.
\eeq
Note that using the variant Eq.~\eqref{eq:RCmode} $U_{\rm ABF}(N,\xi_0)$ is either $0$ or $\infty$ depending
on the sign of the real part of $\gamma_+$.
This implies that the two formulations Eqs.~\eqref{eq:Cres1} and \eqref{eq:Cres2} will give in general different results,
due to the neglected non-zero boundary term. We have indeed verified this numerically.
Despite the fact that this difference is subleading log, and hence either result is formally equally valid,
this difference in the formulations is undesirable.

In this work we propose to solve the ambiguity by modifying the evolution function with suitable higher-twist terms
such that we always have $U(N,\xi_0)=0$. To do so, we use the evolution function
\beq\label{eq:Unew}
U(N,\xi) = U_{\rm ABF}(N,\xi)\, D_\text{higher-twist}(\xi),
\eeq
with
\beq
D_\text{higher-twist}(\xi) =
\begin{cases}
\[1-\(\frac{\log\xi}{\log\xi_0}\)^{1+\frac1{\as\beta_0}}\]\qquad & \xi<1 \\
1 & \xi>1.
\end{cases}
\eeq
It is easy to verify that the damping function $D_\text{higher-twist}(\xi)$ vanishes in $\xi_0$ and smoothly tends to $1$
in $\xi=1$, with all derivatives vanishing in $\xi=1$.
Moreover, it is clearly higher-twist, i.e.\ non analytical in the coupling $\as$,
so it does not influence the perturbative expansion of the evolution factor (which is used for the matching of the resummed expressions to fixed-order).

Using this new damped evolution function, we find that the results obtained using Eq.~\eqref{eq:Cres2} are indistinguishable from
those obtained with the undamped function, which confirms that the results of Ref.~\cite{Bonvini:2016wki}
are unaffected (from the point of view of $U$).
On the other hand, results obtained using Eq.~\eqref{eq:Cres1} are now identical to those obtained using Eq.~\eqref{eq:Cres2},
as they must, since now the boundary term is identically zero by construction.

From the point of view of the numerical implementation, we observe that the $N$ dependence of the resummed coefficient functions
is all contained in $U$, Eqs.~\eqref{eq:Cres1} or \eqref{eq:Cres2}. Therefore, we can first compute the inverse Mellin transform of
$U(N,\xi)$ Eq.~\eqref{eq:Unew} as function of $\xi$, $U(x,\xi)$, which we tabulate for various values of $\as$, $x$ and $\xi$,
and we subsequently use it to compute the $\xi$ integration for each observable.
In \texttt{HELL v2.0} both the old methodology (which integrates first in $\xi$ and then in $N$) and the new one (integration order inverted)
are implemented, and give of course identical results (within numerical integration errors).
The new implementation is faster.

\subsection{Implementation of kinematic theta functions at resummed level}
\label{sec:theta}

In massive coefficient functions, kinematic constraints for the production
of the massive final state are implemented through theta functions appearing in the coefficient functions.
For instance, in the case of DIS neutral-current structure functions, the theta function
has the form $\theta(X-x)$, with $X = 1/(1+4m^2/Q^2)$ and where $x$ is the Mellin integration variable.\footnote
{In the charged-current case, when the mass of the quark before and after hitting the $W$ is different,
the form of $X$ generalizes to $X=1/(1+(m_1+m_2)^2/Q^2)$.}
The very same theta function appears also in the off-shell coefficient, as it depends only on the kinematics of the final state.
In the resummation procedure, the off-shell coefficient function is Mellin transformed with respect to $x$ and then the result is evaluated in Mellin moment $N=0$.
The last step (which is strictly speaking not necessary) loses track
of the theta function, and the inverse Mellin transform of the resummed on-shell coefficient
is non-zero also in kinematically unaccessible regions.

A possible solution to restore the kinematic theta function is simply to avoid computing the off-shell coefficient
in $N=0$. This is possible, however, it is not convenient for at least two reasons:
the first is that all expressions and calculations become significantly more complicated,
and the second is that for consistency this should be done also in the massless case.
The latter requirement is necessary in the construction of the collinearly resummed coefficient functions,
otherwise the massless limit of the (collinear subtracted) resummed massive on-shell coefficients would not tend to the massless ones, Eq.~\eqref{eq:massless_limit}.

Therefore, we seek a solution which restores the theta function in the resummed approach,
while keeping using the off-shell coefficient in $N=0$.
The implementation must satisfy three requirements:
\begin{itemize}
\item the theta function should be restored without affecting the logarithmic accuracy of the result;
\item the $x\to X$ limit must be smooth;
\item in the massless $X\to1$ limit the effect must disappear completely.
\end{itemize}
The first requirement is obvious.
The second one is perhaps not mandatory, but it is satisfied in fixed-order results,
and avoids sharp transitions between results.
The latter requirement is instead needed for a correct implementation of the resummation of collinear logarithms at small-$x$ resummed level.

We have investigated different options for the restoration of the theta function such that the requirements above are satisfied.
We report here the two main alternatives that we consider, which act on $N$ space and on $x$ space, respectively.
The $N$-space approach consists in multiplying the integrand of Eqs.~\eqref{eq:C2gmassive_tilde}, \eqref{eq:C2gmassive} and \eqref{eq:CLgmassive}
by a term of the form
\beq\label{eq:ThetaN}
\Theta_N(N,X) = \frac{X^N}{(1-X)N+1} = 1+\Ord(N)
\eeq
As explicitly indicated, this term is manifestly subleading, and reproduces the theta function $\theta(X-x)$ thanks to the $X^N$ term.
In the massless limit it reduces to $\Theta_N(N,1)=1$, as required.
It can be also verified that, in full generality, the inverse Mellin transform of the resummed coefficient function obtained with this extra function
vanishes smoothly as $x\to X$.
The alternative implementation in $x$ space is obtained by multiplying the final resummed coefficient function in $x$ space by the function
\beq\label{eq:Thetax}
\Theta_x(x,X) = \theta(X-x) \[1-\(\frac{x}{X}\)^{\frac1{1-X}}\].
\eeq
The function in squared brackets ensures smooth $x\to X$ limit, and it is clearly subleading.
In the massless limit $X\to1$ it reduces to $\Theta_x(x,1) = \theta(1-x)$, as required.

The two alternatives are formally equally valid, but lead in general to different numerical results.
For practical reasons, we opt for the $x$-space implementation, Eq.~\eqref{eq:Thetax}.
In this way restoring the theta function can be done at the very end, giving full flexibility
for the implementation of the resummation.
For instance, it is possible to precompute the inverse Mellin transform of the evolution function, $U(x,\xi)$,
as described in Sect.~\ref{sec:ht}, speeding up the computation of resummed massive coefficient functions.
This would not be possible using the $N$-space formulation, Eq.~\eqref{eq:ThetaN}, as in this case
the $N$ dependence of $\xi$-integrand of resummed coefficient functions would include the $\Theta_N(N,X)$ term,
so the Mellin inversion would not act on $U(N,\xi)$ only.

\subsection{A convenient approximate form for the fixed-order anomalous dimension}
\label{sec:gammaapprox}

In the construction of the off-shell kernel for LL and NLL resummation, we need the dual of the
fixed LO or NLO anomalous dimension, denoted $\chi_s$ and $\chi_{s,\rm NLO}$, respectively.
These two functions provide the resummation of collinear (and anticollinear) contributions
in the DL BFKL kernel. If the duals are computed from approximate expressions of the fixed-order
anomalous dimensions, the resummation in the BFKL kernel is only approximate, and one cannot claim to have exact
leading or next-to-leading logarithmic accuracy in BFKL.
However, our goal is to reach LL and NLL accuracy in the resummed anomalous dimension, not in the BFKL kernel.
For this, the BFKL kernel has just to be correct at fixed LO or NLO, since by duality this fully determines the LL and NLL contributions of the DGLAP anomalous dimension.
The reason why also the BFKL kernel is being resummed is just that the resummation stabilizes
its perturbative expansion, which is otherwise highly unstable close to the collinear and anticollinear poles.
Therefore, from the point of view of the accuracy of the result,
it is well possible to use an approximate expression for the LO and NLO anomalous dimensions,
provided their LL and NLL parts are correct, as they correspond by duality to contributions of $\Ord(\as)$ and $\Ord(\as^2)$ in the BFKL kernel,
which need to be correct.
Then, once the fixed-order part of the resummed (N)LO+(N)LL anomalous dimension is subtracted,
the resummed contributions Eqs.~\eqref{eq:deltaLL}, \eqref{eq:deltaNLL} can be added to the exact
fixed-order anomalous dimension, restoring the correct fixed-order part.

This approach was already considered in both Refs.~\cite{Altarelli:2008aj,Bonvini:2016wki},
with two different implementations.
The basic motivation was that the exact fixed-order anomalous dimension $\gamma_+$, being the eigenvalue of a matrix,
contains a square-root branch-cut, which is inherited by the DL anomalous dimension and would give rise to a spurious
oscillating behaviour when performing an inverse Mellin transformation.
The approximate $\gamma_+$ implemented in Ref.~\cite{Bonvini:2016wki}, which is a somewhat simplified
version of the one originally proposed in Ref.~\cite{Altarelli:2008aj},
was simply obtained by taking $\gamma_{gg}$ computed for $n_f=0$
(which is then also the eigenvalue, as there are no quarks and the matrix reduces to a single entry)
and adding the $n_f$-dependent contributions of the exact $\gamma_+$ restricted to LL and NLL.
This procedure ensures that the resulting anomalous dimension reproduces the LL and NLL behaviour
of the exact one, but it behaves as $\gamma_{gg}$ elsewhere in the $N$ plane.
This implies in particular that the anomalous dimension grows (negatively) as $\log N$ at large $N$.
Note also that this construction violates momentum.

We observe that the large-$N$ logarithmic growth of $\gamma_+$ is problematic.
Indeed, the dual function $\chi_s$ (or $\chi_{s,\rm NLO}$) grows exponentially for negative $M$ as $|M|$ gets larger.
The DL kernel, by duality, should then be able to reproduce the logarithmic growth at large $N$.
However, the DL kernel does not only contain $\chi_s$, but also the fixed-order BFKL kernel,
which contains poles for all integer values of $M$.
Therefore, by duality, the large $N$ behaviour of the DL anomalous dimension is determined by the rightmost $M$
pole for negative $M$, which is in $M=-1$, which implies that the DL anomalous dimension
tends to $-1$ as $N\to\infty$.
This problem was ignored in previous works, as the $M=-1$ pole represents a higher twist contribution,
and the practical effect is almost negligible.
However, it would be ideal to avoid this issue.
One option would be to act on the DL BFKL kernel, hacking it such that the poles for negative $M$ are no longer present.
While we have tried this solution, we think that it is not the best approach.
A significantly better solution is obtained if the anomalous dimension used in the duality relation
does not grow at large $N$, but rather it goes as a constant larger than $-1$, such that
$\chi_s$ never hits the rightmost negative $M$ pole: in this way, the large-$N$ behaviour of the DL anomalous dimension
is determined by $\chi_s$ itself, and hence corresponds to the one of the input anomalous dimension.

Thus, here we propose a new approximation for the fixed-order anomalous dimension.
We require that the LL and NLL behaviour is reproduced,
that momentum conservation is preserved exactly, and that at large $N$ it behaves as a constant greater than $-1$.
Given that the LO and NLO anomalous dimensions behave close to $N=0$ as
\begin{align}
  \gamma_0(N) &= \frac{a_{11}}{N} + a_{10} +\Ord(N) \nonumber\\
  \gamma_1(N) &= \frac{a_{22}}{N^2} + \frac{a_{21}}{N} + a_{20} +\Ord(N)
\end{align}
where $a_{22}=0$ (accidental zero), we propose the following approximate expression,
\beq\label{eq:BestgammaApprox}
\gamma(N) = \frac{a_1}N + a_0 - (a_1+a_0)\frac{2N}{N+1},
\eeq
which is valid both at LO and NLO with appropriate coefficients. At NLO they are given by
\begin{align}
  a_1 &= \as a_{11} + \as^2 a_{21}, \nonumber\\
  a_0 &= \as a_{10} + \as^2 a_{20},
\end{align}
and at LO one simply neglects the $\Ord(\as^2)$ terms; the coefficients are given by
\begin{align}
  a_{11} &= \frac{C_A}\pi, \nonumber\\
  a_{21} &= n_f\frac{26C_F-23C_A}{36\pi^2}, \nonumber\\
  a_{10} &= 
           -\frac{11C_A + 2n_f(1-2C_FC_A+4C_F^2)}{12\pi}, \nonumber\\
  a_{20} &= \frac1{\pi^2}\[\frac{1643}{24} - \frac{33}2 \zeta_2 - 18 \zeta_3 + n_f\(\frac49\zeta_2-\frac{68}{81}\) + n_f^2\frac{13}{2187}\].
\end{align}
Note that $a_{20}$ is formally NNLL, so it could in principle be ignored (similarly, for LL resummation one could ignore $a_{10}$);
in practice, including both $a_{11}$ and $a_{10}$ at LO and both $a_{21}$ and $a_{20}$ at NLO
provides an excellent approximation of the anomalous dimension in the small-$N$ region.
The last term in Eq.~\eqref{eq:BestgammaApprox} is a subleading $\Ord(N)$ contribution
which ensures momentum conservation $\gamma(1)=0$.
At large $N$, Eq.~\eqref{eq:BestgammaApprox} behaves as a constant,
\beq
\gamma(N) \overset{N\to\infty}{\to} -2a_1-a_0.
\eeq
We verified that $-2a_1-a_0>-1$ for all values of $\as,n_f$ that we consider.
In particular, the worst case is obtained at NLO with $n_f=3$, where the condition $\gamma(N\to\infty)>-1$ is satisfied for $\as< 0.558$.
For all other values of $n_f$, and at LO, the $\as$ range of validity is larger.
Clearly, this is more than enough, as for such values of $\as$ the perturbative hypothesis is lost.
Indeed, in our current numerical implementation we never consider energy scales for which $\as>0.35$.

Another very important advantage of the approximation Eq.~\eqref{eq:BestgammaApprox} is that its inverse function (i.e.\ the dual) can be computed analytically:
\begin{align}\label{eq:chisanalytic}
\chi_s(M)
&= \frac{a_1+a_0-M + \sqrt{(M+a_1-a_0)^2 + 8a_1(a_1+a_0)}}{2(2a_1+a_0+M)}
\end{align}
(here we generically use the name $\chi_s$ for representing both the dual to the LO anomalous dimension and the dual to the NLO one, previously called $\chi_{s,\rm NLO}$).
This represents a great advantage from the point of view of the numerical implementation,
as in the general case one would have to compute the inverse function by means of zero-finding routines,
which are typically slow and do not ensure convergence, especially when working in the complex plane.
Consider also that the DL anomalous dimension from Eq.~\eqref{eq:duality} is itself obtained by means of zero-finding routines,
applied to the DL kernel which is given in terms of $\chi_s$, giving rise to nested zero-finding which clearly cannot guarantee best performance.
Therefore, using the analytic expression Eq.~\eqref{eq:chisanalytic} for $\chi_s$ allows to have a single layer of zero-finding routines,
improving significantly the numerical stability and the speed of the code.\footnote
{A word of caution is needed in the choice of the branch of the square-root.
Along the Mellin inversion integration path, which is the only place in $N$ space where the DL anomalous dimension
is computed, the standard branch of the square-root is suitable for the collinear $\chi_s$. However,
for the anti-collinear $\chi_s$, a different branch is needed, where the cut is placed on the negative imaginary axis,
which avoids crossing the cut during integration.}
The expansion of $\chi_s$ in power of $\as$ is given by
\beq
\chi_s(M) = a_{11}\frac{\as}M + a_{11}a_{10}\frac{\as^2}{M^2} + a_{21}\as \frac{\as}M + \Ord(\as^3),
\eeq
which implies, according to Eq.~\eqref{eq:chisNLO},
\begin{align}
  \chi_{01} &= a_{11} = \frac{C_A}\pi, \nonumber\\
  \chi_{02} &= a_{11}a_{10} = -\frac{11C_A^2 + 2n_f(C_A-2C_F)}{12\pi^2}, \nonumber\\
  \chi_{11} &= a_{21} = n_f\frac{26C_F-23C_A}{36\pi^2}.
\end{align}
Up to this order, these are identical to the dual of the exact anomalous dimension,
as an obvious consequence to the fact that the approximate anomalous dimension is constructed to preserve LL and NLL accuracy.

\subsection{Inverse Mellin transforms}
\label{sec:invMellin}

Here we compute the various ingredients for the inverse Mellin transforms of Eqs.~\eqref{eq:LLexp}--\eqref{eq:NLLexp}.
Using the approximate form of $\gamma_0$ given in App.~\ref{sec:gammaapprox}, we have
\begin{align}
\gamma_0 (N) &= \frac{a_{11}}{N} -(a_{10}+2a_{11}) +\frac{2(a_{11}+a_{10})}{N+1}, \nonumber\\
\gamma_0'(N) &= -\frac{a_{11}}{N^2} -\frac{2(a_{11}+a_{10})}{(N+1)^2}.
\end{align}
We can also write the function Eq.~\eqref{eq:chimom} as
\beq
f_{\rm mom}(N) = \frac{4}{N+1} - \frac{4}{(N+1)^2}.
\eeq
In Eq.~\eqref{eq:NLLexp} a number of products appear. When a power of $1/N$ multiplies a power of $1/(N+1)$,
it is always possible to write it as a sum of powers of individual poles in either $N$ or $N+1$.
Therefore, most terms can be computed by means of the following inverse Mellin transforms,
\begin{align}
{\cal M}^{-1}\[\frac1{N^{k+1}}\] &= \frac{(-1)^k}{k!} \frac{\log^kx}{x}, \\
{\cal M}^{-1}\[\frac1{(N+1)^{j+1}}\] &= \frac{(-1)^j}{j!} \log^jx,
\end{align}
where the $\Mell^{-1}$ symbol is a shorthand notation for representing the Mellin inversion,
\beq
\Mell^{-1}\[f(N)\] = \int_{\frac12-i\infty}^{\frac12+i\infty} \frac{dN}{2\pi i} \, x^{-N} f(N).
\eeq
Additionally, the function $\psi_1(N+1)$, multiplied by powers of $1/N$ or $1/(N+1)$, appears.
The computation of these inverse Mellin is trickier. We start from
\beq \label{eq:invMellpsi1}
{\cal M}^{-1}\[\psi_1(N+1)\] = \frac{\log x}{x-1}.
\eeq
To compute inverse Mellin of $\psi_1(N+1)$ with products of $1/N$ and $1/(N+1)$,
we compute consecutive convolutions of Eq.~\eqref{eq:invMellpsi1} with the inverse Mellin
of a single power of $1/N$ and $1/(N+1)$, which are given respectively by $1/x$ and $1$.
Starting from
\begin{align}
{\cal M}^{-1}\[\frac1{N}  \psi_1(N+1)\] &= \int_x^1\frac{dz}{x}\frac{\log z}{z-1} = \frac{\Li_2(1-x)}x, \\
{\cal M}^{-1}\[\frac1{N+1}\psi_1(N+1)\] &= \int_x^1\frac{dz}{z}\frac{\log z}{z-1} = \zeta_2 - \Li_2(x) + \frac12\log^2x-\log(1-x) \log x,
\end{align}
we can easily compute successive integrals by just integrating these results (as functions of $z$) in $dz/x$ or $dz/z$.
The relevant results are
\begin{align}
{\cal M}^{-1}\[\frac1{N^2} \psi_1(N+1)\] &= \frac{2[\Li_3(x)-\zeta_3]-[\Li_2(x)+\zeta_2]\log x}x, \\
{\cal M}^{-1}\[\frac1{(N+1)^2}\psi_1(N+1)\] &= 2[\Li_3(x)-\zeta_3]-[\Li_2(x)+\zeta_2]\log x-\frac16\log^3x, \\
{\cal M}^{-1}\[\frac1{N^3} \psi_1(N+1)\] &= \frac{3[\zeta_4-\Li_4(x)]+[\Li_3(x)+2\zeta_3]\log x + \frac12\zeta_2\log^2x}x, \\
{\cal M}^{-1}\[\frac1{(N+1)^3} \psi_1(N+1)\] &= 3[\zeta_4-\Li_4(x)]+[\Li_3(x)+2\zeta_3]\log x + \frac12\zeta_2\log^2x + \frac1{24}\log^4x.
\end{align}
Because the \texttt{HELL-x} code, where these expressions are implemented, has to be fast,
as it is meant to be used in PDF fits, the appearance of polylogarithms is not ideal.
Therefore, we can consider a small-$x$ approximation of these expressions.
After all, the complicated structure of the $\Ord(\as^3)$ contribution in Eq.~\eqref{eq:NLLexp} comes from
the complicated all-order structure of the resummed result, but what really matters in the resummation
is the prediction of small-$x$ contributions, while uncontrolled terms which vanish as $x\to0$ are irrelevant.
Hence, we approximate the expressions above as
\begin{align}
{\cal M}^{-1}\[\frac1{N^2} \(\psi_1(N+1)-\zeta_2\)\]
&= -\frac{2\zeta_3}x + 2 - \log x +\Ord(x) \nonumber \\
&= {\cal M}^{-1}\[-\frac{2\zeta_3}N + \frac2{N+1} + \frac1{(N+1)^2} \] +\Ord(x)\\
{\cal M}^{-1}\[\frac1{(N+1)^2} \(\psi_1(N+1)-\zeta_2\)\]
&= -2\zeta_3 - \frac16\log^3 x +\Ord(x) \nonumber \\
&= {\cal M}^{-1}\[-\frac{2\zeta_3}{N+1} + \frac1{(N+1)^4} \] +\Ord(x)\\
{\cal M}^{-1}\[\frac1{N^3} \(\psi_1(N+1)-\zeta_2\)\]
&= \frac{2\zeta_3}x \log x + \frac{3\zeta_4}x -3 + \log x +\Ord(x) \nonumber \\
&= {\cal M}^{-1}\[-\frac{2\zeta_3}{N^2}+\frac{3\zeta_4}N - \frac3{N+1} - \frac1{(N+1)^2} \] +\Ord(x)\\
{\cal M}^{-1}\[\frac1{(N+1)^3} \(\psi_1(N+1)-\zeta_2\)\]
&= 3\zeta_4 + 2\zeta_3\log x + \frac1{24}\log^4 x +\Ord(x) \nonumber \\
&= {\cal M}^{-1}\[\frac{3\zeta_4}{N+1}-\frac{2\zeta_3}{(N+1)^2} + \frac1{(N+1)^5} \] +\Ord(x)
\end{align}
In these equations we have also provided the Mellin transform of the approximate expressions, which is needed for
the analytic computation of the momentum conservation, Eq.~\eqref{eq:momconsc}.
We verified that using these approximate expressions leads only to tiny deviations with respect to the exact expressions,
and all in a region of $x$ which is not under control of small-$x$ resummation (specifically $x>10^{-2}$).
On the other hand, the speed-up is significant, fully justifying their use.

\phantomsection
\addcontentsline{toc}{section}{References}

\bibliographystyle{jhep}
\bibliography{references}

\providecommand{\href}[2]{#2}\begingroup\raggedright\begin{thebibliography}{100}

\bibitem{Anastasiou:2015ema}
C.~Anastasiou, C.~Duhr, F.~Dulat, F.~Herzog, and B.~Mistlberger, {\it {Higgs
  Boson Gluon-Fusion Production in QCD at Three Loops}},  {\em Phys. Rev.
  Lett.} {\bf 114} (2015) 212001, [\href{http://arxiv.org/abs/1503.06056}{{\tt
  arXiv:1503.06056}}].

\bibitem{Dreyer:2016oyx}
F.~A. Dreyer and A.~Karlberg, {\it {Vector-Boson Fusion Higgs Production at
  Three Loops in QCD}},  {\em Phys. Rev. Lett.} {\bf 117} (2016), no.~7 072001,
  [\href{http://arxiv.org/abs/1606.00840}{{\tt arXiv:1606.00840}}].

\bibitem{Boughezal:2017nla}
R.~Boughezal, A.~Guffanti, F.~Petriello, and M.~Ubiali, {\it {The impact of the
  LHC Z-boson transverse momentum data on PDF determinations}},
  \href{http://arxiv.org/abs/1705.00343}{{\tt arXiv:1705.00343}}.

\bibitem{Czakon:2016olj}
M.~Czakon, N.~P. Hartland, A.~Mitov, E.~R. Nocera, and J.~Rojo, {\it {Pinning
  down the large-x gluon with NNLO top-quark pair differential distributions}},
   {\em JHEP} {\bf 04} (2017) 044, [\href{http://arxiv.org/abs/1611.08609}{{\tt
  arXiv:1611.08609}}].

\bibitem{Currie:2016bfm}
J.~Currie, E.~W.~N. Glover, and J.~Pires, {\it {Next-to-Next-to Leading Order
  QCD Predictions for Single Jet Inclusive Production at the LHC}},  {\em Phys.
  Rev. Lett.} {\bf 118} (2017), no.~7 072002,
  [\href{http://arxiv.org/abs/1611.01460}{{\tt arXiv:1611.01460}}].

\bibitem{Currie:2017eqf}
J.~Currie, A.~Gehrmann-De~Ridder, T.~Gehrmann, E.~W.~N. Glover, A.~Huss, and
  J.~Pires, {\it {Precise predictions for dijet production at the LHC}},
  \href{http://arxiv.org/abs/1705.10271}{{\tt arXiv:1705.10271}}.

\bibitem{Moch:2005ba}
S.~Moch, J.~Vermaseren, and A.~Vogt, {\it {Higher-order corrections in
  threshold resummation}},  {\em Nucl.Phys.} {\bf B726} (2005) 317--335,
  [\href{http://arxiv.org/abs/hep-ph/0506288}{{\tt hep-ph/0506288}}].

\bibitem{Bonvini:2016frm}
M.~Bonvini, S.~Marzani, C.~Muselli, and L.~Rottoli, {\it {On the Higgs cross
  section at N$^{3}$LO+N$^{3}$LL and its uncertainty}},  {\em JHEP} {\bf 08}
  (2016) 105, [\href{http://arxiv.org/abs/1603.08000}{{\tt arXiv:1603.08000}}].

\bibitem{Bonvini:2014joa}
M.~Bonvini and S.~Marzani, {\it {Resummed Higgs cross section at N$^{3}$LL}},
  {\em JHEP} {\bf 1409} (2014) 007, [\href{http://arxiv.org/abs/1405.3654}{{\tt
  arXiv:1405.3654}}].

\bibitem{Catani:2014uta}
S.~Catani, L.~Cieri, D.~de~Florian, G.~Ferrera, and M.~Grazzini, {\it
  {Threshold resummation at N$^3$LL accuracy and soft-virtual cross sections at
  N$^3$LO}},  {\em Nucl.Phys.} {\bf B888} (2014) 75--91,
  [\href{http://arxiv.org/abs/1405.4827}{{\tt arXiv:1405.4827}}].

\bibitem{Bizon:2017rah}
W.~Bizon, P.~F. Monni, E.~Re, L.~Rottoli, and P.~Torrielli, {\it
  {Momentum-space resummation for transverse observables and the Higgs
  $p_\perp$ at N$^3$LL+NNLO}},  \href{http://arxiv.org/abs/1705.09127}{{\tt
  arXiv:1705.09127}}.

\bibitem{Abbate:2010xh}
R.~Abbate, M.~Fickinger, A.~H. Hoang, V.~Mateu, and I.~W. Stewart, {\it {Thrust
  at N$^3$LL with Power Corrections and a Precision Global Fit for
  alphas(mZ)}},  {\em Phys. Rev.} {\bf D83} (2011) 074021,
  [\href{http://arxiv.org/abs/1006.3080}{{\tt arXiv:1006.3080}}].

\bibitem{Hoang:2015hka}
A.~H. Hoang, D.~W. Kolodrubetz, V.~Mateu, and I.~W. Stewart, {\it {Precise
  determination of $\alpha_s$ from the $C$-parameter distribution}},  {\em
  Phys. Rev.} {\bf D91} (2015), no.~9 094018,
  [\href{http://arxiv.org/abs/1501.04111}{{\tt arXiv:1501.04111}}].

\bibitem{Bonvini:2014tea}
M.~Bonvini and L.~Rottoli, {\it {Three loop soft function for N$^3$LL$^\prime$
  gluon fusion Higgs production in soft-collinear effective theory}},  {\em
  Phys.Rev.} {\bf D91} (2015), no.~5 051301,
  [\href{http://arxiv.org/abs/1412.3791}{{\tt arXiv:1412.3791}}].

\bibitem{Ahmed:2015qda}
T.~Ahmed, M.~C. Kumar, P.~Mathews, N.~Rana, and V.~Ravindran, {\it
  {Pseudo-scalar Higgs boson production at threshold N$^3$ LO and N$^3$ LL
  QCD}},  {\em Eur. Phys. J.} {\bf C76} (2016), no.~6 355,
  [\href{http://arxiv.org/abs/1510.02235}{{\tt arXiv:1510.02235}}].

\bibitem{Ahmed:2016otz}
T.~Ahmed, M.~Bonvini, M.~C. Kumar, P.~Mathews, N.~Rana, V.~Ravindran, and
  L.~Rottoli, {\it {Pseudo-scalar Higgs boson production at N$^3$ LO$_{\text
  {A}}$ +N$^3$ LL $'$}},  {\em Eur. Phys. J.} {\bf C76} (2016), no.~12 663,
  [\href{http://arxiv.org/abs/1606.00837}{{\tt arXiv:1606.00837}}].

\bibitem{Bonvini:2015ira}
M.~Bonvini, S.~Marzani, J.~Rojo, L.~Rottoli, M.~Ubiali, R.~D. Ball, V.~Bertone,
  S.~Carrazza, and N.~P. Hartland, {\it {Parton distributions with threshold
  resummation}},  {\em JHEP} {\bf 09} (2015) 191,
  [\href{http://arxiv.org/abs/1507.01006}{{\tt arXiv:1507.01006}}].

\bibitem{Corcella:2005us}
G.~Corcella and L.~Magnea, {\it {Soft-gluon resummation effects on parton
  distributions}},  {\em Phys. Rev.} {\bf D72} (2005) 074017,
  [\href{http://arxiv.org/abs/hep-ph/0506278}{{\tt hep-ph/0506278}}].

\bibitem{Beenakker:2015rna}
W.~Beenakker, C.~Borschensky, M.~KrŠmer, A.~Kulesza, E.~Laenen, S.~Marzani, and
  J.~Rojo, {\it {NLO+NLL squark and gluino production cross-sections with
  threshold-improved parton distributions}},  {\em Eur. Phys. J.} {\bf C76}
  (2016), no.~2 53, [\href{http://arxiv.org/abs/1510.00375}{{\tt
  arXiv:1510.00375}}].

\bibitem{Korchemsky:1988si}
G.~P. Korchemsky, {\it {Asymptotics of the Altarelli-Parisi-Lipatov Evolution
  Kernels of Parton Distributions}},  {\em Mod. Phys. Lett.} {\bf A4} (1989)
  1257--1276.

\bibitem{Albino:2000cp}
S.~Albino and R.~D. Ball, {\it {Soft resummation of quark anomalous dimensions
  and coefficient functions in MS-bar factorization}},  {\em Phys. Lett.} {\bf
  B513} (2001) 93--102, [\href{http://arxiv.org/abs/hep-ph/0011133}{{\tt
  hep-ph/0011133}}].

\bibitem{Lipatov:1976zz}
L.~N. Lipatov, {\it {Reggeization of the Vector Meson and the Vacuum
  Singularity in Nonabelian Gauge Theories}},  {\em Sov. J. Nucl. Phys.} {\bf
  23} (1976) 338--345.

\bibitem{Fadin:1975cb}
V.~S. Fadin, E.~Kuraev, and L.~Lipatov, {\it {On the Pomeranchuk Singularity in
  Asymptotically Free Theories}},  {\em Phys.Lett.} {\bf B60} (1975) 50--52.

\bibitem{Kuraev:1976ge}
E.~A. Kuraev, L.~N. Lipatov, and V.~S. Fadin, {\it {Multi - Reggeon Processes
  in the Yang-Mills Theory}},  {\em Sov.Phys.JETP} {\bf 44} (1976) 443--450.

\bibitem{Kuraev:1977fs}
E.~A. Kuraev, L.~N. Lipatov, and V.~S. Fadin, {\it {The Po\-me\-ran\-chuk
  Singularity in Nonabelian Gauge Theories}},  {\em Sov. Phys. JETP} {\bf 45}
  (1977) 199--204.

\bibitem{Balitsky:1978ic}
I.~I. Balitsky and L.~N. Lipatov, {\it {The Pomeranchuk Singularity in Quantum
  Chromodynamics}},  {\em Sov. J. Nucl. Phys.} {\bf 28} (1978) 822--829.

\bibitem{Fadin:1998py}
V.~S. Fadin and L.~Lipatov, {\it {BFKL pomeron in the next-to-leading
  approximation}},  {\em Phys.Lett.} {\bf B429} (1998) 127--134,
  [\href{http://arxiv.org/abs/hep-ph/9802290}{{\tt hep-ph/9802290}}].

\bibitem{Salam:1998tj}
G.~Salam, {\it {A Resummation of large subleading corrections at small x}},
  {\em JHEP} {\bf 9807} (1998) 019,
  [\href{http://arxiv.org/abs/hep-ph/9806482}{{\tt hep-ph/9806482}}].

\bibitem{Ciafaloni:1999yw}
M.~Ciafaloni, D.~Colferai, and G.~Salam, {\it {Renormalization group improved
  small x equation}},  {\em Phys.Rev.} {\bf D60} (1999) 114036,
  [\href{http://arxiv.org/abs/hep-ph/9905566}{{\tt hep-ph/9905566}}].

\bibitem{Ciafaloni:2003kd}
M.~Ciafaloni, D.~Colferai, G.~P. Salam, and A.~M. Stasto, {\it {The Gluon
  splitting function at moderately small x}},  {\em Phys. Lett.} {\bf B587}
  (2004) 87--94, [\href{http://arxiv.org/abs/hep-ph/0311325}{{\tt
  hep-ph/0311325}}].

\bibitem{Ciafaloni:2003rd}
M.~Ciafaloni, D.~Colferai, G.~Salam, and A.~Stasto, {\it {Renormalization group
  improved small x Green's function}},  {\em Phys.Rev.} {\bf D68} (2003)
  114003, [\href{http://arxiv.org/abs/hep-ph/0307188}{{\tt hep-ph/0307188}}].

\bibitem{Ciafaloni:2007gf}
M.~Ciafaloni, D.~Colferai, G.~Salam, and A.~Stasto, {\it {A Matrix formulation
  for small-$x$ singlet evolution}},  {\em JHEP} {\bf 0708} (2007) 046,
  [\href{http://arxiv.org/abs/0707.1453}{{\tt arXiv:0707.1453}}].

\bibitem{Ball:1995vc}
R.~D. Ball and S.~Forte, {\it {Summation of leading logarithms at small x}},
  {\em Phys.Lett.} {\bf B351} (1995) 313--324,
  [\href{http://arxiv.org/abs/hep-ph/9501231}{{\tt hep-ph/9501231}}].

\bibitem{Ball:1997vf}
R.~D. Ball and S.~Forte, {\it {Asymptotically free partons at high-energy}},
  {\em Phys.Lett.} {\bf B405} (1997) 317--326,
  [\href{http://arxiv.org/abs/hep-ph/9703417}{{\tt hep-ph/9703417}}].

\bibitem{Altarelli:2001ji}
G.~Altarelli, R.~D. Ball, and S.~Forte, {\it {Factorization and resummation of
  small x scaling violations with running coupling}},  {\em Nucl.Phys.} {\bf
  B621} (2002) 359--387, [\href{http://arxiv.org/abs/hep-ph/0109178}{{\tt
  hep-ph/0109178}}].

\bibitem{Altarelli:2003hk}
G.~Altarelli, R.~D. Ball, and S.~Forte, {\it {An Anomalous dimension for small
  x evolution}},  {\em Nucl.Phys.} {\bf B674} (2003) 459--483,
  [\href{http://arxiv.org/abs/hep-ph/0306156}{{\tt hep-ph/0306156}}].

\bibitem{Altarelli:2005ni}
G.~Altarelli, R.~D. Ball, and S.~Forte, {\it {Perturbatively stable resummed
  small x evolution kernels}},  {\em Nucl.Phys.} {\bf B742} (2006) 1--40,
  [\href{http://arxiv.org/abs/hep-ph/0512237}{{\tt hep-ph/0512237}}].

\bibitem{Altarelli:2008aj}
G.~Altarelli, R.~D. Ball, and S.~Forte, {\it {Small x Resummation with Quarks:
  Deep-Inelastic Scattering}},  {\em Nucl.Phys.} {\bf B799} (2008) 199--240,
  [\href{http://arxiv.org/abs/0802.0032}{{\tt arXiv:0802.0032}}].

\bibitem{Thorne:1999sg}
R.~S. Thorne, {\it {Explicit calculation of the running coupling BFKL anomalous
  dimension}},  {\em Phys. Lett.} {\bf B474} (2000) 372--384,
  [\href{http://arxiv.org/abs/hep-ph/9912284}{{\tt hep-ph/9912284}}].

\bibitem{Thorne:1999rb}
R.~S. Thorne, {\it {NLO BFKL equation, running coupling and renormalization
  scales}},  {\em Phys. Rev.} {\bf D60} (1999) 054031,
  [\href{http://arxiv.org/abs/hep-ph/9901331}{{\tt hep-ph/9901331}}].

\bibitem{Thorne:2001nr}
R.~S. Thorne, {\it {The Running coupling BFKL anomalous dimensions and
  splitting functions}},  {\em Phys. Rev.} {\bf D64} (2001) 074005,
  [\href{http://arxiv.org/abs/hep-ph/0103210}{{\tt hep-ph/0103210}}].

\bibitem{White:2006yh}
C.~D. White and R.~S. Thorne, {\it {A Global Fit to Scattering Data with NLL
  BFKL Resummations}},  {\em Phys. Rev.} {\bf D75} (2007) 034005,
  [\href{http://arxiv.org/abs/hep-ph/0611204}{{\tt hep-ph/0611204}}].

\bibitem{Rothstein:2016bsq}
I.~Z. Rothstein and I.~W. Stewart, {\it {An Effective Field Theory for Forward
  Scattering and Factorization Violation}},
  \href{http://arxiv.org/abs/1601.04695}{{\tt arXiv:1601.04695}}.

\bibitem{Catani:1990xk}
S.~Catani, M.~Ciafaloni, and F.~Hautmann, {\it {Gluon contributions to
  small-$x$ heavy flavor production}},  {\em Phys.Lett.} {\bf B242} (1990) 97.

\bibitem{Catani:1990eg}
S.~Catani, M.~Ciafaloni, and F.~Hautmann, {\it {High energy factorization and
  small-$x$ heavy flavour production}},  {\em Nucl. Phys.} {\bf B366} (1991)
  135--188.

\bibitem{Collins:1991ty}
J.~C. Collins and R.~K. Ellis, {\it {Heavy quark production in very high energy
  hadron collisions}},  {\em Nucl. Phys.} {\bf B360} (1991) 3--30.

\bibitem{Catani:1993ww}
S.~Catani, M.~Ciafaloni, and F.~Hautmann, {\it {High-energy factorization in
  QCD and minimal subtraction scheme}},  {\em Phys.Lett.} {\bf B307} (1993)
  147--153.

\bibitem{Catani:1993rn}
S.~Catani and F.~Hautmann, {\it {Quark anomalous dimensions at small x}},  {\em
  Phys.Lett.} {\bf B315} (1993) 157--163.

\bibitem{Catani:1994sq}
S.~Catani and F.~Hautmann, {\it {High-energy factorization and small x deep
  inelastic scattering beyond leading order}},  {\em Nucl.Phys.} {\bf B427}
  (1994) 475--524, [\href{http://arxiv.org/abs/hep-ph/9405388}{{\tt
  hep-ph/9405388}}].

\bibitem{Ball:2007ra}
R.~D. Ball, {\it {Resummation of Hadroproduction Cross-sections at High
  Energy}},  {\em Nucl.Phys.} {\bf B796} (2008) 137--183,
  [\href{http://arxiv.org/abs/0708.1277}{{\tt arXiv:0708.1277}}].

\bibitem{Caola:2010kv}
F.~Caola, S.~Forte, and S.~Marzani, {\it {Small x resummation of rapidity
  distributions: The Case of Higgs production}},  {\em Nucl.Phys.} {\bf B846}
  (2011) 167--211, [\href{http://arxiv.org/abs/1010.2743}{{\tt
  arXiv:1010.2743}}].

\bibitem{Ball:2001pq}
R.~Ball and R.~K. Ellis, {\it {Heavy quark production at high-energy}},  {\em
  JHEP} {\bf 0105} (2001) 053, [\href{http://arxiv.org/abs/hep-ph/0101199}{{\tt
  hep-ph/0101199}}].

\bibitem{Diana:2009xv}
G.~Diana, {\it {High-energy resummation in direct photon production}},  {\em
  Nucl. Phys.} {\bf B824} (2010) 154--167,
  [\href{http://arxiv.org/abs/0906.4159}{{\tt arXiv:0906.4159}}].

\bibitem{Diana:2010ef}
G.~Diana, J.~Rojo, and R.~D. Ball, {\it {High energy resummation of direct
  photon production at hadronic colliders}},  {\em Phys.Lett.} {\bf B693}
  (2010) 430--437, [\href{http://arxiv.org/abs/1006.4250}{{\tt
  arXiv:1006.4250}}].

\bibitem{Marzani:2008uh}
S.~Marzani and R.~D. Ball, {\it {High Energy Resummation of Drell-Yan
  Processes}},  {\em Nucl.Phys.} {\bf B814} (2009) 246--264,
  [\href{http://arxiv.org/abs/0812.3602}{{\tt arXiv:0812.3602}}].

\bibitem{Hautmann:2002tu}
F.~Hautmann, {\it {Heavy top limit and double logarithmic contributions to
  Higgs production at $m_H^2/s$ much less than 1}},  {\em Phys.Lett.} {\bf
  B535} (2002) 159--162, [\href{http://arxiv.org/abs/hep-ph/0203140}{{\tt
  hep-ph/0203140}}].

\bibitem{Marzani:2008az}
S.~Marzani, R.~D. Ball, V.~Del~Duca, S.~Forte, and A.~Vicini, {\it {Higgs
  production via gluon-gluon fusion with finite top mass beyond next-to-leading
  order}},  {\em Nucl.Phys.} {\bf B800} (2008) 127--145,
  [\href{http://arxiv.org/abs/0801.2544}{{\tt arXiv:0801.2544}}].

\bibitem{Caola:2011wq}
F.~Caola and S.~Marzani, {\it {Finite fermion mass effects in pseudoscalar
  Higgs production via gluon-gluon fusion}},  {\em Phys.Lett.} {\bf B698}
  (2011) 275--283, [\href{http://arxiv.org/abs/1101.3975}{{\tt
  arXiv:1101.3975}}].

\bibitem{Forte:2015gve}
S.~Forte and C.~Muselli, {\it {High energy resummation of transverse momentum
  distributions:Higgs in gluon fusion}},
  \href{http://arxiv.org/abs/1511.05561}{{\tt arXiv:1511.05561}}.

\bibitem{Marzani:2015oyb}
S.~Marzani, {\it {Combining $Q_T$ and small-$x$ resummations}},  {\em Phys.
  Rev.} {\bf D93} (2016), no.~5 054047,
  [\href{http://arxiv.org/abs/1511.06039}{{\tt arXiv:1511.06039}}].

\bibitem{Bonvini:2016wki}
M.~Bonvini, S.~Marzani, and T.~Peraro, {\it {Small-$x$ resummation from HELL}},
   {\em Eur. Phys. J.} {\bf C76} (2016), no.~11 597,
  [\href{http://arxiv.org/abs/1607.02153}{{\tt arXiv:1607.02153}}].

\bibitem{Abramowicz:2015mha}
{\bf ZEUS, H1} Collaboration, H.~Abramowicz et~al., {\it {Combination of
  measurements of inclusive deep inelastic ${e^{\pm }p}$ scattering cross
  sections and QCD analysis of HERA data}},  {\em Eur. Phys. J.} {\bf C75}
  (2015), no.~12 580, [\href{http://arxiv.org/abs/1506.06042}{{\tt
  arXiv:1506.06042}}].

\bibitem{White:2006xv}
C.~D. White and R.~S. Thorne, {\it {A Variable flavor number scheme for heavy
  quark production at small x}},  {\em Phys. Rev.} {\bf D74} (2006) 014002,
  [\href{http://arxiv.org/abs/hep-ph/0603030}{{\tt hep-ph/0603030}}].

\bibitem{Caola:2009iy}
F.~Caola, S.~Forte, and J.~Rojo, {\it {Deviations from NLO QCD evolution in
  inclusive HERA data}},  {\em Phys. Lett.} {\bf B686} (2010) 127--135,
  [\href{http://arxiv.org/abs/0910.3143}{{\tt arXiv:0910.3143}}].

\bibitem{Dittmar:2005ed}
M.~Dittmar et~al., {\it {Working Group I: Parton distributions: Summary report
  for the HERA LHC Workshop Proceedings}},
  \href{http://arxiv.org/abs/hep-ph/0511119}{{\tt hep-ph/0511119}}.

\bibitem{Rottoli:2017ifw}
L.~Rottoli and M.~Bonvini, {\it {Towards small-$x$ resummed parton distribution
  functions}},  2017.
\newblock \href{http://arxiv.org/abs/1707.01535}{{\tt arXiv:1707.01535}}.

\bibitem{CooperSarkar:2011pa}
A.~Cooper-Sarkar, P.~Mertsch, and S.~Sarkar, {\it {The high energy neutrino
  cross-section in the Standard Model and its uncertainty}},  {\em JHEP} {\bf
  08} (2011) 042, [\href{http://arxiv.org/abs/1106.3723}{{\tt
  arXiv:1106.3723}}].

\bibitem{Gauld:2015kvh}
R.~Gauld, J.~Rojo, L.~Rottoli, S.~Sarkar, and J.~Talbert, {\it {The prompt
  atmospheric neutrino flux in the light of LHCb}},  {\em JHEP} {\bf 02} (2016)
  130, [\href{http://arxiv.org/abs/1511.06346}{{\tt arXiv:1511.06346}}].

\bibitem{Gauld:2015yia}
R.~Gauld, J.~Rojo, L.~Rottoli, and J.~Talbert, {\it {Charm production in the
  forward region: constraints on the small-x gluon and backgrounds for neutrino
  astronomy}},  {\em JHEP} {\bf 11} (2015) 009,
  [\href{http://arxiv.org/abs/1506.08025}{{\tt arXiv:1506.08025}}].

\bibitem{Buza:1995ie}
M.~Buza, Y.~Matiounine, J.~Smith, R.~Migneron, and W.~L. van Neerven, {\it
  {Heavy quark coefficient functions at asymptotic values $Q^2 \gg m^2$}},
  {\em Nucl. Phys.} {\bf B472} (1996) 611--658,
  [\href{http://arxiv.org/abs/hep-ph/9601302}{{\tt hep-ph/9601302}}].

\bibitem{Blumlein:1995jp}
J.~Bl{\"u}mlein and A.~Vogt, {\it {On the behavior of nonsinglet structure
  functions at small x}},  {\em Phys. Lett.} {\bf B370} (1996) 149--155,
  [\href{http://arxiv.org/abs/hep-ph/9510410}{{\tt hep-ph/9510410}}].

\bibitem{Moch:2004xu}
S.~Moch, J.~A.~M. Vermaseren, and A.~Vogt, {\it {The Longitudinal structure
  function at the third order}},  {\em Phys. Lett.} {\bf B606} (2005) 123--129,
  [\href{http://arxiv.org/abs/hep-ph/0411112}{{\tt hep-ph/0411112}}].

\bibitem{Vermaseren:2005qc}
J.~A.~M. Vermaseren, A.~Vogt, and S.~Moch, {\it {The Third-order QCD
  corrections to deep-inelastic scattering by photon exchange}},  {\em Nucl.
  Phys.} {\bf B724} (2005) 3--182,
  [\href{http://arxiv.org/abs/hep-ph/0504242}{{\tt hep-ph/0504242}}].

\bibitem{Moch:2008fj}
S.~Moch, J.~A.~M. Vermaseren, and A.~Vogt, {\it {Third-order QCD corrections to
  the charged-current structure function F(3)}},  {\em Nucl. Phys.} {\bf B813}
  (2009) 220--258, [\href{http://arxiv.org/abs/0812.4168}{{\tt
  arXiv:0812.4168}}].

\bibitem{Ball:2015dpa}
R.~D. Ball, M.~Bonvini, and L.~Rottoli, {\it {Charm in Deep-Inelastic
  Scattering}},  {\em JHEP} {\bf 11} (2015) 122,
  [\href{http://arxiv.org/abs/1510.02491}{{\tt arXiv:1510.02491}}].

\bibitem{Aivazis:1993kh}
M.~A.~G. Aivazis, F.~I. Olness, and W.-K. Tung, {\it {Leptoproduction of heavy
  quarks. 1. General formalism and kinematics of charged current and neutral
  current production processes}},  {\em Phys. Rev.} {\bf D50} (1994)
  3085--3101, [\href{http://arxiv.org/abs/hep-ph/9312318}{{\tt
  hep-ph/9312318}}].

\bibitem{Aivazis:1993pi}
M.~A.~G. Aivazis, J.~C. Collins, F.~I. Olness, and W.-K. Tung, {\it
  {Leptoproduction of heavy quarks. 2. A Unified QCD formulation of charged and
  neutral current processes from fixed target to collider energies}},  {\em
  Phys.Rev.} {\bf D50} (1994) 3102--3118,
  [\href{http://arxiv.org/abs/hep-ph/9312319}{{\tt hep-ph/9312319}}].

\bibitem{Collins:1997sr}
J.~C. Collins, {\it {Proof of factorization for diffractive hard scattering}},
  {\em Phys. Rev.} {\bf D57} (1998) 3051--3056,
  [\href{http://arxiv.org/abs/hep-ph/9709499}{{\tt hep-ph/9709499}}]. [Erratum:
  Phys. Rev. D61 (2000) 019902].

\bibitem{Kramer:2000hn}
M.~Kramer, F.~I. Olness, and D.~E. Soper, {\it {Treatment of heavy quarks in
  deeply inelastic scattering}},  {\em Phys. Rev.} {\bf D62} (2000) 096007,
  [\href{http://arxiv.org/abs/hep-ph/0003035}{{\tt hep-ph/0003035}}].

\bibitem{Thorne:1997ga}
R.~S. Thorne and R.~G. Roberts, {\it {An Ordered analysis of heavy flavor
  production in deep inelastic scattering}},  {\em Phys. Rev.} {\bf D57} (1998)
  6871--6898, [\href{http://arxiv.org/abs/hep-ph/9709442}{{\tt
  hep-ph/9709442}}].

\bibitem{Thorne:2006qt}
R.~Thorne, {\it {A Variable-flavor number scheme for NNLO}},  {\em Phys.Rev.}
  {\bf D73} (2006) 054019, [\href{http://arxiv.org/abs/hep-ph/0601245}{{\tt
  hep-ph/0601245}}].

\bibitem{Buza:1996wv}
M.~Buza, Y.~Matiounine, J.~Smith, and W.~L. van Neerven, {\it {Charm
  electroproduction viewed in the variable flavor number scheme versus fixed
  order perturbation theory}},  {\em Eur. Phys. J.} {\bf C1} (1998) 301--320,
  [\href{http://arxiv.org/abs/hep-ph/9612398}{{\tt hep-ph/9612398}}].

\bibitem{Cacciari:1998it}
M.~Cacciari, M.~Greco, and P.~Nason, {\it {The $p_T$ spectrum in heavy-flavour
  hadroproduction}},  {\em JHEP} {\bf 05} (1998) 007,
  [\href{http://arxiv.org/abs/hep-ph/9803400}{{\tt hep-ph/9803400}}].

\bibitem{Forte:2010ta}
S.~Forte, E.~Laenen, P.~Nason, and J.~Rojo, {\it {Heavy quarks in
  deep-inelastic scattering}},  {\em Nucl. Phys.} {\bf B834} (2010) 116--162,
  [\href{http://arxiv.org/abs/1001.2312}{{\tt arXiv:1001.2312}}].

\bibitem{Collins:1998rz}
J.~C. Collins, {\it {Hard scattering factorization with heavy quarks: A General
  treatment}},  {\em Phys. Rev.} {\bf D58} (1998) 094002,
  [\href{http://arxiv.org/abs/hep-ph/9806259}{{\tt hep-ph/9806259}}].

\bibitem{Guzzi:2011ew}
M.~Guzzi, P.~M. Nadolsky, H.-L. Lai, and C.-P. Yuan, {\it {General-Mass
  Treatment for Deep Inelastic Scattering at Two-Loop Accuracy}},  {\em
  Phys.Rev.} {\bf D86} (2012) 053005,
  [\href{http://arxiv.org/abs/1108.5112}{{\tt arXiv:1108.5112}}].

\bibitem{Thorne:2008xf}
R.~S. Thorne and W.~K. Tung, {\it {PQCD Formulations with Heavy Quark Masses
  and Global Analysis}},  \href{http://arxiv.org/abs/0809.0714}{{\tt
  arXiv:0809.0714}}.

\bibitem{Bonvini:2015pxa}
M.~Bonvini, A.~S. Papanastasiou, and F.~J. Tackmann, {\it {Resummation and
  Matching of $b$-quark Mass Effects in $b\bar{b}H$ Production}},
  \href{http://arxiv.org/abs/1508.03288}{{\tt arXiv:1508.03288}}.

\bibitem{Ball:2015tna}
R.~D. Ball, V.~Bertone, M.~Bonvini, S.~Forte, P.~G. Merrild, J.~Rojo, and
  L.~Rottoli, {\it {Intrinsic charm in a matched general-mass scheme}},
  \href{http://arxiv.org/abs/1510.00009}{{\tt arXiv:1510.00009}}.

\bibitem{Catani:1992zc}
S.~Catani, M.~Ciafaloni, and F.~Hautmann, {\it {Production of heavy flavors at
  high-energies}},  in {\em {Workshop on Physics at HERA Hamburg, Germany,
  October 29-30, 1991}}, pp.~0690--711, 1992.

\bibitem{Bertone:2013vaa}
V.~Bertone, S.~Carrazza, and J.~Rojo, {\it {APFEL: A PDF Evolution Library with
  QED corrections}},  {\em Comput.Phys.Commun.} {\bf 185} (2014) 1647--1668,
  [\href{http://arxiv.org/abs/1310.1394}{{\tt arXiv:1310.1394}}].

\bibitem{Ciafaloni:2005cg}
M.~Ciafaloni and D.~Colferai, {\it {Dimensional regularisation and
  factorisation schemes in the BFKL equation at subleading level}},  {\em JHEP}
  {\bf 09} (2005) 069, [\href{http://arxiv.org/abs/hep-ph/0507106}{{\tt
  hep-ph/0507106}}].

\bibitem{Marzani:2007gk}
S.~Marzani, R.~D. Ball, P.~Falgari, and S.~Forte, {\it {BFKL at
  next-to-next-to-leading order}},  {\em Nucl. Phys.} {\bf B783} (2007)
  143--175, [\href{http://arxiv.org/abs/0704.2404}{{\tt arXiv:0704.2404}}].

\bibitem{Catani:1996ny}
S.~Catani, {\it {$k_t$ factorization and perturbative invariants at small x}},
  in {\em {Proceedings, 4th International Workshop on Deep inelastic scattering
  and related phenomena (DIS 96)}}, 1996.
\newblock \href{http://arxiv.org/abs/hep-ph/9608310}{{\tt hep-ph/9608310}}.

\bibitem{Bottazzi:1998rs}
G.~Bottazzi, G.~Marchesini, G.~P. Salam, and M.~Scorletti, {\it {Small x one
  particle inclusive quantities in the CCFM approach}},  {\em JHEP} {\bf 12}
  (1998) 011, [\href{http://arxiv.org/abs/hep-ph/9810546}{{\tt
  hep-ph/9810546}}].

\bibitem{Ablinger:2014nga}
J.~Ablinger, A.~Behring, J.~Bl{\"u}mlein, A.~De~Freitas, A.~von Manteuffel, and
  C.~Schneider, {\it {The 3-loop pure singlet heavy flavor contributions to the
  structure function $F_2(x,Q^2)$ and the anomalous dimension}},  {\em Nucl.
  Phys.} {\bf B890} (2014) 48--151, [\href{http://arxiv.org/abs/1409.1135}{{\tt
  arXiv:1409.1135}}].

\bibitem{Behring:2014eya}
A.~Behring, I.~Bierenbaum, J.~Bl{\"u}mlein, A.~De~Freitas, S.~Klein, and
  F.~Wi§brock, {\it {The logarithmic contributions to the $O(\alpha^3_s)$
  asymptotic massive Wilson coefficients and operator matrix elements in deeply
  inelastic scattering}},  {\em Eur. Phys. J.} {\bf C74} (2014), no.~9 3033,
  [\href{http://arxiv.org/abs/1403.6356}{{\tt arXiv:1403.6356}}].

\bibitem{Tung:2001mv}
W.-K. Tung, S.~Kretzer, and C.~Schmidt, {\it {Open heavy flavor production in
  QCD: Conceptual framework and implementation issues}},  {\em J. Phys.} {\bf
  G28} (2002) 983--996, [\href{http://arxiv.org/abs/hep-ph/0110247}{{\tt
  hep-ph/0110247}}].

\bibitem{Salam:1999cn}
G.~P. Salam, {\it {An Introduction to leading and next-to-leading BFKL}},  {\em
  Acta Phys. Polon.} {\bf B30} (1999) 3679--3705,
  [\href{http://arxiv.org/abs/hep-ph/9910492}{{\tt hep-ph/9910492}}].

\bibitem{Bonvini:2012sh}
M.~Bonvini, {\em {Resummation of soft and hard gluon radiation in perturbative
  QCD}}.
\newblock PhD thesis, Genoa U., 2012.
\newblock \href{http://arxiv.org/abs/1212.0480}{{\tt arXiv:1212.0480}}.

\bibitem{Vogt:2004mw}
A.~Vogt, S.~Moch, and J.~A.~M. Vermaseren, {\it {The three-loop splitting
  functions in QCD: The singlet case}},  {\em Nucl. Phys.} {\bf B691} (2004)
  129--181, [\href{http://arxiv.org/abs/hep-ph/0404111}{{\tt hep-ph/0404111}}].

\bibitem{Altarelli:1999vw}
G.~Altarelli, R.~D. Ball, and S.~Forte, {\it {Resummation of singlet parton
  evolution at small x}},  {\em Nucl. Phys.} {\bf B575} (2000) 313--329,
  [\href{http://arxiv.org/abs/hep-ph/9911273}{{\tt hep-ph/9911273}}].

\bibitem{Forte:2010dt}
S.~Forte, {\it {Parton distributions at the dawn of the LHC}},  {\em Acta Phys.
  Polon.} {\bf B41} (2010) 2859--2920,
  [\href{http://arxiv.org/abs/1011.5247}{{\tt arXiv:1011.5247}}].

\bibitem{Forte:2013wc}
S.~Forte and G.~Watt, {\it {Progress in the Determination of the Partonic
  Structure of the Proton}},  {\em Ann.Rev.Nucl.Part.Sci.} {\bf 63} (2013)
  291--328, [\href{http://arxiv.org/abs/1301.6754}{{\tt arXiv:1301.6754}}].

\bibitem{Ball:2017nwa}
{\bf NNPDF} Collaboration, R.~D. Ball et~al., {\it {Parton distributions from
  high-precision collider data}},  \href{http://arxiv.org/abs/1706.00428}{{\tt
  arXiv:1706.00428}}.

\bibitem{Ball:2015oha}
R.~D. Ball, {\it {Global Parton Distributions for the LHC Run II}},  in {\em
  {29th Rencontres de Physique de La Vallee d'Aoste La Thuile, Aosta, Italy,
  March 1-7, 2015}}, 2015.
\newblock \href{http://arxiv.org/abs/1507.07891}{{\tt arXiv:1507.07891}}.

\bibitem{Caola:2016upw}
F.~Caola, S.~Forte, S.~Marzani, C.~Muselli, and G.~Vita, {\it {The Higgs
  transverse momentum spectrum with finite quark masses beyond leading order}},
   {\em JHEP} {\bf 08} (2016) 150, [\href{http://arxiv.org/abs/1606.04100}{{\tt
  arXiv:1606.04100}}].

\bibitem{Buza:1997mg}
M.~Buza and W.~L. van Neerven, {\it {$ O(\alpha_s^2)$ contributions to charm
  production in charged-current deep-inelastic lepton hadron scattering}},
  {\em Nucl. Phys.} {\bf B500} (1997) 301--324,
  [\href{http://arxiv.org/abs/hep-ph/9702242}{{\tt hep-ph/9702242}}].

\bibitem{Blumlein:2014fqa}
J.~Bl{\"u}mlein, A.~Hasselhuhn, and T.~Pfoh, {\it {The $O(\alpha_s^2)$ heavy
  quark corrections to charged current deep-inelastic scattering at large
  virtualities}},  {\em Nucl. Phys.} {\bf B881} (2014) 1--41,
  [\href{http://arxiv.org/abs/1401.4352}{{\tt arXiv:1401.4352}}].

\bibitem{Gluck:1987uk}
M.~Gluck, R.~M. Godbole, and E.~Reya, {\it {HEAVY FLAVOR PRODUCTION AT
  HIGH-ENERGY e p COLLIDERS}},  {\em Z. Phys.} {\bf C38} (1988) 441. [Erratum:
  Z. Phys.C39,590(1988)].

\bibitem{Gluck:1996ve}
M.~Gluck, S.~Kretzer, and E.~Reya, {\it {The Strange Sea Density and Charm
  Production in Deep Inelastic Charged Current Processes}},  {\em Phys. Lett.}
  {\bf B380} (1996) 171--176, [\href{http://arxiv.org/abs/hep-ph/9603304}{{\tt
  hep-ph/9603304}}].

\bibitem{Laenen:1992zk}
E.~Laenen, S.~Riemersma, J.~Smith, and W.~L. van Neerven, {\it {Complete $O
  (\alpha_s)$ corrections to heavy flavor structure functions in
  electroproduction}},  {\em Nucl. Phys.} {\bf B392} (1993) 162--228.

\bibitem{Berger:2016inr}
E.~L. Berger, J.~Gao, C.~S. Li, Z.~L. Liu, and H.~X. Zhu, {\it {Charm-Quark
  Production in Deep-Inelastic Neutrino Scattering at Next-to-Next-to-Leading
  Order in QCD}},  {\em Phys. Rev. Lett.} {\bf 116} (2016), no.~21 212002,
  [\href{http://arxiv.org/abs/1601.05430}{{\tt arXiv:1601.05430}}].

\end{thebibliography}\endgroup

\end{document}